\documentclass[aps,twocolumn,prd,floatfix,preprintnumbers, nofootinbib,10.5pt]{revtex4-2}
\usepackage{xpatch}
\makeatletter
\patchcmd{\@ssect@ltx}
    {\addcontentsline{toc}{#1}{\protect\numberline{}#8}}
    {}
    {}
    {}
\makeatother
\pdfoutput=1
\usepackage{color}
\usepackage{enumitem}
\usepackage{booktabs}
\usepackage{mathtools}
\usepackage[dvipsnames]{xcolor}
\usepackage{float}
\usepackage{stmaryrd}

\usepackage{array}[=2016-10-06]
\usepackage{hyperref}
\usepackage{cleveref}
\definecolor{amaranth}{rgb}{0.9, 0.17, 0.31}
\definecolor{forestForestGreen(web)}{rgb}{0.13, 0.55, 0.13}
\definecolor{SymTFTColores}{rgb}{0, 0.33, 70.1}
\definecolor{blue(munsell)}{HTML}{005567}
\definecolor{oxfordblue}{rgb}{0.0, 0.2, 0.4}
\definecolor{bblue}{rgb}{0.0, 0.58, 0.71}
\hypersetup{
pdfstartview={FitH}, 
pdftitle={}, 
colorlinks=true, 
linkcolor=oxfordblue, 
citecolor=oxfordblue, 
filecolor=oxfordblue, 
urlcolor=oxfordblue
}
\usepackage{pgfplots}
\pgfplotsset{compat=1.18}
\usepackage{amssymb}
\usepackage{amsfonts}
\usepackage{graphicx}
\usepackage{epstopdf}
\usepackage{dcolumn}
\usepackage{amsmath}
\usepackage{latexsym,bm}
\usepackage{amsthm}
\usepackage{slashed}
\usepackage{color}
\usepackage{url}
\usepackage{rotating}
\usepackage[normalem]{ulem}
\usepackage{faktor}
\usepackage{tikz-cd}
\usepackage{tensor}
\usepackage{array}
\usepackage{multirow}
\usepackage{comment}

\usepackage{tabularx}
\usepackage{makecell}
\usepackage[normalem]{ulem}
\usepackage{changepage}
\numberwithin{equation}{section}

\usepackage{makecell}

\usepackage{tikz}
\usepackage{diagbox}
\usetikzlibrary{positioning}
\usetikzlibrary{calc}
\usetikzlibrary{decorations.pathreplacing,calligraphy}

\usepackage{xstring}
\usetikzlibrary{decorations.pathmorphing} 
\usetikzlibrary{decorations.markings} 
\usetikzlibrary{arrows} 
\usetikzlibrary{shapes} 
\usetikzlibrary{matrix} 
\usetikzlibrary{positioning} 
\usepackage[english]{babel} 
\usepackage[autostyle]{csquotes}
\usepackage{pifont}
\usetikzlibrary{shapes.multipart}

\tikzset{->-/.style={decoration={
  markings,
  mark=at position .6 with {\arrow{Latex[length=1.5mm,width=1.5mm]}}},postaction={decorate}}}

\usepackage{quantikz}

\usepackage{graphicx}
\usepackage{dcolumn}
\usepackage{bm}
\usepackage[dvipsnames]{xcolor}
\usepackage{comment}
\usepackage{bbm}

\tikzset{
  qgparent/.style={fill=blue!13},
  qgupper/.style={fill=red!15},
  qglower/.style={fill=green!17},
  qgline/.style={line width=.65pt,black},
  qgfused/.style={line width=1.1pt,densely dashed,blue!80!black},
  qgtext/.style={font=\footnotesize,inner sep=1pt},
  qgsmall/.style={font=\scriptsize,inner sep=1pt},
  qgrg/.style={circle,inner sep=2.1pt,fill=orange!90!black},
  qgmeet/.style={circle,inner sep=1.85pt,fill=black}
}

\newcommand{\Z}{\mathbb{Z}}
\newcommand{\C}{\mathbb{C}}
\newcommand{\R}{\mathbb{R}}

\DeclareMathOperator{\Hom}{Hom}
\DeclareMathOperator{\End}{End}
\DeclareMathOperator{\Rep}{Rep}
\DeclareMathOperator{\Vect}{Vec}
\DeclareMathOperator{\Mod}{Mod}
\DeclareMathOperator{\Bimod}{Bimod}

\DeclareMathOperator{\Fun}{Fun}
\DeclareMathOperator{\Mat}{Mat}

\newcommand{\one}{\mathbf{1}}
\newcommand{\Zt}{\mathbb{Z}_2}
\newcommand{\Zth}{\mathbb{Z}_3}

\makeatother

\makeatletter

\renewcommand\subsubsection{\@startsection{subsubsection}{3}{\z@}%
                                     {-3.25ex\@plus -1ex \@minus -.2ex}%
                                     {2.5ex \@plus .2ex}%
                                     {\centering\itshape\bfseries\small}}
\makeatother

\usepackage{physics}

\makeatletter
\newcommand\xlabel[2][]{\phantomsection\def\@currentlabelname{#1}\label{#2}}
\makeatother

\makeatletter
\def\l@subsubsection#1#2{}
\makeatother

\begin{document}

\title{Topological interfaces between universes and generalized families of 2d QFTs}

\author{Pinak Banerjee$^1$}
\author{Yuhan Gai$^3$}
\author{Alonso Perez-Lona$^4$}
\author{Daniel Robbins$^2$}
\author{Subham Roy$^2$}
\author{Eric Sharpe$^1$}
\author{with an appendix by Xingyang Yu$^{1, 5}$}

\affiliation{\it $^1$Department of Physics MC 0435,\\
850 West Campus Drive, Virginia Tech,  Blacksburg, VA  24061, USA \\
$^2$ Department of Physics, \\
1400 Washington Ave, SUNY Albany, NY 12222, USA \\
$^3$Mathematical Institute, University of Oxford, Woodstock Road, Oxford, OX2 6GG, United Kingdom
\\
$^4$Max Planck Institute for Mathematics in the Sciences, Inselstraße 22, 04103 Leipzig, Germany 
\\
$^5$ FirstPrinciples, Toronto, Ontario, Canada
}

\begin{abstract}
\noindent

Two-dimensional quantum field theories with topological point operators decompose into universes. Just as fusion categories have played an important role in understanding non-invertible symmetries, multi-fusion categories play an analogous role for the symmetries of such decomposing theories. We describe a finite symmetry as a multi-fusion category together with its action functor, which keeps track of trivially acting symmetries and of the topological interfaces between universes. Relative to a chosen symmetry, we classify pairs of universes by the topological interfaces it contains, which reflect the structure of the multi-fusion category: universes in different indecomposable summands are joined by no interface (Type~III), while within a summand they are related by invertible (Type~I) or only non-invertible (Type~II) interfaces. For orbifolds with a trivially acting abelian normal subgroup, we derive the full inherited symmetry, which characterizes the operators between universes found case by case by Robbins, Sharpe and Vandermeulen and yields pairs of Types~I and~II but never of Type~III. We construct Type~III pairs instead from extensions of $2$-group type and from renormalization group flows out of a common ultraviolet theory. Finally, we propose a correspondence with the generalized families of quantum field theories of Brennan and Intriligator, in which equivariant families correspond to Type~I pairs and categorical ones to Type~II pairs whose gauging preserves the symmetry, family anomalies appear as universes carrying different symmetry-protected topological phases, and Type~III pairs can realize families over a discrete parameter. An appendix by Xingyang Yu works out the higher-charge Schwinger model as a Lagrangian example.

\end{abstract}


\maketitle
\tableofcontents

\section{\label{sec:intro}Introduction}

Global symmetries are among the most powerful tools for studying quantum field theories (QFTs). They impose selection rules on correlation functions and, through 't Hooft anomaly matching, they constrain the renormalization group flows. Over the past decade the notion of a global symmetry has been generalized substantially. In the modern formulation \cite{Gaiotto:2014kfa}, symmetries correspond to topological operators. This generalization of symmetries in terms of topological operators has led to the notion of higher-form, higher-group and non-invertible symmetries \cite{Cordova:2022ruw, Schafer-Nameki:2023jdn, Brennan:2023mmt,Bhardwaj:2023ayw, Freed:2022iao,Shao:2023gho,Bah:2022xfv,Bhardwaj:2023kri,McGreevy:2022oyu,Chang:2018iay,Bhardwaj:2022yxj,Iqbal:2024pee, Costa:2024wks}.

Beyond these more familiar cases, a $d$-dimensional theory can also have a $(d-1)$-form symmetry, which is generated by topological point operators. Such a symmetry is associated with a phenomenon known as
\emph{decomposition}, see e.g.~\cite{Hellerman:2006zs, Caldararu:2010ljp, Sharpe:2014tca, Sharpe:2019ddn, Sharpe:2021srf, Sharpe:2022ene, Tanizaki:2019rbk, Nguyen:2021yld, Nguyen:2021naa, Komargodski:2020mxz, Cherman:2020cvw, Robbins:2020msp, Robbins:2021xce, Honda:2021ovk, Pantev:2022kpl, Pantev:2022pbf, Lin:2022xod, Robbins:2022wlr, Lin:2025oml,Perez-Lona:2023llv, Pantev:2023dim, Sharpe:2023lfk, Sharpe:2024ujm, Santilli:2024dyz, Pantev:2024kva, Yu:2024jtk, Najjar:2024vmm, Damia:2024kyt,Robbins:2025urk, Perez-Lona:2025add,Banerjee:2026ywy}, or \cite{Sharpe:2022ene} for a recent review. A local QFT $\mathfrak{T}$ with a $(d-1)$-form symmetry $K^{(d-1)}$ is equivalent to a direct sum of QFTs, also known as \textit{universes},
\begin{equation} \label{eq:decomp}
  \mathfrak{T} \;=\; \bigsqcup_{\chi} {\cal T}_\chi .
\end{equation}
Suitable linear combinations of the topological point operators define orthogonal projectors onto the individual universes \cite{Sharpe:2019ddn}.

One way to engineer decomposition is through symmetries that act trivially. Gauging a trivially acting $(d-2)$-form symmetry in a $d$-dimensional QFT produces a $(d-1)$-form symmetry \cite{Robbins:2022wlr, Pantev:2022kpl, Robbins:2021xce, Robbins:2021lry, Robbins:2020msp, Sharpe:2021srf, Lin:2025oml}. This connection between decomposition and trivially acting symmetries is by now very well established. Trivial actions of symmetries often show up along a renormalization group flow, deep in the infrared (IR). The operators charged under the symmetry can be absent from the low-energy description of the quantum field theory. In that case, part of the ultraviolet (UV) symmetry acts trivially on the IR degrees of freedom of the theory. In favorable finite semisimple symmetry examples, the scenario can then be captured by a short exact sequence, 
\begin{equation}
    \mathrm{Vec} \rightarrow \mathcal{A} \rightarrow \mathcal{B} \rightarrow \mathcal{B}_{\rm eff} \rightarrow \mathrm{Vec},
\end{equation}
where $\mathcal{B}$ is the UV symmetry, whose action factors through the effective IR symmetry $\mathcal{B}_{\rm eff}$, and $\mathcal{A}$ is the trivially acting kernel. The objects in the sequence can be, for example, fusion categories. Such categorical short exact sequences have already appeared in the literature \cite{Perez-Lona:2025ncg, Robbins:2025puq, Antinucci:2025fjp, Lu:2025yru}. For ordinary group-like symmetries, the data of the exact sequence of fusion categories reduce to that of a short exact sequence of finite groups. In terms of topological defects, the trivially acting symmetries can be thought of as topological defects that terminate on a higher-codimension topological operator. For example, a trivially acting $0$-form symmetry in $2d$ can be visualized as being generated by line operators that terminate on topological point operators \cite{Robbins:2022wlr, Lin:2025oml}. We use both descriptions. They are related by a trivialization, a monoidal isomorphism between the image of the trivially acting symmetry under the action functor and copies of the identity line. Its components are the topological junctions at which the trivially acting lines end. 

In a different direction, it has long been appreciated that QFTs can be organized into families, or more generally into a \textit{space of theories} \cite{Kutasov:1988xb, Douglas:2010ic, Yamazaki:2017cqu}. A QFT typically depends on parameters, such as coupling constants or $\theta$-angles. These couplings can be viewed as coordinates on the space of theories, and in this way they define families of theories. Recently, Brennan and Intriligator (BI) introduced a notion of \textit{generalized families} of QFTs \cite{Brennan:2026ira}. The idea is that a symmetry which is explicitly broken by the couplings is not lost if the couplings transform as spurions. Instead of acting on a single theory, it then relates the theories at different points of the space of theories, and in this way organizes them into a family. The members of a family can be related by invertible or by non-invertible operations. Our proposal is that the members of a finite (sub)family can appear as the universes of a single theory, related by lower-form symmetries \cite{Aloni:2024jpb, Yu:2024jtk, Najjar:2024vmm, Lin:2025oml, Robbins:2025urk, Banerjee:2026ywy, Perez-Lona:2025add, Perez-Lona:2026mlj, Khlaif:2025jnx, Najjar:2025htp, Heckman:2024oot}, and each member is recovered by undoing decomposition~\cite{Sharpe:2019ddn}.

It is known that, in a two-dimensional theory with topological point operators, the finite symmetry categories that contain these point operators are multi-fusion \cite{Bhardwaj:2023ayw,Bhardwaj:2024qrf,Perez-Lona:2025wup,Damia:2024kyt}. This is because the point operators split the identity line into the identity lines of the universes. Our starting point is then to treat a symmetry as a multi-fusion category \textit{together with} its action functor on the theory. This is what allows us to make precise both trivially acting symmetries and statements about which interfaces exist.

We then sort pairs of universes by the topological interfaces between them.\footnote{A symmetry sees only the point operators in its image, so it may split $\mathfrak T$ only into \textit{sectors}, each a sum of universes, as the center symmetry of two-dimensional $SU(2)$ Yang--Mills theory below does. The types are defined for pairs of sectors, and we call sectors universes whenever the symmetry resolves them.} \textit{Relative to a chosen symmetry}, we say that a pair of universes is of
\begin{itemize}
    \item \textbf{Type I} if the symmetry contains an invertible topological interface between them: the two universes are then isomorphic, and the interface is the isomorphism.
    \item \textbf{Type IIa} if the symmetry contains topological interfaces between them, none of them invertible, and the two universes carry tensor-equivalent symmetry categories. One universe is then a gauging of the other that returns a tensor-equivalent symmetry category.
    \item \textbf{Type IIb} if the symmetry contains topological interfaces between them but the two universes carry inequivalent symmetry categories: one universe is then a gauging of the other that changes the symmetry. The interfaces make the two symmetry categories Morita equivalent, so the symmetry topological field theory (SymTFT)\footnote{The SymTFT is a $(d+1)$-dimensional topological field theory we can associate with a $d$-dimensional quantum field theory, which encodes all the symmetry-related data of the QFT \cite{Freed:2012bs, Apruzzi:2021nmk, vanBeest:2022fss, Kaidi:2022cpf, Kaidi:2023maf, Bhardwaj:2023bbf, Bhardwaj:2023ayw, Cvetic:2024dzu, Baume:2023kkf,Heckman:2022muc, Heckman:2022xgu, Bergman:2026lnz, Robbins:2025puq}.} is shared.
    \item \textbf{Type III} if the symmetry contains no topological interface connecting them.
\end{itemize}
The type is a property of a pair, not of an individual theory, so one theory can contain pairs of every type. The types measure how universes are related, not their anomalies.

The rest of this article is organized as follows. In Section~\ref{sec:mfc}, we describe symmetries of decomposing theories as multi-fusion categories with an action functor, and in Section~\ref{sect:org} we classify pairs of universes. In Section~\ref{sec:trivial}, we derive the symmetry of orbifolds with a trivially acting abelian normal subgroup and show that, relative to it, they give Types~I and~II but never Type~III. Section~\ref{sec:typesI-II} works out examples of Types~I and~II, and Section~\ref{sec:typeIII} constructs Type~III pairs from extensions of $2$-group type, by cohomological base change, and from exact sequences of tensor categories. In Section~\ref{sec:families}, we relate the types to the families of Brennan--Intriligator \cite{Brennan:2026ira}, and we conclude in Section~\ref{sec:discussion}. 
Appendix~\ref{Karoubi} briefly reviews Karoubi completion.
Appendix~\ref{app: details_der.} collects details of derivations used in the main text.
Appendix~\ref{app:qg} reviews quarter gauging in SymTFT.
Finally, Appendix~\ref{app:schwinger} (written by X.~Yu) works out the higher-charge Schwinger model.

\section{Decomposition and multi-fusion categories}\label{sec:mfc} 

In this section, we describe the symmetry structure of a decomposing two-dimensional QFT, to which we restrict from now on. Topological lines, together with their junctions, form a rigid monoidal category. We first show how topological point operators split the identity line, and then review the resulting multi-fusion categories. We then define a symmetry as a multi-fusion category with an action functor and set up the dictionary between categorical data and universes and their interfaces.

\subsection{Decomposition}

Consider a $2d$ QFT $\mathfrak{T}$ with a finite $1$-form symmetry $K^{(1)} = BK$, generated by topological point operators $U_k$, $k\in K$. We assume for the moment that the $U_k$ generate all topological point operators of $\mathfrak T$. These operators span the group algebra $\mathbb{C}[K]$, which contains the projectors, 
\begin{equation}\label{eq:projectors}
    \Pi_\chi = \frac{1}{|K|} \sum_{k \in K} \chi(k)^{-1}U_k, \qquad  \chi \in \widehat{K},
	\end{equation}
	where $\widehat{K}={\rm Hom}(K,U(1))$, satisfying 
\begin{equation}
    \Pi_\chi\Pi_{\chi'} = \delta_{\chi\chi'} \Pi_\chi, \qquad \sum_\chi\Pi_\chi=1. 
\end{equation}
The presence of the point operators signals decomposition,
	\begin{equation}
		\mathfrak{T} = \bigsqcup_{\chi \in \widehat K} {\cal T}_\chi ,
	\end{equation}
	where $\Pi_\chi$ projects onto the universe ${\cal T}_\chi$.

A genuine topological point operator can be viewed as a topological junction of the identity line with itself, i.e.\ an element of $\mathrm{End}(\mathbbm{1})$. In a decomposing theory, the projectors $\Pi_\chi$ split the identity line itself,
\begin{equation}\label{eq:unit-splitting}
    \mathbbm{1} = \bigoplus_{\chi\in\widehat K} \mathbbm{1}_\chi, \qquad
    \mathbbm{1}_\chi := \Pi_\chi \mathbbm{1},
\end{equation}
where $\mathbbm{1}_\chi$, the identity line with the projector $\Pi_\chi$ inserted on it, is the identity line of the universe ${\cal T}_\chi$:
\begin{equation}
    \mathrm{End}(\mathbbm{1}) \cong \bigoplus_\chi \mathrm{End}(\mathbbm{1}_\chi) \cong \mathbb{C}^{\widehat{K}}\cong\mathbb{C}[K].
\end{equation}
In a non-decomposing theory, $\mathrm{End}(\mathbbm 1)=\mathbb C$ and the finite symmetries form a fusion category. In a decomposing theory, the unit is not simple, so the finite symmetry categories that contain the projectors $\Pi_\chi$ are multi-fusion but not fusion \cite[Definition 4.1.1]{etingof2015tensor}.

The observation above, that the existence of the topological point projection operators implies the splitting of the identity line operator, has a very nice mathematical description in terms of {\it{Karoubi completion}}, reviewed in Appendix~\ref{Karoubi}.  The projectors are idempotents in $\mathrm{End}(\mathbbm 1)$ and Karoubi completeness then requires that the object $\mathbbm 1$ splits.  Another perspective~\cite{Lin:2022xod} is that we can $1$-gauge the $1$-form symmetry generated by the topological point operators to create condensation defects which are the topological line operators $\mathbbm{1}_\chi$, and fusion of the point operator $\Pi_\chi$ on any line operator $L$ is equivalent to the line operator fusion of $\mathbbm{1}_\chi$ with $L$. See e.g. \cite{Gaiotto:2019xmp,Bah:2025oxi} for more on the role of Karoubi completion in the context of generalized symmetries.

An aside on notation:  we will use $\sqcup$ to denote disjoint unions of quantum field theories, $\oplus$ to denote sums of objects, and $\boxplus$ to denote analogous sums of categories.

\subsection{Review of multi-fusion categories}\label{sec:review}

A \textit{multi-fusion category} $\mathcal{C}$ is a rigid semisimple $\mathbb{C}$-linear monoidal category with finitely many simple objects and finite-dimensional hom-spaces \cite[Definition 4.1.1]{etingof2015tensor}. In a general multi-fusion category, the monoidal unit decomposes as
\begin{equation}
\mathbbm{1}\cong \oplus_i \mathbbm{1}_i,
\end{equation} where the $\mathbbm{1}_i$ are simple. If the monoidal unit $\mathbbm{1}$ is a simple object, then the multi-fusion category is a fusion category. 

As reviewed in \cite[Section 4.3]{etingof2015tensor}, because the $\mathbbm{1}_i$ are orthogonal idempotents for the tensor product, any multi-fusion category $\mathcal{C}$ admits a matrix-like decomposition. As an abelian category, $\cal C$ can be written as 
\begin{equation}
    \mathcal{C}=\boxplus_{i,j}\mathcal{C}_{ij},
\end{equation}
where
\begin{equation}
    \mathcal{C}_{ij} := \mathbbm{1}_i\otimes \mathcal{C}\otimes \mathbbm{1}_j,
\end{equation}
and we have the fusion rule 
\begin{equation}
   \mathcal{C}_{ij} \otimes \mathcal{C}_{kl}\subseteq \mathcal{C}_{il}, \, \text{when} \,\, j=k, \, \text{otherwise} \, 0.
\end{equation}  
The multi-fusion category $\mathcal{C}$ can be thought of as a finite matrix of abelian categories
\begin{equation}
    \mathcal{C}=\begin{pmatrix}
    \mathcal{C}_{11} & \mathcal{C}_{12} & \mathcal{C}_{13} & \cdots \\
    \mathcal{C}_{21} & \mathcal{C}_{22} & \mathcal{C}_{23} & \cdots \\
    \mathcal{C}_{31} & \mathcal{C}_{32} & \mathcal{C}_{33} & \cdots \\
    \cdots  &\cdots & \cdots & \cdots
    \end{pmatrix},
\end{equation}
where each diagonal entry $\mathcal{C}_{ii}$ corresponds to a fusion category, and the off-diagonal ones $\mathcal{C}_{ij}$ are $(\mathcal{C}_{ii},\mathcal{C}_{jj})$-\textit{bimodule} categories.\footnote{A left $\mathcal C$-module category ${}_{\mathcal C}\mathcal M$ is a category $\mathcal M$ with a functor $\triangleright:\mathcal C\times\mathcal M\to\mathcal M$, $(X,m)\mapsto X\triangleright m$ \cite[Definition 7.1.1]{etingof2015tensor}, and isomorphisms
\begin{equation}
(X\otimes Y)\triangleright m\cong X\triangleright(Y\triangleright m),\qquad \mathbbm 1_{\mathcal C}\triangleright m\cong m,
\end{equation}
satisfying the usual coherence conditions. Right $\mathcal D$-module categories are defined analogously. A $(\mathcal C,\mathcal D)$-bimodule category ${}_{\mathcal C}\mathcal M_{\mathcal D}$ carries both actions together with isomorphisms
\begin{equation}
(X\triangleright m)\triangleleft Y\cong X\triangleright(m\triangleleft Y),\qquad X\in\mathcal C,\ Y\in\mathcal D,\ m\in\mathcal M.
\end{equation}
It is invertible if there is a bimodule category ${}_{\mathcal D}\mathcal N_{\mathcal C}$ with
\begin{equation}
\mathcal M\boxtimes_{\mathcal D}\mathcal N\simeq\mathcal C,\qquad \mathcal N\boxtimes_{\mathcal C}\mathcal M\simeq\mathcal D,
\end{equation}
where $\boxtimes_{\mathcal D}$ is the relative Deligne product \cite{RelativeDeligne:2014}, the relative version of the Deligne tensor product \cite[Section 1.11]{etingof2015tensor}, and $\mathcal C$, $\mathcal D$ on the right-hand sides are the regular bimodule categories. Fusion categories related by an invertible bimodule category are called \textit{Morita equivalent}.}
For more on the theory of fusion and multi-fusion categories, see \cite{Etingof:2002vpdn,Kong:2017}. 

If a multi-fusion category $\mathcal{C}$ can be written as 
\begin{equation}
\mathcal{C}=\mathcal{C'}\boxplus\mathcal{C''}
\end{equation}
for $\mathcal{C'},\mathcal{C''}$ nonzero multi-fusion categories, then it is decomposable. Otherwise, it is indecomposable. The relation $i\leftrightarrow j \Leftrightarrow \mathcal{C}_{ij}\neq0$ is an equivalence relation whose classes are the indecomposable summands. Within a summand, every $\mathcal{C}_{ij}$ is nonzero and an invertible $(\mathcal{C}_{ii},\mathcal{C}_{jj})$-bimodule category, so an indecomposable multi-fusion category is like a Morita equivalence class, see e.g.\ \cite[Remark 4.3.4]{etingof2015tensor} and \cite[Section 2.4]{Etingof:2002vpdn}.

\paragraph{Examples.} Finite groupoids give the basic examples of multi-fusion categories. For a finite groupoid $\mathsf G$, the linearization $\mathrm{Vec}(\mathsf G)$ is a multi-fusion category whose units correspond to the objects of $\mathsf G$ and whose indecomposable summands correspond to $\pi_0(\mathsf G)$: the pair groupoid on $n$ objects gives $\mathrm{Mat}_n(\mathrm{Vec})$, the groupoid $*\sqcup *$ gives $\mathrm{Vec}\boxplus\mathrm{Vec}$, and $BG$ gives $\mathrm{Vec}(G)$. Physically, a finite groupoid acting on the target of a $\sigma$-model is an (electric) $(-1)$-form symmetry \cite{Perez-Lona:2025xxm},~\cite[Remark 2.1]{Perez-Lona:2026mlj} (see also \cite[Example 1.4]{Stockall:2025ngz}). Similarly, $2$-groups built from $1$-form and $0$-form symmetries in two dimensions have Karoubi-completed linearizations that are multi-fusion categories (Section~\ref{sec:typeIII}).  

Finally, let us review gradings of fusion categories. A fusion category $\mathcal D$ is \textit{graded} by a finite group $\Upsilon$ if $\mathcal D=\bigoplus_{u\in\Upsilon}\mathcal D_u$ as an abelian category with $\mathcal D_u\otimes\mathcal D_v\subseteq\mathcal D_{uv}$, and the grading is \textit{faithful} if every component is nonzero. The \textit{adjoint subcategory} $\mathcal D_{\rm ad}$ is the fusion subcategory generated by the objects $X\otimes X^\vee$. Every fusion category is faithfully graded by its \textit{universal grading group} $U(\mathcal D)$, with trivial component $\mathcal D_{\rm ad}$, and every faithful grading of $\mathcal D$ by a group $\Upsilon$ is induced from it through a surjection $U(\mathcal D)\to\Upsilon$ \cite[Section 4.14]{etingof2015tensor}, \cite{Gelaki:2006}. For example, $U(\mathrm{Vec}(G))=G$ and $U(\mathrm{Rep}(G))=\widehat{Z(G)}$, while the Fibonacci category, with $\tau\otimes\tau=\mathbbm 1\oplus\tau$, has trivial universal grading group. 

\subsection{Symmetries as multi-fusion categories with an action functor}\label{sec:framework}

In this subsection, we define a symmetry of a decomposing theory, explain how it sees universes and when part of it acts trivially, and check that gauging is compatible with these notions.

\paragraph{Action functor.} We describe a finite symmetry of a two-dimensional theory $\mathfrak T$ by a pair $(\mathcal C,F)$ of a multi-fusion category $\mathcal C$ and a tensor functor $F:\mathcal C\to\mathrm{Top}(\mathfrak T)$ into the topological defects of $\mathfrak T$, which we call the \textit{action functor}. The functor $F$ need not be full, injective on simple objects, or essentially surjective. In general, $\mathrm{Top}(\mathfrak T)$ is not even finite (the compact boson already has continuous families of topological lines), so there is no canonical largest choice of $\mathcal C$, and every statement below is relative to a chosen pair. We merely assume that $\mathrm{Top}(\mathfrak T)$ is a multi-tensor category (locally finite, but not necessarily finite), so that tensor functors are defined. This excludes two-dimensional Yang--Mills theory, whose topological point operators split it into infinitely many universes \cite{Nguyen:2021naa}, and we treat that theory only through its finite center symmetry.

\paragraph{Universes and sectors.} The individual \textit{universes} of $\mathfrak T$ are labeled by the primitive idempotents of the semisimple commutative algebra $\mathrm{End}(\mathbbm 1)$ of topological point operators in $\mathrm{Top}(\mathfrak T)$. 
A symmetry $(\mathcal C,F)$ sees only the point operators in $F(\mathrm{End}_{\mathcal C}(\mathbbm 1))$, so it splits $\mathfrak T$ into \textit{sectors}\footnote{Not to be confused with superselection sectors.}, 
each a sum of universes.  We can label sectors by an index set $I$, with each sector having an identity line $F(\mathbbm 1_i)$, $i\in I$ (which may in turn split further into simple lines).
We say that $(\mathcal C,F)$ \textit{resolves} the universes when every sector is a single universe. Sectors are relative in the same way as interfaces: a symmetry that ``omits'' point operators sees coarser sectors, just as one that omits lines sees fewer interfaces. This happens below for a presentation through $\Gamma/K'$ with $K'$ smaller than the full trivially acting subgroup (the $Q_8$ example of Section~\ref{sec:typesI-II}), and for the center symmetry of two-dimensional Yang--Mills theory. The classification into Types~I--III is defined for pairs of sectors. Outside these two examples, we assume that the symmetry resolves the universes, 
and we refer to sectors as universes. 

We do not require a group structure on $I$: when a group-like $K^{(1)}$ is present, its characters provide one labeling, but nonbanded decompositions such as $S_3\supset\mathbb Z_3$ have two universes rather than $|\widehat{\mathbb Z}_3|=3$. Every sector is nonzero, since tensor functors are faithful by definition \cite[Definition 4.2.5]{etingof2015tensor}: for simple $X\in\mathcal C_{ij}$, $F(X)\otimes F(X^\vee)\supset F(\mathbbm 1_i)\neq0$, so every nonzero block $\mathcal C_{ij}$ yields topological interfaces of $\mathfrak T$.

\paragraph{Trivially acting symmetries.} A trivially acting symmetry is invisible in $\mathrm{Top}(\mathfrak T)$ itself, where a trivially acting line is simply isomorphic to a sum of copies of the identity line, but it is visible in the action functor. Consider a symmetry $(\mathcal B,F)$ with $\mathcal B$ fusion. We say that $F\cong F_{\rm eff}\circ q$ is an \textit{effective presentation} if $q:\mathcal B\to\mathcal B_{\rm eff}$ is a dominant tensor functor and $F_{\rm eff}$ has trivial kernel \cite{Bruguieres2011}, that is, only direct sums of copies of the unit are sent to direct sums of copies of the identity line. A fusion subcategory $\mathcal B'\subseteq\mathcal B$ \textit{acts trivially} if $F|_{\mathcal B'}\cong u\circ\varphi$ for a fiber functor $\varphi:\mathcal B'\to\mathrm{Vec}$, where $u(V)=\mathbbm 1^{\oplus \dim(V)}$ with $\mathbbm 1$ the trivial topological defect \cite[Section 4.5]{Robbins:2022wlr},~\cite[Section 5.1]{Perez-Lona:2023djo},~\cite[Section 6.1]{Perez-Lona:2024sds},~\cite[Section VI]{Bhardwaj:2024qrf}. The pair of $\varphi$ and the monoidal isomorphism realizing this, which we call the \textit{trivialization}, is part of the trivial action data. For $\mathcal B'=\mathrm{Vec}(K)$ with $K$ abelian, fiber functors are the rank-one module categories $\mathcal M(K,\psi)$ \cite[Theorem 3.1]{Ostrik:2002ohv}, so the choice of $\varphi$ is a class $\psi\in H^2(K,U(1))$ \cite[Theorem 1.1]{Natale:2016}. The action functor then sends the regular algebra of $\mathrm{Vec}(K)$ to $\mathbb C^{\psi}[K]\otimes\mathbbm 1$, and the choice matters: for $\psi\neq0$, the center of $\mathbb C^{\psi}[K]$, and hence the number of resulting universes, shrinks, reflecting the projective fusion of point operators \cite[Section 4.1]{Robbins:2020msp}.

\paragraph{Action on a single universe.} Given a tensor functor $\Phi:\mathcal B\to\mathcal C$ into a multi-fusion category, we can ask how $\mathcal B$ acts on a single universe $i$, but only for the part of $\mathcal B$ that does not move it. 
The projection $X\mapsto\mathbbm 1_i\otimes X\otimes\mathbbm 1_i$ is not monoidal on objects that change universe: in $\mathrm{Mat}_2(\mathrm{Vec})$, $E_{12}\otimes E_{21}=E_{11}$, while the projections of $E_{12}$ and $E_{21}$ vanish. On the full subcategory of objects $s$ with  
\begin{equation}
    \Phi(s)\otimes\mathbbm 1_i\cong\mathbbm 1_i\otimes\Phi(s) ,
\end{equation}
that is, those whose image has no off-diagonal entries in row or column $i$, the projection is monoidal by block multiplication and defines the \textit{action on universe $i$}. A subsymmetry can act trivially on one universe and faithfully on another.

\paragraph{Gauging.} Effective presentations are compatible with gauging. A tensor functor $q:\mathcal B\to\mathcal B_{\rm eff}$ carries a separable algebra $A$ to a separable algebra $q(A)$ \cite[Remark 2.6.ii]{davydov2011wittgroupnondegeneratebraided},\footnote{Throughout this article, we call a Frobenius algebra $A$ \textit{special} when $m\circ\Delta$ is a nonzero multiple of the identity and $\varepsilon(1_A)\neq0$, without normalizing these constants. Different normalizations differ by an Euler counterterm.} and it carries $A$-bimodules and $\otimes_A$ along because it is exact and monoidal. Gauging $A$ therefore endows $\mathfrak T/A$ with the effective symmetry $\mathrm{Bimod}_{q(A)}(\mathcal B_{\rm eff})$, and the quantum symmetry is the composite
\begin{equation}
    \mathrm{Bimod}_A(\mathcal B)\to\mathrm{Bimod}_{q(A)}(\mathcal B_{\rm eff})\to\mathrm{Top}(\mathfrak T/A),
\end{equation}
which in general is neither full nor injective on simple objects.

\subsection{Dictionary}\label{sec:dictionary}

Having set up the framework, we now turn to the physical meaning of these category-theoretic data. We denote the decomposing theory as 
\begin{equation}
    \mathfrak T=\bigsqcup_{i=1}^{n}\mathcal T_i ,
\end{equation}
where $\mathcal T_i$ are the individual universes and we denote as $\mathcal C$ a multi-fusion category describing symmetries of $\mathfrak T$.

The physical meaning of a diagonal block $\mathcal C_{ii}$, which is a fusion category, is particularly simple. It describes the symmetry lines of the $i$-th universe, that is, the topological lines of $\mathcal T_i$ in the chosen symmetry. Its tensor unit is the identity line $\mathbbm{1}_i$ of that universe. The topological point operators seen by the symmetry form $\mathrm{End}_{\mathcal C}(\mathbbm 1)\cong\mathbb C^n$, whose primitive idempotents project onto the components $\mathbbm{1}_i$ and hence, physically, onto the sectors. Wrapped on a circle, the identity line $\mathbbm 1_i$ acts as the projector onto the Hilbert space $\mathcal H_i$ of sector $i$.

The off-diagonal blocks may be nonzero, in which case $\mathcal C$ contains topological interfaces between universes. We use the convention that $X\in\mathcal C_{ij}$ has universe $i$ on its left and universe $j$ on its right. Under $F$, such an $X$ is a topological interface between $\mathcal T_i$ and $\mathcal T_j$. Conversely, every topological line $L$ of $\mathfrak T$ decomposes into such interfaces as
\begin{equation}
    L=\bigoplus_{i,j}\mathbbm 1_i\otimes L\otimes\mathbbm 1_j.
\end{equation}
Notably, $\mathcal C_{ij}=0$ does not imply that no topological interface exists in $\mathfrak T$, only that none lies in the image of $F$. The bimodule structure of $\mathcal C_{ij}$ records how an interface fuses with the lines of the two adjacent universes.

The correspondence between the physical and categorical data is summarized in Table~\ref{tab:multifusion-dictionary}.

\begin{table}[htbp]
    \centering
    \renewcommand{\arraystretch}{1.2}
    \begin{tabularx}{\linewidth}{@{}>{\raggedright\arraybackslash}X>{\raggedright\arraybackslash}X@{}}
        \toprule
        \textbf{Decomposition, relative to $(\mathcal C,F)$} & \textbf{Multi-fusion category $\mathcal C$} \\
        \midrule
        Identity line of sector $i$ (of universe $\mathcal T_i$ when $(\mathcal C,F)$ resolves the universes)
        & Simple summand $\mathbbm 1_i$ of the unit \\
        Projection onto sector $i$
        & Primitive idempotent $\Pi_i\in\mathrm{End}_{\mathcal C}(\mathbbm 1)$ \\
        Topological point operators seen by the symmetry
        & $\mathrm{End}_{\mathcal C}(\mathbbm 1)\cong\mathbb C^n$ (commutative, semisimple) \\
        Symmetry lines of universe $\mathcal T_i$
        & Objects of the diagonal fusion category $\mathcal C_{ii}$ \\
        Topological interface with $\mathcal T_i$ on its left and $\mathcal T_j$ on its right
        & Objects of the off-diagonal block $\mathcal C_{ij}$ \\
        Isomorphism $\mathcal T_i\cong\mathcal T_j$ realized by the symmetry
        & Invertible object of $\mathcal C_{ij}$ \\
        Universes connected by topological interfaces
        & Units in one indecomposable summand $\mathcal C^{(a)}$ \\
        \bottomrule
    \end{tabularx}
    \caption{Physical interpretation of the multi-fusion structure of a decomposing two-dimensional QFT with symmetry $(\mathcal C,F)$. Interfaces are those in the image of $F$. Here, $\mathfrak{T} = \sqcup_{i}\mathcal{T}_i$ is the decomposing theory and $\mathcal{T}_i$ are the universes, $F:\mathcal{C} \rightarrow \operatorname{Top}(\mathfrak{T})$ is the action functor of Section~\ref{sec:framework}, and $\mathcal{C} = \boxplus_{a} \mathcal{C}^{(a)}$ is the decomposition of $\mathcal{C}$ into indecomposable summands.}
    \label{tab:multifusion-dictionary}
\end{table}
To see how Table~\ref{tab:multifusion-dictionary} is read in practice, we work through two two-dimensional gauge theories, $SU(2)$ Yang--Mills theory and $\mathbb{Z}_2$ gauge theory. Both have a $\mathbb{Z}_2$ center $1$-form symmetry, and, relative to the symmetries specified below, the two theories agree in the first four rows of the table. They are told apart by the last three rows, which concern interfaces. Pure $SU(2)$ Yang--Mills theory decomposes into
\begin{equation}
    SU(2) = SO(3)_+ \sqcup SO(3)_- ,
\end{equation}
where $SO(3)_\pm$ denotes $SO(3)$ gauge theory with discrete theta angle, while in $\mathbb{Z}_2$ gauge theory both universes are the trivial theory,
\begin{equation}
    \mathfrak T_{\mathbb{Z}_2} = \mathcal T_1 \sqcup \mathcal T_2 , \qquad \mathcal T_1 \cong \mathcal T_2 \cong \text{trivial theory}.
\end{equation}
In both cases, the generator $U$ of the $1$-form symmetry obeys $U^2 = 1$, and the projectors onto the two sectors are $\Pi_\pm = \tfrac12(1 \pm U)$ (we write $\Pi_{1,2}$ for the $\mathbb{Z}_2$ gauge theory). In the language of the table, the $1$-form symmetry gives a $\mathbb{C}^2$ worth of point operators in both theories.

Relative to the symmetry generated by these point operators and, for the $\mathbb{Z}_2$ gauge theory, the Wilson line, the diagonal entries are
\begin{equation}
    \mathcal{C}_{++} = \mathcal{C}_{--} = \mathrm{Vec}, \qquad \mathcal{C}_{11} = \mathcal{C}_{22} = \mathrm{Vec}.
\end{equation}
In both theories, the Wilson line $W$ in a representation charged under the center anticommutes with $U$, so it is an interface between the two sectors. 

The two theories differ in whether this interface is topological. In $SU(2)$ Yang--Mills theory, the fundamental Wilson line connects the two center sectors but is not topological, and in fact no topological interface exists. 
We can see this as follows.
\begin{enumerate}
\item Suppose a topological interface $X$ exists.  Let ${\cal H}_{\pm}$ be the Hilbert spaces of $SO(3)_{\pm}$. On a circle, the states are labeled by spins $j$, with energies proportional to $j(j+1)$, where ${\cal H}_{\pm}$ consists of, separately, integer or half-integer spin $j$.  Note that  the center generator acts on them by $(-1)^{2j}$ \cite[Equation 19]{Nguyen:2021naa}.  
Let $\hat{X}$ be the topological interface $X$ wrapped on the spatial circle. Then, $\hat{X}$ maps one of ${\cal H}_{\pm}$ into the other and, because it is topological, commutes with the Hamiltonian.  Physically, no such energy-preserving operator exists as the energies on either side are different. 

Note that this is not quite yet a contradiction, as it is a statement about the operator in the untwisted sector.

\item Consider a torus with interfaces $X$, $X^{\vee}$ on parallel spatial circles.  Its partition function 
is of the form $\mathrm{Tr}(\hat{X} e^{-\beta_1 H} \hat{X}^{\dag} e^{-\beta_2 H})$, which vanishes since $\hat X=0$. In the crossed channel, however, it is a trace over the Hilbert space twisted by $X \otimes X^{\vee}$.
Since $X \otimes X^{\vee} \supset \mathbbm 1$, this Hilbert space contains the untwisted sector, so the partition function is 
strictly positive by reflection positivity. 
\end{enumerate}
Hence, no topological interface exists:
\begin{equation}
    \mathcal{C}_{+-} = \mathcal{C}_{-+} =0 ,
\end{equation}
and the center symmetry of $SU(2)$ Yang--Mills theory is $(\mathcal C_{SU(2)},F)$ with
\begin{equation}
\mathcal{C}_{SU(2)}=\begin{pmatrix} \text{Vec} & 0\\ 0 & \text{Vec} \end{pmatrix} ,
\end{equation}
a decomposable multi-fusion category. The center symmetry does not resolve the universes, so $SO(3)_\pm$ are sectors rather than single universes.\footnote{The non-invertible $1$-form symmetry of \cite{Nguyen:2021naa} splits each $SO(3)_\pm$ into infinitely many universes labeled by spins, which lie outside the finite setting (Section~\ref{sec:discussion}). The spectral argument above shows that no two of them are joined by a topological interface.}

In the $\mathbb Z_2$ gauge theory, by contrast, the Wilson line is topological and swaps the two universes, so the symmetry is the indecomposable multi-fusion category
\begin{equation}
\mathcal{C}_{\Z_2}=\begin{pmatrix} \text{Vec} & \text{Vec}\\ \text{Vec} & \text{Vec} \end{pmatrix},
\end{equation}
whose off-diagonal blocks contain the swap. This is the case $n=2$, with $\mathcal T$ trivial, of the identical universes of Section~\ref{sec:typesI-II}.

\section{Types of decomposition}\label{sect:org}

In this section, we sort pairs of universes into types relative to a chosen symmetry (Table~\ref{tab:types}) and collect the structure of the indecomposable summands that underlies them. Throughout, $(\mathcal C,F)$ is a symmetry of $\mathfrak T$ as in Section~\ref{sec:framework}, with $\mathbbm 1=\bigoplus_{i\in I}\mathbbm 1_i$.

\paragraph{Types of pairs of universes.} Relative to $(\mathcal C,F)$, we compare two universes $i,j$ through three \textit{relations}, each finer than the previous one:
\begin{equation}
\begin{aligned}
i\leftrightarrow j &:\ \mathcal{C}_{ij}\neq 0 \\
  &\quad\text{(connected by a topological interface)}, \\[2pt]
i\sim j &:\ i\leftrightarrow j \ \text{and}\ \mathcal{C}_{ii}\simeq\mathcal{C}_{jj} \text{ as fusion categories} \\
  &\quad\text{(same symmetry)}, \\[2pt]
i\approx j &:\ \mathcal{C}_{ij}\ \text{contains an invertible object} \\
  &\quad\text{(isomorphic through the symmetry)}.
\end{aligned}
\end{equation}

Here an isomorphism $\mathcal T_i\cong\mathcal T_j$ is an invertible topological interface, which is a line of $\mathfrak T$. Accordingly, an object $X\in\mathcal C_{ij}$ is \textit{invertible} if there exists an object $X^\vee \in \mathcal{C}_{ji}$ such that $X\otimes X^\vee\cong\mathbbm 1_i$ and $X^\vee\otimes X\cong\mathbbm 1_j$. The relation $\approx$ records only those in the image of $F$, so universes in a pair of any type can still be isomorphic through an interface outside the chosen symmetry, possibly one that does not intertwine their symmetries. All three are equivalence relations, and $i\approx j\Rightarrow i\sim j\Rightarrow i\leftrightarrow j$: invertible interfaces compose, and an invertible $X\in\mathcal C_{ij}$ gives the tensor equivalence $Y\mapsto X^\vee\otimes Y\otimes X$ from $\mathcal C_{ii}$ to $\mathcal C_{jj}$. The types are the successive differences between the three relations:
\begin{equation}\label{eq:types}
\begin{aligned}
\text{I}&:\ i\approx j, \\
\text{IIa}&:\ i\sim j,\ i\not\approx j, \\
\text{IIb}&:\ i\leftrightarrow j,\ i\not\sim j, \\
\text{III}&:\ i\not\leftrightarrow j.
\end{aligned}
\end{equation}
Every pair of distinct universes has exactly one type, for any multi-fusion category $\mathcal C$: the types sort pairs, not categories.

Table~\ref{tab:types} summarizes the types and their SymTFT counterpart, which is derived in Section~\ref{sec:IIa-IIb} for Types~I and~II and in item~(iii) below for Type~III, and Table~\ref{tab:type-examples} collects examples of each type.

We emphasize two points that prevent potential misreadings. First, the types measure how universes are related, not what their symmetries are. For example, a Type~III pair need not carry different anomalies: its diagonal blocks can be equivalent, as for the center sectors of $SU(2)$ Yang--Mills theory. Only the IIa/IIb split compares the symmetry categories themselves. Second, enlarging the symmetry without splitting its units, through a tensor functor $\Phi:\mathcal C\to\mathcal C'$ with $F\cong F'\circ\Phi$ that sends each $\mathbbm 1_i$ to a simple unit of $\mathcal C'$, can create interfaces, turning Type~III into Type~II or~I, or Type~II into Type~I, but never the reverse: $\Phi$ is faithful and monoidal, so it sends nonzero objects of $\mathcal C_{ij}$ to nonzero objects of $\mathcal C'_{ij}$ and invertible ones to invertible ones. 

For example, the two trivial universes of two-dimensional $\Z_2$ gauge theory can be described by a Type~III pair relative to $\mathrm{Vec}\boxplus\mathrm{Vec}$, which retains only their identity lines, or as a Type~I pair relative to $\mathrm{Mat}_2(\mathrm{Vec})$, which also contains the quantum $\Z_2$ line resolved by the universe projectors. 
This reflects the fact that the multi-fusion category ${\cal C}$ is ambiguous, as described earlier
	in Section~\ref{sec:framework}.
More generally, whether a decomposing theory was assembled by hand or obtained by gauging is not a property of the theory: for $A$ a separable algebra in a fusion symmetry $\mathcal B$ of $\mathcal T$, $\mathcal T \sqcup \mathcal T/A\cong\mathcal T/(\mathbbm 1\oplus A)$ (up to Euler counterterms), which carries the indecomposable symmetry $\mathrm{Bimod}_{\mathbbm 1\oplus A}(\mathcal B)$.

The IIa/IIb split, by contrast, is not monotone: enlarging the symmetry can turn Type~IIa into Type~IIb. For $\mathcal T$ with a non-anomalous $S_3$ symmetry, the pair $\mathcal T$, $\mathcal T/\mathbb Z_2$ has diagonal blocks $\mathrm{Vec}(\mathbb Z_2)$ and $\mathrm{Rep}(\mathbb Z_2)$ relative to the $\mathbb Z_2$ symmetry (Type~IIa), but $\mathrm{Vec}(S_3)$ and $\mathrm{Rep}(S_3)$ relative to the $S_3$ symmetry (Type~IIb) \cite{Bhardwaj:2017xup}.

\begin{table*}[t]
\centering
\small
\renewcommand{\arraystretch}{1.25}
\begin{tabular}{|>{\raggedright\arraybackslash}p{0.06\textwidth}|>{\raggedright\arraybackslash}p{0.16\textwidth}|>{\raggedright\arraybackslash}p{0.2\textwidth}|>{\raggedright\arraybackslash}p{0.19\textwidth}|>{\raggedright\arraybackslash}p{0.25\textwidth}|}
\hline
\textbf{Type} & \textbf{Interfaces in $\mathcal C_{ij}$} & \textbf{Relation between universes} & \textbf{Symmetries $\mathcal C_{ii},\mathcal C_{jj}$}
& \textbf{SymTFT} \\ \hline
I & an invertible one & isomorphic, through an interface in the image of $F$ & equivalent, via the interface & same symmetry boundary \\ \hline
IIa & nonzero, none invertible & gauging that returns a tensor-equivalent symmetry category & equivalent & boundaries exchanged by an invertible
bulk surface \\ \hline
IIb & nonzero, none invertible & gauging that changes the symmetry & inequivalent, Morita equivalent & boundaries in different
orbits of bulk symmetries \\ \hline
III & none & no topological interface in the image of $F$, though possibly isomorphic & arbitrary, possibly equivalent & decomposed bulk, with no line of $\mathcal C$ between boundaries over different components \\ \hline
\end{tabular}
\caption{Types of pairs of universes relative to a symmetry $(\mathcal C,F)$, defined in \eqref{eq:types}.}
\label{tab:types}
\end{table*}

\begin{table}[htbp]
\centering
\small
\renewcommand{\arraystretch}{1.25}
\begin{tabularx}{\linewidth}{|l|X|}
\hline
\textbf{Type} & \textbf{Examples} \\ \hline
I & $n$ identical universes, including the $\mathbb Z_2$ gauge theory \newline
the two $[\mathcal T/\mathbb Z_2]$ in $[\mathcal T/Q_8]$, $K=\mathbb Z_4$ \newline
the two $[\mathcal T/\mathbb Z_4]$ in $[\mathcal T/((\mathbb Z_2^2)\rtimes\mathbb Z_4)]$ \\ \hline
IIa & $S_3\supset\mathbb Z_3$ trivially acting subsymmetry (Section~\ref{sec:TypeII})\newline
$S_4\supset\mathbb Z_2^2$ trivially acting subsymmetry (Section~\ref{sec:TypeII})\newline
$\mathcal T$ and $[\mathcal T/\mathbb Z_2]$ in $[\mathcal T/Q_8]$, $K=\mathbb Z_4$ \newline
$[\mathcal T/\mathbb Z_2^2]$ and $[\mathcal T/\mathbb Z_2^2]_{\rm d.t.}$ in $[\mathcal T/D_4]$, $K$ the center \newline
$\mathcal T\sqcup \mathcal T/\mathbb Z_2$ relative to a $\mathbb Z_2$ symmetry \newline
$\mathrm{Ising}_m\sqcup \mathrm{Ising}_{-m}$ \\ \hline
IIb & $[\mathcal T/\mathbb Z_4]$ and $[\mathcal T/\mathbb Z_2]$ in $[\mathcal T/((\mathbb Z_2^2)\rtimes\mathbb Z_4)]$ \newline
$\mathcal T\sqcup \mathcal T/\mathbb Z_2$ relative to an $S_3$ symmetry \\ \hline
III & center sectors of $SU(2)$ Yang--Mills \newline
$\mathrm{Vec}(\mathbb Z_2)\boxplus\mathrm{Vec}(\mathbb Z_2,\omega)$ \newline
$\mathrm{Rep}(D_4)\boxplus\mathrm{Rep}(Q_8)$ \newline
trivial$\oplus$trivial with $\mathrm{Vec}\boxplus\mathrm{Vec}$ \newline
$\mathcal T\sqcup \mathcal T/G$ with $\mathrm{Vec}(G)\boxplus\mathrm{Rep}(G)$ \\ \hline
\end{tabularx}
\caption{Examples of the types of Table~\ref{tab:types} in this article. Each example is relative to the symmetry stated for it in the text, and the same theory can appear under different types for different symmetries.}
\label{tab:type-examples}
\end{table}

\paragraph{Structure of the summands.} We write $\mathcal C=\boxplus_a\mathcal C^{(a)}$ for the unique decomposition of a multi-fusion category into indecomposable multi-fusion summands \cite[Remark 4.3.4]{etingof2015tensor}, \cite[Section 2.4]{Etingof:2002vpdn}, and denote by $I_a\subset I$ the set of universes whose units lie in $\mathcal C^{(a)}$. The following list collects standard results \cite{etingof2015tensor,etingof2010fusion,Frohlich:2006ch} together with their physical meaning.

\begin{enumerate}[label=(\roman*)]
\item \textit{Universes in one summand.} For $i,j\in I_a$, the diagonal blocks $\mathcal C_{ii}$ and $\mathcal C_{jj}$ are Morita equivalent through the invertible bimodule category $\mathcal C_{ij}$, so they share the Drinfeld center $\mathcal Z(\mathcal C_{ii})\simeq\mathcal Z(\mathcal C^{(a)})$: the universes of a summand share their SymTFT. Every simple $X\in\mathcal C_{ij}$ relates the two universes by a gauging,
\begin{equation}\label{eq:gauging-interface}
\mathcal T_i\cong\mathcal T_j/A_X,\qquad A_X:=X^\vee\otimes X\in\mathcal C_{jj}.
\end{equation}
Here $A_X$ is a separable algebra, and a special Frobenius algebra once $\mathcal C$ carries a pivotal structure, with structure maps built from the evaluation and coevaluation of $X$ and that pivotal structure. It is haploid since $\mathrm{Hom}(\mathbbm 1_j,A_X)\cong\mathrm{End}(X)=\mathbb C$. To derive \eqref{eq:gauging-interface}, consider the left $\mathcal C$-module category 
\begin{equation}
    \mathcal C\otimes\mathbbm 1_i=\bigoplus_{k\in I_a}\mathcal C_{ki}
\end{equation}
of right $\mathbbm 1_i$-modules. The object $X^\vee\in\mathcal C_{ji}$ generates it, since every simple $Z\in\mathcal C_{ki}$ is a summand of 
\begin{equation}
    (Z\otimes X)\otimes X^\vee\supseteq Z\otimes\mathbbm 1_i ,
\end{equation}
and its internal endomorphism algebra is $X^\vee\otimes X$. By \cite[Theorem 7.10.1]{etingof2015tensor}, this module category is equivalent to the category of right $A_X$-modules in $\mathcal C$, so $A_X$ and $\mathbbm 1_i$ are Morita-equivalent algebras in $\mathcal C$. Gauging depends only on the Morita class of the algebra \cite[Section 3.3]{Frohlich:2006ch}, \cite{Bhardwaj:2017xup}. Gauging $\mathbbm 1_i$ projects onto universe $i$, while gauging $A_X$, whose unit lies in universe $j$, gauges within $\mathcal T_j$, which gives \eqref{eq:gauging-interface}. Physically, $X$ is the interface obtained by gauging $A_X$ on half of spacetime, and $A_X$ is recovered by folding $X$ with its dual.

Conversely, label the universes of the summand so that $1\in I_a$, and choose a simple $X_j\in\mathcal C_{j1}$ for each $j\in I_a$, with $X_1=\mathbbm 1_1$. Since gauging a direct sum of algebras gives the direct sum of the gauged theories, 
\begin{equation}
    \bigsqcup_{j\in I_a}\mathcal T_j\cong\mathcal T_1/A
\end{equation}
for the separable algebra 
\begin{equation}
    A=\bigoplus_j X_j^\vee\otimes X_j
\end{equation}
in $\mathcal C_{11}$.
A summand is thus a set of theories obtained from any one of them by gauging.

Two notions of invertibility are at play here. The bimodule category $\mathcal C_{ij}$ is always invertible, while its objects need not be. Whether the gauging \eqref{eq:gauging-interface} is trivial, that is, whether $A_X$ is Morita equivalent to $\mathbbm 1_j$ in $\mathcal C_{jj}$, depends on $\mathcal C_{ij}$ and not on $X$. Indeed, the argument above, applied to $\mathcal C_{jj}$ acting on $\mathcal C_{ji}$, identifies the right $A_X$-modules in $\mathcal C_{jj}$ with the left $\mathcal C_{jj}$-module category $\mathcal C_{ji}$ for every simple $X$, and this module category is the regular one precisely when $\mathcal C_{ij}$ contains an invertible object (item~(ii)). 
For instance, for $\mathcal T\sqcup \mathcal T$ with $\mathcal T$ carrying $\mathrm{Rep}(S_3)$ and the symmetry $\mathrm{Mat}_2(\mathrm{Rep}(S_3))$ of item~(ii), the non-invertible object $X=E_{12}\otimes\rho$ gives $A_X\cong E_{22}\otimes(\mathbbm 1\oplus\mathrm{sgn}\oplus\rho)$. This is the internal endomorphism algebra of $\rho$ in the regular module category, so gauging it returns $\mathcal T$, as it must, since $\mathcal C_{12}$ also contains the invertible object $E_{12}\otimes\mathbbm 1$.
\item \textit{Universes isomorphic through the symmetry.} If $\mathcal C_{ij}$ contains an invertible object $X$, then $F(X)$ is an isomorphism $\mathcal T_i\cong\mathcal T_j$. This happens precisely when $\mathcal C_{ij}$ is the regular bimodule category twisted by a tensor equivalence $\mathcal C_{ii}\simeq\mathcal C_{jj}$, namely $Y\mapsto X^\vee\otimes Y\otimes X$. If it happens for every pair in $I_a$ (by transitivity, equivalently for every pair containing one fixed universe),\footnote{One isomorphic pair does not suffice. In $\mathcal T \sqcup \mathcal T \sqcup \mathcal T/\mathbb Z_2$, with $\mathcal T$ carrying a non-anomalous $\mathbb Z_2$ and the symmetry $\mathrm{Bimod}_{\mathbbm 1\oplus\mathbbm 1\oplus\mathbb C[\mathbb Z_2]}(\mathrm{Vec}(\mathbb Z_2))$, all three diagonal blocks are equivalent to $\mathrm{Vec}(\mathbb Z_2)$, since $\mathrm{Rep}(\mathbb Z_2)\simeq\mathrm{Vec}(\mathbb Z_2)$. The block $\mathcal C_{12}$ contains an invertible object, while $\mathcal C_{13}$ has a single simple object, of quantum dimension $\sqrt2$, so the summand is not of matrix form.} then
\begin{equation}
\begin{aligned}
\mathcal{C}^{(a)} &\simeq \mathrm{Mat}_{n_a}(\mathcal{D}_a) \\
&\simeq \mathcal{D}_a\boxtimes\mathrm{Mat}_{n_a}(\mathrm{Vec}), \qquad n_a=|I_a|,
\end{aligned}
\end{equation}
with $\mathcal D_a=\mathcal C_{ii}$ for any $i\in I_a$, where $\mathrm{Mat}_n(\mathcal D)$ has $n\times n$ matrices of objects of $\mathcal D$ as objects and matrix multiplication as tensor product. Its invertible objects are the monomial matrices with invertible entries, which form the group $\mathrm{Inv}(\mathcal D_a)\wr S_{n_a}$ and realize every permutation of the universes. This group is in general larger than a given group symmetry permuting the universes, such as the quantum $\mathbb Z_n$ of $[\mathcal T/\mathbb Z_n]$ for $n\geq3$ (Section~\ref{sec:typesI-II}). A $G^{(0)}$-family structure is therefore additional data: a tensor functor $\mathrm{Vec}(G,\omega)\to\mathcal C$, possibly with nontrivial $\omega$, permuting the units.
\item \textit{Universes in different summands.} For $i\in I_a$, $j\in I_b$ with $a\neq b$, no topological interface in the image of $F$ connects the two universes, and the SymTFT decomposes,
\begin{equation}
    \mathcal Z(\mathcal C)=\boxplus_a\mathcal Z(\mathcal C^{(a)}),
\end{equation}
with the symmetry boundary of each universe over a single component. Bulk surfaces between components, the $(\mathcal C^{(a)},\mathcal C^{(b)})$-bimodule categories, can nevertheless exist. Such a surface is a wall of the SymTFT, not a topological interface of $\mathfrak T$. Totally reflective walls always exist, since each component $\mathcal Z(\mathcal C^{(a)})$ has a topological boundary, while invertible ones exist precisely when the summands are Morita equivalent. Otherwise, the SymTFTs are inequivalent. For the center sectors of $SU(2)$ Yang--Mills theory, with $\mathcal C=\mathrm{Vec}\boxplus\mathrm{Vec}$, the SymTFT is the three-dimensional $BF$ theory of a $\mathbb Z_2$-valued $0$-form field and a $\mathbb Z_2$-valued $2$-form field $B$. Its two components are exchanged by the invertible wall $\exp(i\pi\oint B)$, which links with the point operator. Fused onto the symmetry boundary, this wall shifts the discrete theta angle, exchanging $SO(3)_+$ and $SO(3)_-$. It cannot end on that boundary, since $\mathcal C_{+-}=0$.

Morita-equivalent but inequivalent summands also occur, as for $\mathcal T \sqcup \mathcal T/G$, $G$ nonabelian, with the symmetry $\mathrm{Vec}(G)\boxplus\mathrm{Rep}(G)$, which omits the gauging interfaces. Different associators need not signal different anomalies: $\mathrm{Rep}(D_4)$ and $\mathrm{Rep}(Q_8)$ differ only in their associator and are not even Morita equivalent, yet neither is anomalous (Section~\ref{sec:typeIII}). Gauging an algebra in the image of $F$ can still relate universes in different summands, as $\mathcal T/G$ is obtained from $\mathcal T$ by gauging $\mathbb C[G]\in\mathrm{Vec}(G)$, but no interface in the image of $F$ realizes it. Within the theory, the summands are combined by the direct sum, which is the flat gauging of a $(-1)$-form symmetry. For instance, $SU(2)$ Yang--Mills theory sums $SO(3)_\pm$ over the discrete theta angle \cite{Sharpe:2014tca,Lin:2025oml,Perez-Lona:2025add,Perez-Lona:2026mlj,PerezLona:inprep}.
\end{enumerate}

\paragraph{Types within a summand.} Fix $k\in I_a$, and let $\mathcal D:=\mathcal C_{kk}$ act on $\mathcal M_j:=\mathcal C_{jk}$ from the right, 
so that $X\in\mathcal C_{ij}$ acts by left multiplication $\mathcal M_j\to\mathcal M_i$. 
This identifies $\mathcal C^{(a)}$ with 
\begin{equation}
    \mathrm{Fun}_{\mathcal D} \left(\boxplus_{j\in I_a}\mathcal M_j,\boxplus_{j\in I_a}\mathcal M_j \right) ,
\end{equation}
with $\mathcal C_{ij}\simeq\mathrm{Fun}_{\mathcal D}(\mathcal M_j,\mathcal M_i)$ and $\mathcal C_{jj}\simeq\mathcal D^*_{\mathcal M_j}$ \cite[Remark 7.12.5, Theorem 7.12.11]{etingof2015tensor}, and each $\mathcal M_j$ is indecomposable, since $\mathcal C_{jj}$ has a simple unit. For $i,j\in I_a$,
\begin{equation}\label{eq:types-module}
\begin{aligned}
i\approx j &\iff \mathcal M_i\simeq\mathcal M_j, \\[2pt]
i\sim j &\iff \mathcal M_j\simeq\mathcal M_i\boxtimes_{\mathcal D}\mathcal P,\ \ \mathcal P\in\mathrm{BrPic}(\mathcal D),
\end{aligned}
\end{equation}
where $\mathrm{BrPic}(\mathcal D)$ is the group of invertible $\mathcal D$-bimodule categories. The first line holds because an invertible object of $\mathrm{Fun}_{\mathcal D}(\mathcal M_j,\mathcal M_i)$ is an equivalence of module categories. For the second, $-\boxtimes_{\mathcal D}\mathcal P$ induces $\mathcal D^*_{\mathcal M_i}\simeq\mathcal D^*_{\mathcal M_i\boxtimes_{\mathcal D}\mathcal P}$. Conversely, a tensor equivalence $f:\mathcal C_{ii}\simeq\mathcal C_{jj}$ makes $\mathcal M_j$ a $(\mathcal C_{ii},\mathcal D)$-bimodule category ${}^f\mathcal M_j$, and $\mathcal P:=\mathcal M_i^{-1}\boxtimes_{\mathcal C_{ii}}{}^f\mathcal M_j$ is invertible with $\mathcal M_i\boxtimes_{\mathcal D}\mathcal P\simeq\mathcal M_j$. Type~I pairs thus have equivalent module categories, Type~IIa pairs have inequivalent module categories in one $\mathrm{BrPic}(\mathcal D)$-orbit, and Type~IIb pairs have module categories in different orbits.

\section{Trivially acting symmetries and decomposition}\label{sec:trivial}

In this section, we derive the symmetry that the orbifold of a theory by a group with a trivially acting normal abelian subgroup inherits from the $\mathrm{Vec}(\Gamma)$ symmetry of the parent. The result upgrades the decomposition formula \cite[Equation 2.3]{Robbins:2022wlr} to the level of symmetries: each universe carries exactly the symmetry it inherits as an orbifold, and the interfaces between universes are partial-gauging interfaces. It covers Types~I and~II at once. After fixing the setup, we decompose the gauged algebra into twisted group algebras of stabilizers, which gives the main result \eqref{eq:trivial-main}. We then read off the symmetry of each universe, the interfaces, the quantum symmetry and the types of pairs, and we interpret the result as a partial gauging. Section~\ref{sec:typesI-II} illustrates the result in examples. An analysis to the same effect for certain kinds of non-invertible symmetries is in \cite[Sections 4, 5]{Perez-Lona:2025wup}.

\paragraph{Setup.} Let a finite group $\Gamma$ act on a two-dimensional theory $\mathcal T$ without anomaly, and let $K\trianglelefteq\Gamma$ be an abelian normal subgroup acting trivially, with $G=\Gamma/K$ and projection $\pi$. We keep the hypotheses of \cite[Equation 2.3]{Robbins:2022wlr}: there is no discrete torsion in the $\Gamma$-gauging (the cocycles $\nu_\alpha$ below are induced by the extension instead), and the trivialization of $K$ is untwisted ($\psi=0$) and compatible with $\Gamma$, that is, $\Gamma$-equivariant, so that the action of $\Gamma$ factors through $\pi$ as a tensor functor. The action of $\Gamma$ is then $F_G\circ\pi_*$, with $F_G$ the action of $\mathrm{Vec}(G)$ and $\pi_*:\mathrm{Vec}(\Gamma)\to\mathrm{Vec}(G)$ the strict pushforward. When $G$ acts faithfully, this is an effective presentation in the sense of Section~\ref{sec:framework}. The subgroup $K$ need be neither central nor split. Gauging $\Gamma$ means gauging
\begin{equation}
A:=\pi_*\mathbb C[\Gamma]\in\mathrm{Vec}(G),\quad A_g=\mathbb C[\pi^{-1}(g)],\quad A_{1_G}=\mathbb C[K],
\end{equation}
and the gauged theory inherits $\mathcal C=\mathrm{Bimod}_A(\mathrm{Vec}(G))$ as a symmetry category. We fix a section $s:G\to\Gamma$ with $s(1)=1$ and let $c(g,h)=s(g)s(h)s(gh)^{-1}\in K$. Because $K$ is abelian, $g\cdot k=s(g)ks(g)^{-1}$ is an action of $G$ on $K$ independent of $s$, with dual action $(g\cdot\alpha)(k)=\alpha(g^{-1}\cdot k)$ on $\widehat K$. We denote by $G_\alpha$ the stabilizer of $\alpha$.

\paragraph{Topological point operators.} The trivial-degree part $A_{1_G}=\mathbb C[K]$ contains the point operators at the ends of the $K$ lines. Its idempotents are the projectors \eqref{eq:projectors} of the partial gauging $\mathcal T/K$ \cite[Equation 2.6]{Robbins:2022wlr}, which we denote by $e_\alpha$, $\alpha\in\widehat K$,
\begin{equation}\label{eq:trivial-idempotents}
\begin{aligned}
e_\alpha &= \frac{1}{|K|}\sum_{k\in K}\alpha(k)^{-1}k, \\[2pt]
e_\alpha\,k &= \alpha(k)\,e_\alpha, \qquad s(g)\,e_\alpha\,s(g)^{-1}=e_{g\cdot\alpha}.
\end{aligned}
\end{equation}
The second identity follows by relabeling the sum, and the third from the definition of the dual action.

\paragraph{The corner algebra.} By \eqref{eq:trivial-idempotents},
\begin{equation}
\begin{aligned}
e_\alpha A_g e_\alpha &= \mathbb{C}\,e_\alpha\,s(g)\,e_\alpha = \mathbb{C}\,e_\alpha e_{g\cdot\alpha}\,s(g) \\
&= \delta_{g\cdot\alpha,\alpha}\,\mathbb{C}\,w_g, \qquad w_g := e_\alpha s(g),
\end{aligned}
\end{equation}
and for $g,h\in G_\alpha$
\begin{equation}
w_gw_h=e_\alpha\,s(g)s(h)=e_\alpha\,c(g,h)\,s(gh)=\alpha\big(c(g,h)\big)\,w_{gh}.
\end{equation}
The corner of $A$ at $e_\alpha$, i.e., $e_{\alpha} A e_{\alpha}$, is therefore a twisted group algebra of the stabilizer, with $w_g$ in degree $g$,
\begin{equation}\label{eq:corner}
e_\alpha Ae_\alpha\cong\mathbb C^{\nu_\alpha}[G_\alpha],\qquad \nu_\alpha:=\alpha\circ c\,\big|_{G_\alpha\times G_\alpha}.
\end{equation}
Since $\alpha$ is $G_\alpha$-invariant, applying it to the identity 
\begin{equation}
    c(g,h)\,c(gh,\ell)=(g\cdot c(h,\ell))\,c(g,h\ell) ,
\end{equation}
which follows from associativity in $\Gamma$, shows that $\nu_\alpha$ is a 2-cocycle, and a change of section changes it by a coboundary. The convention of \eqref{eq:projectors} fixes whether one obtains $\nu_\alpha$ or its inverse. Replacing $\alpha$ by $\bar\alpha$ inverts $\nu_\alpha$, so the list of sectors below does not depend on that convention.

\paragraph{Decomposition of the algebra.} The orbit sums $e_{[\alpha]}=\sum_{\alpha'\in G\cdot\alpha}e_{\alpha'}$ commute with $\mathbb C[K]$ and with every $s(g)$, which permutes the orbit, 
so they are central idempotents of trivial degree and $A=\bigoplus_{[\alpha]}Ae_{[\alpha]}$ \textit{as algebras} in $\mathrm{Vec}(G)$. They are also all the topological point operators that $\mathcal C$ sees: $\mathrm{End}_{\mathcal C}(\mathbbm 1)$ consists of the central elements of $A$ of trivial degree, which form $\mathbb C[K]^G$, spanned by the $e_{[\alpha]}$. 
An individual $e_\alpha$ with a nontrivial orbit is not gauge invariant, and the sectors of $(\mathcal C,F)$ are labeled by $\widehat K/G$. 
Choosing one $\alpha$ per orbit, we obtain an idempotent $e=\sum_{[\alpha]}e_\alpha$ that is full, since $AeA$ contains every $s(g)e_\alpha s(g)^{-1}=e_{g\cdot\alpha}$ and therefore $\sum_{\beta\in\widehat K}e_\beta=1$. Corners at different orbits do not mix, since 
\begin{equation}
    e_\alpha A_ge_{\alpha'}\propto e_\alpha e_{g\cdot\alpha'}=0
\end{equation}
for $[\alpha]\neq[\alpha']$. 
As $e$ has trivial degree, $M\mapsto Me$ is an equivalence of $\mathrm{Vec}(G)$-module categories and $M\mapsto eMe$ a monoidal equivalence of categories of bimodules: $A$ is Morita equivalent in $\mathrm{Vec}(G)$ to $\bigoplus_{[\alpha]}\mathbb C^{\nu_\alpha}[G_\alpha]$. Gauging depends only on the Morita class of the algebra, and gauging $\mathbb C^{\nu}[H]$ in $\mathrm{Vec}(G)$ is gauging the subgroup $H$ with discrete torsion $\nu$. Hence
\begin{equation}\label{eq:trivial-main}
\boxed{
\begin{aligned}
[\mathcal{T}/\Gamma] &\cong \bigsqcup_{[\alpha]\in\widehat{K}/G}[\mathcal{T}/G_\alpha]_{\nu_\alpha}, \\[4pt]
\mathcal{C} &\simeq \mathrm{Bimod}_{\bigoplus_{[\alpha]}\mathbb{C}^{\nu_\alpha}[G_\alpha]}\big(\mathrm{Vec}(G)\big).
\end{aligned}
}
\end{equation}
The first statement is the decomposition formula \cite[Equation 2.3]{Robbins:2022wlr}, and the second is its refinement to symmetry categories, since conjugating a line $M$ by the invertible interface $Ae$ realizing the first statement gives $eA\otimes_A M\otimes_A Ae\cong eMe$. Another representative $g\cdot\alpha$ of the orbit gives the conjugate data $G_{g\cdot\alpha}=gG_\alpha g^{-1}$ and, up to a coboundary, $\nu_{g\cdot\alpha}=\nu_\alpha^{\,g}$, where $\nu^{g}(x,y)=\nu(g^{-1}xg,g^{-1}yg)$, hence an isomorphic sector. The counit inherited by the factors $Ae_{[\alpha]}$ fixes the relative Euler counterterms of the sectors (Section~\ref{sec:framework}).

\paragraph{The symmetry of each universe.} The diagonal block of universe $[\alpha]$ is
\begin{equation}\label{eq:trivial-diagonal}
\mathcal C_{[\alpha][\alpha]}\simeq\mathrm{Bimod}_{\mathbb C^{\nu_\alpha}[G_\alpha]}\big(\mathrm{Vec}(G)\big),
\end{equation}
the group-theoretical fusion category \cite[Section 9.7]{etingof2015tensor} that $[\mathcal T/G_\alpha]_{\nu_\alpha}$ inherits from the $\mathrm{Vec}(G)$ symmetry of $\mathcal T$. It is the dual of $\mathrm{Vec}(G)$ with respect to the module category $\mathcal M(G_\alpha,\nu_\alpha)$, where $\mathcal M(H,\mu):=\mathrm{Mod}_{\mathbb C^{\mu}[H]}(\mathrm{Vec}(G))$ denotes the indecomposable module categories of $\mathrm{Vec}(G)$ \cite[Theorem 3.1]{Ostrik:2002ohv}. Its simple lines are labeled by a double coset $G_\alpha gG_\alpha$ and an irreducible projective representation of $G_\alpha\cap gG_\alpha g^{-1}$ with cocycle $\nu_\alpha\,(\nu_\alpha^{\,g})^{-1}$ \cite[Proposition 3.1]{Ostrik:2002ohv}, \cite[Lemmas 3.3 and 3.5]{Natale:2016}. On the trivial double coset this cocycle is trivial, and these lines form the quantum symmetry $\mathrm{Rep}(G_\alpha)$ independently of $\nu_\alpha$: they form the dual of $\mathrm{Vec}(G_\alpha)$ with respect to a fiber functor, which is the representation category of a Drinfeld twist of $\mathbb C[G_\alpha]$ and hence tensor-equivalent to $\mathrm{Rep}(G_\alpha)$. The other double cosets are the remnants of the ungauged $G$ lines, in general non-invertible. In the extreme cases $G_\alpha=1$ and $G_\alpha=G$, one finds $\mathrm{Vec}(G)$ and $\mathrm{Rep}(G)$.

\paragraph{Interfaces.} Likewise, $\mathcal C_{[\alpha][\alpha']}$ is the category of $\mathbb C^{\nu_\alpha}[G_\alpha]$--$\mathbb C^{\nu_{\alpha'}}[G_{\alpha'}]$ bimodules in $\mathrm{Vec}(G)$, the interfaces obtained by partial gauging between the two orbifolds. Their simples are labeled by a double coset $G_\alpha gG_{\alpha'}$ and an irreducible projective representation of $G_\alpha\cap gG_{\alpha'} g^{-1}$ with cocycle $\nu_\alpha\,(\nu_{\alpha'}^{\,g})^{-1}$ \cite[Lemmas 3.3 and 3.5]{Natale:2016}.

\paragraph{Quantum symmetry and trivially acting lines.} The quantum symmetry of the whole $\Gamma$-gauging is the functor $\mathcal F:\mathrm{Rep}(\Gamma)\to\mathcal C$, $V\mapsto A\otimes V$, where $A$ acts on $A\otimes V$ by left multiplication and from the right by $(x\otimes v)\cdot\gamma=x\gamma\otimes\gamma^{-1}v$, $\gamma\in\Gamma$. For $V$ inflated from $G$, $K$ acts on $A\otimes V$ through $A$ only, so $\mathcal F(V)$ does not connect different orbits, and its component in universe $[\alpha]$, transported through the corner, is
\begin{equation}\label{eq:trivial-quantum}
\mathcal F(V)\big|_{[\alpha]}\simeq\mathbb C^{\nu_\alpha}[G_\alpha]\otimes\mathrm{Res}^G_{G_\alpha}V .
\end{equation}
Thus $V$ acts trivially on universe $[\alpha]$ precisely when its restriction to $G_\alpha$ is a multiple of the trivial representation.

\paragraph{Types of pairs.} Since $\mathrm{Vec}(G)$ is fusion, nonzero module functors connect all the module categories $\mathcal M(G_\alpha,\nu_\alpha)$ of the summands, so $\mathcal C$ is indecomposable: relative to the inherited symmetry, \textit{gauging in a fusion category never produces Type~III}. If $\mathcal T$ does not decompose and $G$ acts faithfully, the sectors $[\alpha]$ are the universes of $[\mathcal T/\Gamma]$. Indeed, a topological point operator of $[\mathcal T/\Gamma]$ ends a line $F_G(\pi(\gamma))$ of $\mathcal T$ for some $\gamma\in\Gamma$, and $\mathrm{Hom}(\mathbbm 1,F_G(\pi(\gamma)))=0$ for $\gamma\notin K$. The topological point operators are therefore the $\Gamma$-invariant combinations of the endpoints of the $K$ lines, which $\Gamma$ permutes without phases by the hypotheses of the setup. Their number is the number of $G$-orbits on $K$, which equals the number of $G$-orbits on $\widehat K$, the dimension of $\mathrm{End}_{\mathcal C}(\mathbbm 1)$, since an automorphism of $K$ and its dual have equally many fixed points. As $(\mathcal C,F)$ already sees all topological point operators, no symmetry extending it splits the universes, so no two universes form a Type~III pair relative to $(\mathcal C,F)$ or any symmetry extending it (Section~\ref{sect:org}). Two universes are of Type~I if and only if $\mathcal M(G_\alpha,\nu_\alpha)$ and $\mathcal M(G_{\alpha'},\nu_{\alpha'})$ are equivalent as $\mathrm{Vec}(G)$-module categories, which for untwisted $\mathrm{Vec}(G)$ means that the pairs $(G_\alpha,[\nu_\alpha])$ and $(G_{\alpha'},[\nu_{\alpha'}])$ are $G$-conjugate \cite[Theorem 1.1]{Natale:2016}. They are of Type~IIa if the categories \eqref{eq:trivial-diagonal} are equivalent without such conjugacy, and of Type~IIb otherwise. Distinct orbits with conjugate data give Type~I pairs, as in the $Q_8$ example of Section~\ref{sec:typesI-II}.

\paragraph{Partial gauging.} Gauging the trivially acting $K$ alone gives $|K|$ copies of $\mathcal T$ permuted by the $G$ lines. The extension enters only through the monoidal structure of this action, which on the stabilizer of copy $\alpha$ is $\nu_\alpha$: the copies form the $\mathrm{Vec}(G)$-module category 
\begin{equation}
\mathrm{Mod}_A(\mathrm{Vec}(G))\simeq\boxplus_{[\alpha]}\mathcal M(G_\alpha,\nu_\alpha),
\end{equation}
and gauging $G$ gives \eqref{eq:trivial-main}.

\section{Types I and II: examples}\label{sec:typesI-II}

In this section, we illustrate Section~\ref{sec:trivial} in examples. We first treat identical universes (Type~I), then the decomposition of $[\mathcal T/S_3]$ and four further orbifolds, which contain pairs of Types~I and~II, and finally the SymTFT distinction between the two kinds of Type~II pairs (Section~\ref{sec:IIa-IIb}). Throughout, $\mathcal T$ does not decompose and, except for the $Q_8$ case of the central $\mathbb Z_2$ example, $G$ acts faithfully, so that by Section~\ref{sec:trivial} the sectors are universes. In each orbifold, $\Gamma$ acts through $\pi:\Gamma\to G$ on a theory $\mathcal T$ with a faithful non-anomalous $G$ symmetry, which provides the compatible trivialization of $K$ assumed in Section~\ref{sec:trivial}.

\subsection{Type I}

In this subsection, we derive the symmetry of decomposing theories whose universes are all identical. Identical universes are a special case of decomposition, but they carry the most symmetry and are a natural starting point. One way to construct them in two dimensions is to gauge a trivially acting $\Z^{(0)}_n$ symmetry of a theory $\mathcal T$,
\begin{equation}\label{eq:identical-universes}
  \boxed{   \mathfrak{T} := [\mathcal{T}/\Z_n] \cong \bigsqcup_{i=1}^{n} \mathcal{T} ,}
\end{equation}
a decomposing theory whose $n$ universes are all copies of $\mathcal{T}$. Since $H^2(\mathbb Z_n,U(1))=0$, neither a twisted trivialization nor discrete torsion can occur, and this is the case $\Gamma=K=\Z_n$, $G=1$ of \eqref{eq:trivial-main}. The examples below derive its symmetry directly, for two, three and $n$ universes.

\subsubsection*{Two identical universes}

Let us start with the simplest case of Equation \eqref{eq:identical-universes}, namely $n=2$,
\begin{equation}
    \mathfrak{T} = [\mathcal{T}/\Z_2] \cong \mathcal{T} \sqcup \mathcal{T} .
\end{equation}
To begin with, we assume that $\mathcal{T}$ has no additional symmetry beyond the trivial symmetry. We first work out the symmetry structure of $\mathfrak{T}$ directly, by physical reasoning, and then derive it from scratch by performing the gauging.

We denote the two identical universes as $1$ and $2$. We should have precisely two topological point operators, $\Pi_1$ and $\Pi_2$, which project onto each universe, and satisfy
\begin{equation}
\Pi_1^2=\Pi_1,\quad\Pi_2^2=\Pi_2,\quad\Pi_1\Pi_2=\Pi_2\Pi_1=0.
\end{equation}
The identity topological point operator on the full theory is given by the sum, 
\begin{equation}
    \mathbbm 1=\Pi_1+\Pi_2.
\end{equation}
What about topological line operators?  This decomposing theory has an invertible $\mathbb{Z}_2$ symmetry generated by swapping the two universes, so we have line operators $\mathbbm 1$ and $S$ satisfying the fusion rules
\begin{equation}
    \mathbbm 1^2=S^2=\mathbbm 1,\quad \mathbbm 1 S=S\mathbbm 1=S.
\end{equation}
It should also be true that $\mathbbm 1^\dagger=\mathbbm 1$ and $S^\dagger=S$, where $\dagger$ indicates a reversal of the orientation of the line operator.

But we have more than this.  We have a line operator that acts as an interface between the universes with universe $1$ on the left (in the direction of the line operator) and universe $2$ on the right, implementing the isomorphism between universes.  We also have line operators that act as the identity line but only in one universe or the other. 

These are obtained by inserting projectors on the line operators $\mathbbm 1$ and $S$. Inserting $\Pi_1$ on the identity line gives a line that acts as the identity line in universe $1$ and as zero in universe $2$. We call this line $E_{11}$, and similarly $E_{22}$ is the identity line of universe $2$. In the notation of Section~\ref{sec:dictionary}, these are the lines $\mathbbm 1_i=(\mathbbm 1,\Pi_i)$. Inserting a projector on $S$ instead gives an interface between the universes. Since $S$ exchanges the projectors, a projector changes its label when it is moved across $S$,
\begin{equation}\label{eq:cross1}
 \Pi_1\,S=S\,\Pi_2,\qquad \Pi_2\,S=S\,\Pi_1 .
\end{equation}
Hence the line
\begin{equation}
 E_{12}:=\Pi_1S=S\,\Pi_2
\end{equation}
has universe $1$ on the left and universe $2$ on the right. Its orientation reversal 
\begin{equation}
    E_{21}:=E_{12}^\dagger=\Pi_2S=S\,\Pi_1
\end{equation}
has universe $2$ on the left and universe $1$ on the right.

Inserting $\mathbbm 1=\Pi_1+\Pi_2$ on $\mathbbm 1$ and on $S$, we find that the invertible symmetry lines are sums of these:
\begin{equation}
 \mathbbm 1=E_{11}\oplus E_{22}, \qquad S=E_{12}\oplus E_{21}.
\end{equation}
So the invertible symmetries are not simple in this theory, a consequence of the decomposition.

The four lines $E_{ij}$ are the simple topological lines of $\mathfrak{T}$. Moreover, taking $E_{ij}$ to be fundamental, they have fusion products
\begin{center}
\begin{tabular}{c|cccc}
\toprule
$\otimes$ & $E_{11}$ & $E_{12}$ & $E_{21}$ & $E_{22}$\\
\midrule
$E_{11}$ & $E_{11}$ & $E_{12}$ & $0$ & $0$\\
$E_{12}$ & $0$ & $0$ & $E_{11}$ & $E_{12}$\\
$E_{21}$ & $E_{21}$ & $E_{22}$ & $0$ & $0$\\
$E_{22}$ & $0$ & $0$ & $E_{21}$ & $E_{22}$\\
\bottomrule
\end{tabular}
\end{center}

If we write the full Hilbert space as $\mathcal{H}_1\oplus\mathcal{H}_2$, then by wrapping any of these line operators on the cylinder we can get a corresponding operator on the Hilbert space, denoted with a hat.  Then in block form we have
\begin{equation}
\begin{aligned}
    \widehat{E}_{11} &= \left(\begin{matrix} 1 & 0 \\ 0 & 0 \end{matrix}\right),\quad &\widehat{E}_{22} &= \left(\begin{matrix} 0 & 0 \\ 0 & 1 \end{matrix}\right), \\
    \widehat{E}_{12} &= \left(\begin{matrix} 0 & 1 \\ 0 & 0 \end{matrix}\right),\quad &\widehat{E}_{21} &= \left(\begin{matrix} 0 & 0 \\ 1 & 0 \end{matrix}\right).
\end{aligned}    
\end{equation}
Of course the invertible operators are simply
\begin{equation}
    \widehat{\mathbbm 1}=\left(\begin{matrix} 1 & 0 \\ 0 & 1 \end{matrix}\right),\qquad\widehat{S}=\left(\begin{matrix} 0 & 1 \\ 1 & 0 \end{matrix}\right).
\end{equation}
We can also contemplate fusing one of our point operators onto a line operator from either the left or the right, and these are given by
\begin{align}
    \Pi_1E_{11} & = E_{11}\Pi_1 = E_{11}, & \Pi_2E_{11} & =E_{11}\Pi_2=0, \nonumber \\
    \Pi_1E_{22}& =E_{22}\Pi_1=0, & \Pi_2E_{22}& =E_{22}\Pi_2=E_{22},\nonumber\\
    \Pi_1E_{12} & =E_{12}\Pi_2=E_{12}, & \Pi_2E_{12} & =E_{12}\Pi_1=0,\nonumber\\
    \Pi_1E_{21}& =E_{21}\Pi_2=0, & \Pi_2E_{21} & =E_{21}\Pi_1=E_{21}.
\end{align}
These imply that the operator $\mathbbm 1$ acts as the identity and the operator $S$ acts as
\begin{equation}
    S\cdot\Pi_1=\Pi_2,\qquad S\cdot\Pi_2=\Pi_1,
\end{equation}
as we would expect from the swap symmetry.

This is the symmetry structure of two identical universes: four simple lines $E_{ij}$, with fusion $E_{ij}\otimes E_{kl}=\delta_{jk}E_{il}$, a unit object that is not simple, $\mathbbm{1}=E_{11}\oplus E_{22}$. All junction spaces are one-dimensional, and the associator is a 3-cocycle on the pair groupoid of the universes, which is contractible. It can therefore be gauged to the identity, and the fusion rules determine the category, as the gauging construction below confirms. Hence the symmetry category of $\mathfrak{T}$ is the multi-fusion category
\begin{equation}
    \text{Mat}_2(\text{Vec})=\begin{pmatrix} \text{Vec} & \text{Vec} \\ \text{Vec} & \text{Vec} \end{pmatrix}.
\end{equation} 

Let us now derive this structure from scratch, by gauging the trivially acting $\Z_2$ symmetry of $\mathcal{T}$. The gauging constructions of \cite{Bhardwaj:2017xup,Perez-Lona:2023djo, Bhardwaj:2024qrf,Perez-Lona:2024sds} allow us to argue at the level of the symmetry and its action: we only need the effective presentation and its trivialization (Section~\ref{sec:framework}), not the theory $\mathcal{T}$ itself. A $\mathbb{Z}_2^{(0)}$ symmetry is described by (the action of) the fusion category $\text{Vec}(\mathbb{Z}_2)$. The specification that it is furthermore a trivially acting symmetry means that its action factors through the forgetful functor $\text{Vec}(\mathbb{Z}_2) \to \text{Vec}$, which forgets the $\mathbb{Z}_2$-grading:
\begin{equation}\label{eq:forgetful-functor-z2}
  q:  \mathrm{Vec}(\mathbb{Z}_2) \longrightarrow \mathrm{Vec}.
\end{equation}
The gaugeable symmetries of a symmetry category are specified by symmetric special Frobenius algebra objects. The algebra object associated with the whole $\Z_2$ symmetry is the regular representation $\C[\Z_2]$ as an algebra object in $\mathrm{Vec}(\mathbb{Z}_2)$. But since the symmetry is trivially acting, the relevant algebra object is more precisely the image under the forgetful functor \eqref{eq:forgetful-functor-z2}, which is the group algebra object $\C[\Z_2]$ but as an object in $\mathrm{Vec}$. Briefly, $A=q(\C[\Z_2])$ with $\Z_2 =\{e,s\}$, a commutative
separable Frobenius algebra with the coalgebra structure 
\begin{equation}
    \varepsilon \left(\sum\lambda_g g \right) = \lambda_e, \quad \Delta(g)=\sum_h h\otimes h^{-1}g.
\end{equation}
We write $\widehat{\Z_2}=\{\one,\epsilon\}$ for its characters, so
$A\cong\C \times\C$ as algebras, with idempotents $e_\pm=\tfrac12(e\pm s)$.

To obtain the symmetry category of the gauged theory, we need to compute the category $\mathrm{Bimod}_{\C[\Z_2]}(\mathrm{Vec})$ of $\C[\Z_2]$-bimodules in $\mathrm{Vec}$ \cite[Section 4.4]{Bhardwaj:2017xup}. This is equivalently the category of $B:=A\otimes A^{\mathrm{op}}$-modules in $\mathrm{Vec}$, where $A^{\mathrm{op}}$ is the opposite algebra. Since $A$ is commutative, $A^{\mathrm{op}}\cong A$, and we have $B=\C[\Z_2\times \Z_2]$. Hence
\begin{equation}
\Bimod_A(\Vect)\;\simeq_{\mathrm{ab.cat.}}\;\Rep(\Z_2\times\Z_2)
\end{equation}
as an abelian category: four simples, all one-dimensional, labeled
\begin{equation}
S_{(\chi_1,\chi_2)},\quad \chi_1,\chi_2\in\widehat{\Z_2}.
\end{equation}
Here $\chi_1$ records the left $\Z_2$-action and
$\chi_2$ the right one as $a\cdot m=\chi_1(a)m$, $m\cdot b=\chi_2(b)m$. From the group law, we can see that the unit is $A$ itself
\begin{equation}
    \mathbbm{1} \;=\;A\;\cong\;S_{(\one,\one)}\ \oplus\ S_{(\epsilon,\epsilon)} ,
\end{equation}
indicating that the symmetry category is a multi-fusion category. The two remaining simples $S_{(\one,\epsilon)}$, $S_{(\epsilon,\one)}$ must therefore be off-diagonal. This we can verify by computing the fusion rules, which are given by the tensor product over $A$. For $M,N\in\Bimod_A(\Vect)$, $m\in M$, $n\in N$, and $k=(g,h)\in\Z_2\times\Z_2$, the module structure on the tensor product $M\otimes_A N$ is
\begin{equation}
   k\cdot(m\otimes n)=(gm)\otimes(nh^{-1}),
\end{equation}
so the left tensor factor is acted on by $(g,e)$ and the right factor by $(e,h)$. This gives the fusion $S_{(\chi_1,\chi_2)}\otimes_A S_{(\psi_1,\psi_2)}$ by imposing the relations $(m\cdot a)\otimes n-m\otimes(a\cdot n)$, so that
\begin{equation}
S_{(\chi_1,\chi_2)}\otimes_A S_{(\psi_1,\psi_2)}\;=\;\delta_{\chi_2,\psi_1}\,S_{(\chi_1,\psi_2)}.
\end{equation}
The fusion rules are summarized in the following table, for $E_{11}:=S_{(\one,\one)}$, $E_{12}:=S_{(\one,\epsilon)}$, $E_{21}:=S_{(\epsilon,\one)}$ and $E_{22}:=S_{(\epsilon,\epsilon)}$:
\begin{center}
\begin{tabular}{c|cccc}
\toprule
$\otimes_A$ & $E_{11}$ & $E_{12}$ & $E_{21}$ & $E_{22}$\\
\midrule
$E_{11}$ & $E_{11}$ & $E_{12}$ & $0$ & $0$\\
$E_{12}$ & $0$ & $0$ & $E_{11}$ & $E_{12}$\\
$E_{21}$ & $E_{21}$ & $E_{22}$ & $0$ & $0$\\
$E_{22}$ & $0$ & $0$ & $E_{21}$ & $E_{22}$\\
\bottomrule
\end{tabular}
\end{center}
The associator information is inherited from the associator of $\Vect$, which is trivial. Thus we identify the symmetry category of the gauged theory as a multi-fusion category:
\begin{equation}
\Bimod_{\C[\Z_2]}(\Vect)\;\simeq\;\Mat_2(\Vect)
=\begin{pmatrix}\Vect&\Vect\\ \Vect&\Vect\end{pmatrix}, 
\end{equation}
where the unit object splits as
\begin{equation}
\mathbbm{1}=E_{11}\oplus E_{22}.
\end{equation}
Physically, the two diagonal blocks $\Vect$ correspond to the two identical universes. The simple object in each is the identity line $\mathbbm 1_i$ of that universe. Meanwhile, the off-diagonal blocks $\Vect$ correspond to the topological lines that implement the isomorphism between the two universes.

\subsubsection*{Three identical universes}

For three identical universes,
\begin{equation}
\mathfrak{T}=[\mathcal{T}/\Z_3] \cong \mathcal{T} \sqcup \mathcal{T} \sqcup \mathcal{T},
\end{equation}
where, as for two universes, $\mathcal T$ has no topological lines besides its identity line, $\mathrm{Top}(\mathcal T)\simeq\mathrm{Vec}$. The physics is the same as for two universes, with one new feature: the invertible symmetry that permutes identical universes is now $S_3$, which is larger than the quantum symmetry $\widehat{\Z}_3$, whereas for two universes the two coincide, $S_2 \cong \widehat{\Z}_2$. As before, we first work out the symmetry of $\mathfrak{T}$ by physical reasoning, through the point operators, the invertible lines, the simple lines obtained by inserting projectors on them, and their fusion, and then derive it by gauging.

The topological point operators are the projectors $\Pi_i$, $i=1,2,3$, onto the universes, which satisfy
\begin{equation}
\Pi_i\Pi_j =\delta_{ij}\Pi_i, \qquad \mathbbm 1=\Pi_1+\Pi_2+\Pi_3 .
\end{equation}
The permutations of the universes give invertible lines $I$, $R$ and $S$ with the defining relations of $S_3$,
\begin{equation}
    R^3=I,\qquad S^2=I,\qquad SRS=R^{-1}.
\end{equation}
Gauging $\Z_3$ provides the quantum symmetry $\mathrm{Rep}(\mathbb{Z}_3) \simeq \mathrm{Vec}(\mathbb{Z}_3)$, which is generated by a line of order three, so the quantum lines are $I$, $R$ and $R^2$, the cyclic subgroup $\Z_3\subset S_3$. The other permutations, such as $S$, are not in the image of the quantum symmetry. They arise from the Karoubi completion \cite{Gaiotto:2019xmp,Bah:2025oxi}, which inserts projectors on the quantum lines and splits them into interfaces, whose direct sums give arbitrary permutations.

In the Karoubi completion of a category, one adds, for every object $X$ and every idempotent $p\in\mathrm{End}(X)$, an object $(X,p)$ representing the image of $p$, so that $X\cong\bigoplus_j(X,p_j)$ for any orthogonal idempotents $p_j$ with $\sum_jp_j=\mathrm{id}_X$. By itself, this does nothing to $\mathrm{Rep}(\Z_3)$, which is semisimple and therefore already Karoubi complete. What gauging a trivially acting symmetry adds are the point operators $\Pi_j$, which can be inserted on any line. The space of topological junctions of each of the lines $I$, $R$ and $R^2$ with itself is therefore $\mathbb{C}^3$ rather than $\mathbb{C}$, and the Karoubi completion with respect to these idempotents produces the interfaces.

We insert the projectors on the three quantum lines first, since they already produce all simple lines of $\mathfrak{T}$, and recombine the pieces into the rest of $S_3$ afterwards. Inserting $\Pi_j$ on the identity line gives the identity line $E_{jj}$ of universe $j$. Since $R$ maps universe $j$ to universe $j+1$, with indices modulo $3$,
\begin{equation}\label{eq:cross3}
    \Pi_{j+1}\,R=R\,\Pi_j ,
\end{equation}
so inserting $\Pi_j$ on $R$ and on $R^2$ gives the lines
\begin{equation}
    E_{j+1,j} := \Pi_{j+1}\,R=R\,\Pi_j ,\qquad E_{j+2,j} :=\Pi_{j+2}R^2=R^2\Pi_j .
\end{equation}
This yields nine lines $E_{ij}$, $i,j\in\{1,2,3\}$, with $E_{ij}^\dagger=E_{ji}$, and the quantum lines are their sums,
\begin{equation}
    \begin{aligned}
    I&=E_{11}\oplus E_{22}\oplus E_{33},\\
    R&=E_{21}\oplus E_{32}\oplus E_{13},\\
    R^2&=E_{12}\oplus E_{23}\oplus E_{31}.
    \end{aligned}
\end{equation}
The $E_{ij}$ are the simple topological lines of $\mathfrak{T}$: since $\mathrm{Top}(\mathcal T)\simeq\mathrm{Vec}$, every line within universe $j$ is a multiple of $E_{jj}$, and every interface with universe $i$ on its left and universe $j$ on its right is a multiple of $E_{ij}$. Since $S$ exchanges universes $1$ and $2$ and fixes universe $3$, inserting $\mathbbm 1=\Pi_1+\Pi_2+\Pi_3$ on the other permutations gives
\begin{equation}
    \begin{aligned}
    S &=E_{12}\oplus E_{21}\oplus E_{33},\\
    RS &=E_{13}\oplus E_{31}\oplus E_{22},\\
    R^2S &= E_{23}\oplus E_{32}\oplus E_{11},
\end{aligned}
\end{equation}
and these combinations satisfy the $S_3$ relations.

The simple lines fuse as $E_{ij}\otimes E_{kl}=\delta_{jk}E_{il}$ (Table~\ref{tab:fusion-n3}).
\begin{table*}[ht]
\centering
\small
\begin{tabular}{c|ccccccccc}
\toprule
$\otimes$ & $E_{11}$ & $E_{12}$ & $E_{13}$ & $E_{21}$ & $E_{22}$ & $E_{23}$ & $E_{31}$ & $E_{32}$ & $E_{33}$\\
\midrule
$E_{11}$ & $E_{11}$ & $E_{12}$ & $E_{13}$ & $0$ & $0$ & $0$ & $0$ & $0$ & $0$\\
$E_{12}$ & $0$ & $0$ & $0$ & $E_{11}$ & $E_{12}$ & $E_{13}$ & $0$ & $0$ & $0$\\
$E_{13}$ & $0$ & $0$ & $0$ & $0$ & $0$ & $0$ & $E_{11}$ & $E_{12}$ & $E_{13}$\\
$E_{21}$ & $E_{21}$ & $E_{22}$ & $E_{23}$ & $0$ & $0$ & $0$ & $0$ & $0$ & $0$\\
$E_{22}$ & $0$ & $0$ & $0$ & $E_{21}$ & $E_{22}$ & $E_{23}$ & $0$ & $0$ & $0$\\
$E_{23}$ & $0$ & $0$ & $0$ & $0$ & $0$ & $0$ & $E_{21}$ & $E_{22}$ & $E_{23}$\\
$E_{31}$ & $E_{31}$ & $E_{32}$ & $E_{33}$ & $0$ & $0$ & $0$ & $0$ & $0$ & $0$\\
$E_{32}$ & $0$ & $0$ & $0$ & $E_{31}$ & $E_{32}$ & $E_{33}$ & $0$ & $0$ & $0$\\
$E_{33}$ & $0$ & $0$ & $0$ & $0$ & $0$ & $0$ & $E_{31}$ & $E_{32}$ & $E_{33}$\\
\bottomrule
\end{tabular}
\caption{Fusion of the nine simple lines of three identical universes.}
\label{tab:fusion-n3}
\end{table*}
On the Hilbert space $\mathcal H_1\oplus\mathcal H_2\oplus\mathcal H_3$, the lines $E_{ij}$ act as matrix units, and the invertible lines act as the permutation matrices
\begin{equation}
\widehat{R}=\begin{pmatrix}0&0&1\\1&0&0\\0&1&0\end{pmatrix},\qquad
\widehat{S}=\begin{pmatrix}0&1&0\\1&0&0\\0&0&1\end{pmatrix}.
\end{equation}
As for two universes, all junction spaces are one-dimensional and the associator can be gauged away, so the symmetry of $\mathfrak T$ is the multi-fusion category
\begin{equation}
\mathrm{Mat}_3(\mathrm{Vec})=\begin{pmatrix}\mathrm{Vec}&\mathrm{Vec}&\mathrm{Vec}\\ \mathrm{Vec}&\mathrm{Vec}&\mathrm{Vec}\\ \mathrm{Vec}&\mathrm{Vec}&\mathrm{Vec}\end{pmatrix}.
\end{equation}

Similarly to the 2-universe case, we now derive this structure via bimodules. Gauging the trivially acting $\Z_3$ results in the symmetry $\Bimod_{\C[\Z_3]}(\Vect)$. Since $\C[\Z_3]$ is commutative,
\begin{equation}
    \Bimod_{\C[\Z_3]}(\Vect) \simeq_{\text{ab.cat.}}\Mod(\C[\Z_3]^{\otimes 2}) \simeq \Rep(\Z_3\times \Z_3)
\end{equation}
as abelian categories. For $\psi_1,\psi_2\in\widehat{\Z}_3$, let $S_{(\psi_1,\psi_2)}$ be the one-dimensional bimodule with $a\cdot m=\psi_1(a)m$ and $m\cdot b=\psi_2(b)m$. With $\chi$ a generator of $\widehat{\Z}_3$, the nine simple objects are
\begin{equation}
    E_{ij}:=S_{(\chi^i, \chi^j)}, \qquad i,j\in \{1,2,3\}.
\end{equation}
For bimodules $M$ and $N$, the $\C[\Z_3]^{\otimes 2}$-module structure on $M\otimes_{\C[\Z_3]} N$ is $(g,h)\cdot(m\otimes n)=(gm)\otimes(nh^{-1})$, and imposing the relations $(m\cdot a) \otimes n - m\otimes (a \cdot n)$ gives
\begin{equation}
    S_{(\psi_1, \psi_2)} \otimes_{\C[\Z_3]}S_{(\psi'_1, \psi'_2)} = \delta_{\psi_2, \psi'_1}\, S_{(\psi_1, \psi'_2)},
\end{equation}
which is Table~\ref{tab:fusion-n3}. The unit is $\C[\Z_3]\cong E_{11} \oplus E_{22} \oplus E_{33}$, so the category is multi-fusion, and the six simple objects $E_{ij}$ with $i\neq j$ are off-diagonal. The quantum line of $\chi$ is $\C[\Z_3]\otimes\chi\cong\bigoplus_j S_{(\chi^j,\chi^{j-1})}=E_{21}\oplus E_{32}\oplus E_{13}$, which is $R$, so the labels agree with those of the physical derivation. The associator is inherited from $\mathrm{Vec}$ and is trivial, so
\begin{equation}
    \Bimod_{\C[\Z_3]}(\Vect) \simeq \Mat_3(\Vect).
\end{equation}
Physically, the diagonal blocks $\mathcal C_{ii}\simeq\mathrm{Vec}$, each spanned by the identity line $\mathbbm{1}_i=E_{ii}$, correspond to the three identical universes, and the off-diagonal blocks contain the lines $E_{ij}$ that implement the isomorphisms between them.

\subsubsection*{$n$ identical universes}

The same steps for $[\mathcal T/\mathbb Z_n]$ give $n^2$ simple lines $E_{ij}$, obtained by inserting projectors on the quantum $\mathbb Z_n$ lines, with $E_{ij}\otimes E_{kl}=\delta_{jk}E_{il}$ and an associator that can be gauged away. Hence $\mathcal C = \Bimod_{\C[\Z_n]}(\Vect)\simeq\mathrm{Mat}_n(\mathrm{Vec})$. Its invertible objects are the permutation matrices, and the full group $S_n$ that they form acts on the same theory through a tensor functor $\mathrm{Vec}(S_n)\to\mathrm{Mat}_n(\mathrm{Vec})$ that extends the quantum $\mathbb Z_n$ symmetry.

\subsection{Type II}
\label{sec:TypeII}

In this subsection, we work out orbifolds of Section~\ref{sec:trivial} whose universes are related by non-invertible interfaces: $[\mathcal T/S_3]$ in detail, then four further examples. The distinction between equivalent and merely Morita-equivalent diagonal blocks is the subject of Section~\ref{sec:IIa-IIb}.

\subsubsection*{$\Z_3\leq S_3$ trivially acting subsymmetry}

Let $\mathcal{T}$ carry a non-anomalous $S_3$ symmetry whose normal subgroup $\Z_3$ acts trivially. Gauging $S_3$ then produces a decomposing theory with a multi-fusion symmetry, and the decomposition formula \cite[Equation 4.11]{Robbins:2022wlr} reads
\begin{equation}\label{eq:s3-decomp}
    \mathfrak T = [\mathcal T / S_3] \cong [\mathcal{T} / \Z_2] \sqcup \mathcal T .
\end{equation}
This is the case $\Gamma=S_3$, $K=\mathbb Z_3$, $G=\mathbb Z_2$ of Section~\ref{sec:trivial}. The two orbits $\{\one\}$ and $\{\zeta,\zeta^2\}$ on $\widehat{\mathbb Z}_3$ have stabilizers $\mathbb Z_2$ and $1$, and all $[\nu_\alpha]$ are trivial. Hence \eqref{eq:trivial-main} predicts the quantum symmetry $\mathrm{Rep}(\mathbb Z_2)$ in the first universe $[\mathcal T/\mathbb Z_2]$, the effective $\mathrm{Vec}(\mathbb Z_2)$ in the second universe $\mathcal T$, and one simple object in each off-diagonal block. The authors of \cite[Section 4.2]{Robbins:2022wlr} observed line operators connecting the two universes from physical arguments, and we return to their observations below. We now confirm the prediction by a direct computation, with details in Appendix~\ref{app:S3}.

Since $\Z_3$ acts trivially, the action of $S_3$ factors through the pushforward along $\mathrm{sgn}: S_3 \to \mathbb Z_2$,
\begin{equation}
\label{eqn:pushforward_along_sign}
    q:\mathrm{Vec}(S_3)\to\mathrm{Vec}(\Z_2),\quad \C_g\longmapsto\C_{\mathrm{sgn}(g)}.
\end{equation}
The gauged theory carries symmetry $\mathrm{Bimod}_A(\mathrm{Vec}(\Z_2))$ for $A:=q(\C[S_3])$. Unlike $\C[S_3]$ in $\mathrm{Vec}(S_3)$, the algebra $A$ is not haploid. Writing the $\Z_2$ grading additively, its trivial-degree component, $A_{1_G}$ in the notation of Section~\ref{sec:trivial}, is
\begin{equation}
    A_0=q(\C[\Z_3])\cong\C[\Z_3],
\end{equation}
which is three-dimensional, as is the odd component $A_1$. What makes $\mathrm{Bimod}_{A}(\mathrm{Vec}(\Z_2))$ multi-fusion rather than fusion is that $\mathrm{End}_{\mathcal C}(\mathbbm 1)=\C[\Z_3]^{\Z_2}$ is two-dimensional.

The symmetry $\mathcal C=\mathrm{Bimod}_A(\mathrm{Vec}(\mathbb Z_2))$ is the $\mathbb Z_2$-equivariantization $\mathcal C\simeq(\mathrm{Bimod}_{A_0}(\mathrm{Vec}))^{\mathbb Z_2}$ of $\mathrm{Mat}_3(\mathrm{Vec})$ under inversion on $\widehat{\mathbb Z}_3$ (Appendix~\ref{app:S3}). Here $\tau\in A_1$ is a transposition, $\kappa=\mathrm{Ad}_\tau|_{A_0}$ is the inversion automorphism of $A_0=\mathbb C[\mathbb Z_3]$, and $\kappa_*$ denotes the induced $\mathbb Z_2$ action $M\mapsto{}_\kappa M_\kappa$ on $\mathrm{Mat}_3(\mathrm{Vec})$.\footnote{
The action $\kappa_*$ is conjugation by the permutation $P=E_{\one\one}\oplus E_{\zeta\zeta^2}\oplus E_{\zeta^2\zeta}$, which exchanges the copies $\zeta,\zeta^2$ of $[\mathcal T/\mathbb Z_3]\cong \sqcup_3 \mathcal T$ and fixes the copy $\one$. The residual $\mathbb Z_2$ symmetry of $[\mathcal T/\mathbb Z_3]$, however, is the line $P\otimes L$, where $L$ generates the effective $\mathbb Z_2$ of $\mathcal T$: the transposition both permutes the twisted sectors and acts on each copy. Since $L$ commutes with $\mathrm{Mat}_3(\mathrm{Vec})$, both $P$ and $P\otimes L$ induce $\kappa_*$, so the equivariantization does not distinguish them, but the action functor does. Gauging $P\otimes L$ gives $[\mathcal T/\mathbb Z_2]\sqcup \mathcal T$, while gauging $P$ alone would give $\sqcup_3 \mathcal T$. The copies form a $G^{(0)}$-family in the sense of BI \cite{Brennan:2026ira}, with $G=\mathbb Z_2$ acting through $g\mapsto P\otimes L$, and equivariantizing gauges this $\mathbb Z_2$. Its permutation part is a transposition in the $S_3$ of universe permutations of $\mathrm{Mat}_3(\mathrm{Vec})$ found above for three identical universes.
}

In this presentation, the simple objects and their fusion are explicit. The projectors \eqref{eq:projectors} of $\Zth$, $e_\alpha = \tfrac{1}{3}\sum_{x \in \Zth} \alpha(x)^{-1}x$, decompose
\begin{equation}
    A_0 = \C e_{\one} \oplus \C e_{\zeta} \oplus \C e_{\zeta^2},
\end{equation}
which identifies $\mathrm{Bimod}_{A_0}(\mathrm{Vec}) \simeq \mathrm{Mat}_3(\mathrm{Vec})$, with matrix units $E_{ij}$, $i,j \in \{\one, \zeta, \zeta^2\}$, and $E_{ij} \otimes E_{kl} = \delta_{jk} E_{il}$. Since $\kappa$ inverts $\Zth$, it maps $e_\alpha$ to $e_{\alpha^{-1}}$, fixing $e_{\one}$ and exchanging $e_{\zeta}$ and $e_{\zeta^2}$. Denoting this involution on indices by an overline, $\kappa_*$ is $E_{ij} \mapsto E_{\bar\imath\bar\jmath}$.

The equivariantization has one simple object for each of the four free orbits and two for the fixed point $E_{\one\one}$ (Appendix~\ref{app:S3}), six in total:
\begin{gather}
    \mathbbm 1_1 = E_{\one\one}, \quad \mathrm{sgn}_1 = E_{\one\one} \otimes \mathrm{sgn}, \quad \mathbbm 1_2 = E_{\zeta\zeta} \oplus E_{\zeta^2\zeta^2}, \\
    g = E_{\zeta\zeta^2} \oplus E_{\zeta^2\zeta}, \quad X = E_{\one\zeta} \oplus E_{\one\zeta^2}, \quad Y = E_{\zeta\one} \oplus E_{\zeta^2\one}.
\end{gather}
The monoidal unit is $A_0 = \mathbbm 1_1 \oplus \mathbbm 1_2$, whose summands are the identity lines of the universes labeled by the orbits $\{e_{\one}\}$ and $\{e_{\zeta}, e_{\zeta^2}\}$ of $\tau$. The object $g$ is the effective $\mathbb Z_2$ line $L$ of the universe $\mathcal T$. With the fusion rules of Table~\ref{tab:S3-fusion}, derived in Appendix~\ref{app:S3}, the symmetry is
\begin{equation}
   \mathcal C\simeq\begin{pmatrix}\mathrm{Rep}(\Zt) & \langle X\rangle \\ \langle Y\rangle & \mathrm{Vec}(\Zt)\end{pmatrix},
\end{equation}
where $\langle X\rangle$ denotes the additive category generated by $X$.
\begin{table}[htbp]
\centering
\renewcommand{\arraystretch}{1.2}
\begin{tabular}{c|cccccc}
\toprule
$\otimes_A$ & $\mathbbm 1_1$ & $\mathrm{sgn}_1$ & $X$ & $\mathbbm 1_2$ & $g$ & $Y$ \\
\midrule
$\mathbbm 1_1$ & $\mathbbm 1_1$ & $\mathrm{sgn}_1$ & $X$ & $0$ & $0$ & $0$ \\
$\mathrm{sgn}_1$   & $\mathrm{sgn}_1$ & $\mathbbm 1_1$ & $X$ & $0$ & $0$ & $0$ \\
$X$      & $0$ & $0$ & $0$ & $X$ & $X$ & $\mathbbm 1_1 \oplus \mathrm{sgn}_1$ \\
$\mathbbm 1_2$ & $0$ & $0$ & $0$ & $\mathbbm 1_2$ & $g$ & $Y$ \\
$g$      & $0$ & $0$ & $0$ & $g$ & $\mathbbm 1_2$ & $Y$ \\
$Y$      & $Y$ & $Y$ & $\mathbbm 1_2 \oplus g$ & $0$ & $0$ & $0$ \\
\bottomrule
\end{tabular}
\caption{Fusion rules of $\mathrm{Bimod}_{A}(\mathrm{Vec}(\Z_2))$ for $A=q(\C[S_3])$.}
\label{tab:S3-fusion}
\end{table}

The diagonal blocks are equivalent as fusion categories, $\mathrm{Rep}(\Zt) \simeq \mathrm{Vec}(\Zt)$, and the off-diagonal blocks contain no invertible object, since $X\otimes Y\cong\mathbbm 1_1\oplus\mathrm{sgn}_1$ gives $X$ the Frobenius--Perron dimension $\sqrt2$. The pair is therefore of Type~IIa. In the SymTFT, which is the toric code, the two symmetry boundaries are exchanged by the $e\leftrightarrow m$ bulk symmetry (Section~\ref{sec:IIa-IIb}).

\paragraph{Comparison with earlier work.} The authors of \cite[Section 4.2]{Robbins:2022wlr} identified three topological point operators $\Pi_L$, $g_L$ and $\Pi_R$, where $\Pi_R$ is an idempotent orthogonal to $\Pi_L$ and $g_L^2=\Pi_L$. In our framework, the projectors are the identity lines of the two universes,
\begin{equation}
    \Pi_R = e_{\one}\ \leftrightarrow\ \mathbbm 1_1, \qquad \Pi_L = e_{\zeta} + e_{\zeta^2}\ \leftrightarrow\ \mathbbm 1_2,
\end{equation}
and they span the $\tau$-even part of $A_0$,
\begin{equation}
    \mathrm{End}(\mathbbm 1) = \big(\mathrm{End}_{\mathrm{Mat}_3(\mathrm{Vec})}(A_0)\big)^{\Zt} = \C \{e_{\one}, \, e_{\zeta} + e_{\zeta^2}\}.
\end{equation}
The $\tau$-odd combination $g_L\propto e_{\zeta}-e_{\zeta^2}$ (not to be confused with the simple object $g$) is not gauge invariant, so it is not a genuine point operator of $[\mathcal T/S_3]$. It ends the line $\mathcal F(\mathrm{sgn})$ introduced below, and in the universe $\mathcal T$, where $\mathrm{sgn}$ acts trivially, it is the junction that realizes this trivial action (see \eqref{eq:trivial-quantum} and \cite[Equations 4.20 and 4.21]{Robbins:2022wlr}). Once the trivialization identifies $\mathcal F(\mathrm{sgn})$ with the identity line on $\mathcal T$, the operator $g_L$ becomes $\pm\Pi_L$, which explains $g_L^2=\Pi_L$ and shows that it adds no universe. It is accounted for by the symmetry category rather than by $\mathrm{End}(\mathbbm 1)$.

\paragraph{The quantum symmetry.} We now locate the $\mathrm{Rep}(S_3)$ quantum symmetry. The pushforward $q$ of \eqref{eqn:pushforward_along_sign} sends $\mathbb{C}[S_3]$ to $A$ and induces the tensor functor
\begin{equation}
    \begin{split}
        \mathcal{F}:\mathrm{Rep}(S_3) \simeq \mathrm{Bimod}_{\mathbb{C}[S_3]}(\mathrm{Vec}(S_3)) & \rightarrow \mathrm{Bimod}_{A}(\mathrm{Vec}(\mathbb{Z}_2)),\\
        V & \mapsto \mathbb{C}[S_3] \otimes V,
    \end{split}
\end{equation}
where $\mathbb{C}[S_3] \otimes V$ is graded by $\mathrm{sgn}(x)$ and carries the bimodule actions
\begin{equation}
    h\cdot (x\otimes v) = hx \otimes v, \qquad (x\otimes v) \cdot h = xh \otimes h^{-1}v.
\end{equation}
Denote by $\mathbf{1}$, $\mathrm{sgn}$ and $\rho$ the trivial, sign and two-dimensional irreducible representations of $S_3$. On simple objects, $\mathcal F$ acts as follows, where $\mathcal F(\mathbf 1)$ and $\mathcal F(\mathrm{sgn})$ follow from \eqref{eq:trivial-quantum} and $\mathcal F(\rho)$ is determined in Appendix~\ref{app:S3}:
\begin{equation}
\label{eqn:RepS3functor}
\begin{aligned}
    \mathcal{F}(\mathbf 1) &= \begin{pmatrix}
        \mathbbm 1_1 & 0\\
        0 & \mathbbm 1_2
    \end{pmatrix}, \\
    \mathcal{F}(\mathrm{sgn}) &= \begin{pmatrix}
        \mathrm{sgn}_1 & 0\\
        0 & \mathbbm 1_2
    \end{pmatrix}, \\
    \mathcal{F}(\rho) &= \begin{pmatrix}
        0 & X \\
        Y & g
    \end{pmatrix}.
\end{aligned}
\end{equation}
Since $\rho$ restricts nontrivially to $\Z_3$, its image has off-diagonal components, so the quantum symmetry connects the two universes. Moreover, $\mathcal{F}$ is faithful but not full: $\mathrm{End}(\mathcal{F}(\rho)) \cong \mathbb{C}^3$.

\subsubsection*{$\mathbb Z_4\le Q_8$ trivially acting subsymmetry}
Take $\Gamma=Q_8$ with $K=\mathbb Z_4=\langle j\rangle$ acting trivially and $G=\mathbb Z_2$ generated by the image of $i$. Since $iji^{-1}=-j=j^{-1}$, $G$ acts on $\widehat K$ by inversion: the characters with $\alpha(j)=\pm1$ are fixed, and those with $\alpha(j)^2=-1$ are exchanged. All $[\nu_\alpha]$ are trivial because $H^2(\mathbb Z_2,U(1))=0$, and \eqref{eq:trivial-main} gives
\begin{equation}
    [\mathcal T/Q_8]\cong[\mathcal T/\mathbb Z_2] \sqcup [\mathcal T/\mathbb Z_2]\sqcup \mathcal T,
\end{equation}
as in \cite[Equation 4.23]{Robbins:2022wlr}. The two $[\mathcal T/\mathbb Z_2]$ universes have the same stabilizer and torsion, so they form a Type~I pair, related by the invertible quantum line of the one-dimensional representation $\chi$ of $Q_8$ with $\chi(j)=-1$. Each of them forms a Type~IIa pair with $\mathcal T$. By \eqref{eq:trivial-quantum}, the inflated $\mathrm{Rep}(\mathbb Z_2)$ acts trivially exactly on the universe $\mathcal T$, reproducing \cite[Equation 4.32]{Robbins:2022wlr}.

\subsubsection*{Central $\mathbb Z_2\le D_4,\,Q_8$}
Take $K=\mathbb Z_2$ the center of $\Gamma=D_4$ or $Q_8$, so that $G=\mathbb Z_2^2$. The commutator of lifts of the generators of $G$ is the central element, so $[\nu_\alpha]$ is nontrivial for the nontrivial $\alpha$, and \eqref{eq:trivial-main} gives
\begin{equation}
    [\mathcal T/\Gamma]\cong[\mathcal T/\mathbb Z_2^2] \sqcup [\mathcal T/\mathbb Z_2^2]_{\rm d.t.}.
\end{equation}
For $D_4$, with $G$ acting faithfully, these are universes with diagonal blocks $\mathrm{Rep}(\mathbb Z_2^2)\simeq\mathrm{Vec}(\mathbb Z_2^2)$ but non-conjugate data $(\mathbb Z_2^2,0)$ and $(\mathbb Z_2^2,\nu)$, so they form a Type~IIa pair. The off-diagonal block has a single simple object, labeled by the two-dimensional $\nu$-projective representation of $\mathbb Z_2^2$. It is non-invertible and is a component of the quantum line of the two-dimensional irreducible representation of $D_4$. For $Q_8$ as in the previous example, the image of $j$ acts trivially, so the two terms are only sectors, not universes. They split into the three universes found there: the first into $[\mathcal T/\mathbb Z_2] \sqcup [\mathcal T/\mathbb Z_2]$, since the image of $i$ fixes both copies of $\mathcal T$ produced by the image of $j$, and the second into $\mathcal T$, since the torsion phase makes the image of $i$ exchange them.

\subsubsection*{$\mathbb Z_2\times\mathbb Z_2\le S_4$ trivially acting subsymmetry}
Take $\Gamma=S_4$ with $K=\mathbb Z_2\times\mathbb Z_2$ acting trivially and $G=S_3$, which permutes the three nontrivial characters of $K$. The orbits are the trivial character, with stabilizer $S_3$, and the three nontrivial characters, with stabilizer $\mathbb Z_2$. The extension splits, so all $[\nu_\alpha]$ are trivial, and
\begin{equation}
    [\mathcal T/S_4]\cong[\mathcal T/S_3]\sqcup [\mathcal T/\mathbb Z_2].
\end{equation}
The diagonal blocks are $\mathrm{Rep}(S_3)$ and the dual of $\mathrm{Vec}(S_3)$ with respect to gauging $\mathbb Z_2\subset S_3$, which is again $\mathrm{Rep}(S_3)$ \cite[Page 44]{Bhardwaj:2017xup}. The stabilizers are not conjugate, so the pair is not of Type~I, and since the diagonal blocks are equivalent, it is of Type~IIa.

\subsubsection*{$\mathbb Z_2\times\mathbb Z_2\le(\mathbb Z_2\times\mathbb Z_2)\rtimes\mathbb Z_4$ trivially acting subsymmetry}
Take $\Gamma=(\mathbb Z_2\times\mathbb Z_2)\rtimes\mathbb Z_4$, with the generator of $\mathbb Z_4$ swapping the two factors, $K=\mathbb Z_2\times\mathbb Z_2$ acting trivially and $G=\mathbb Z_4$. On $\widehat K$ the generator swaps the two components. The characters $(0,0)$ and $(1,1)$ are fixed, with stabilizer $\mathbb Z_4$, while $(1,0)$ and $(0,1)$ form an orbit with stabilizer $2\mathbb Z_4\cong\mathbb Z_2$. The extension splits, so all $[\nu_\alpha]$ are trivial, and
\begin{equation}
    [\mathcal T/\Gamma]\cong[\mathcal T/\mathbb Z_4] \sqcup [\mathcal T/\mathbb Z_4] \sqcup [\mathcal T/\mathbb Z_2].
\end{equation}
The two $[\mathcal T/\mathbb Z_4]$ universes form a Type~I pair. The diagonal block of $[\mathcal T/\mathbb Z_4]$ is $\mathrm{Rep}(\mathbb Z_4)\simeq\mathrm{Vec}(\mathbb Z_4)$. That of $[\mathcal T/\mathbb Z_2]$ is the dual of $\mathrm{Vec}(\mathbb Z_4)$ with respect to gauging $H=\mathbb Z_2\subset\mathbb Z_4$. By the double-coset description of Section~\ref{sec:trivial}, its simples are labeled by the two cosets of $H$ and the two characters of $H$, all of dimension one. The residual line supported on the odd coset squares to a bimodule supported on $H$, which is the unit: multiplication in $\mathbb C[\mathbb Z_4]$ gives bimodule isomorphisms $\mathbb C[gH]\otimes_{\mathbb C[H]}\mathbb C[g'H]\cong\mathbb C[gg'H]$, and $g^2H=H$ for $g$ in the odd coset. The four invertible lines therefore form $\mathbb Z_2\times\mathbb Z_2$ rather than $\mathbb Z_4$, with the quantum $\mathbb Z_2$ and the residual $\mathbb Z_4/\mathbb Z_2$ carrying a mixed anomaly \cite{Tachikawa:2017gyf}. The fusion rings differ, so these pairs are of Type~IIb: gauging part of an abelian group can change the symmetry.

\subsection{Two kinds of Type~II pairs}\label{sec:IIa-IIb}
In this subsection, we show that the split of Type~II pairs into IIa and IIb has a sharp meaning in the SymTFT, given by \eqref{eq:IIa-IIb-symtft} and illustrated for the $\mathbb Z_4$ toric code, and we relate the two kinds of pairs to the families of BI.

In the SymTFT, a decomposing theory is a sandwich with a decomposable symmetry boundary. For bulk $\mathcal Z(\mathcal D)$ and symmetry boundary $\boxplus_i\mathcal M_i$, with $\mathcal M_i$ indecomposable $\mathcal D$-module categories, the symmetry is 
\begin{equation}
    \mathrm{Fun}_{\mathcal D}(\boxplus_i\mathcal M_i,\boxplus_i\mathcal M_i) ,
\end{equation}
whose blocks $\mathcal C_{ij}=\mathrm{Fun}_{\mathcal D}(\mathcal M_j,\mathcal M_i)$ are the boundary-changing lines. \textit{For $\mathcal D$ fusion, all blocks are nonzero.} Universes engineered from one indecomposable SymTFT with this full symmetry, in particular by gauging in a fusion category, are therefore connected by topological interfaces. Type~III pairs relative to such a symmetry, by contrast, require a decomposable bulk.

\paragraph{The SymTFT criterion.} Fix an indecomposable summand $\mathcal C^{(a)}$ with $\mathcal D=\mathcal C_{kk}$ and $\mathcal M_j=\mathcal C_{jk}$ as in \eqref{eq:types-module}, which identifies $\mathcal C^{(a)}$ with the full symmetry of the previous paragraph. By \eqref{eq:gauging-interface} in item~(i) of Section~\ref{sect:org}, every universe of the summand is a gauging of $\mathcal T_k$, so the universes are sandwiches with the same bulk $\mathcal Z(\mathcal D)$ and the same physical boundary, and they differ only in their symmetry boundaries. These are the indecomposable module categories $\mathcal M_j$, equivalently Lagrangian algebras $L_j\subset\mathcal Z(\mathcal D)$ \cite[Proposition 4.8]{davydov2011wittgroupnondegeneratebraided}. Physically, $L_j$ collects the bulk anyons that condense on the boundary. The symmetry of universe $j$ is $\mathcal C_{jj}=\mathrm{Fun}_{\mathcal D}(\mathcal M_j,\mathcal M_j)$, the category of $L_j$-modules \cite[Remark 4.9]{davydov2011wittgroupnondegeneratebraided}. Invertible bulk surfaces, which are the $0$-form symmetries of the SymTFT, form $\mathrm{Aut}^{\mathrm br}(\mathcal Z(\mathcal D))\cong\mathrm{BrPic}(\mathcal D)$ \cite{etingof2010fusion} and act on Lagrangian algebras. Within the summand,
\begin{equation}\label{eq:IIa-IIb-symtft}
\begin{aligned}
i\approx j &\iff L_i\cong L_j, \\[2pt]
i\sim j &\iff L_i\in\mathrm{Aut}^{\mathrm{br}}(\mathcal{Z}(\mathcal{D}))\cdot L_j.
\end{aligned}
\end{equation}
An invertible object of $\mathcal C_{ij}=\mathrm{Fun}_{\mathcal D}(\mathcal M_j,\mathcal M_i)$ is an equivalence of module categories $\mathcal M_j\simeq\mathcal M_i$, so the first equivalence holds because \cite[Proposition 4.8]{davydov2011wittgroupnondegeneratebraided} is a bijection of equivalence classes. For the second, a bulk symmetry $\varphi$ with $\varphi(L_j)=L_i$ identifies $L_j$- and $L_i$-modules. Conversely, the Morita equivalences $\mathcal D\sim\mathcal C_{ii},\mathcal C_{jj}$ give braided equivalences $\Phi_i,\Phi_j$ out of $\mathcal Z(\mathcal D)$ that carry $L_i,L_j$ to the canonical Lagrangian algebras \cite[Equation 12, Corollary 4.1, and Lemma 3.9]{davydov2011wittgroupnondegeneratebraided}. A tensor equivalence $f:\mathcal C_{ii}\simeq\mathcal C_{jj}$ preserves canonical algebras \cite[Proposition 4.4]{davydov2011wittgroupnondegeneratebraided}, so $\Phi_j^{-1}\mathcal Z(f)\Phi_i$ maps $L_i$ to $L_j$. Thus Type~IIa pairs are distinct symmetry boundaries in one orbit of bulk symmetries, and Type~IIb pairs are boundaries in different orbits.

\paragraph{The $\mathbb Z_4$ toric code.} For $\mathcal D=\mathrm{Vec}(\mathbb Z_4)$, the SymTFT is the $\mathbb Z_4$ toric code, with anyons $e^am^b$ of spin $e^{2\pi i ab/4}$. Its Lagrangian algebras correspond to the subgroups $\langle e\rangle$, $\langle m\rangle$ and $\langle e^2,m^2\rangle$, matching the three module categories of $\mathrm{Vec}(\mathbb Z_4)$, one for each gauged subgroup $1$, $\mathbb Z_4$ or $\mathbb Z_2$. The exchange of electric and magnetic anyons maps $\langle e\rangle$ to $\langle m\rangle$, so for instance $\mathcal T$ and $\mathcal T/\mathbb Z_4$ form a Type~IIa pair. In contrast, $\langle e^2,m^2\rangle$ is the only Lagrangian subgroup without an anyon of order four, so every braided autoequivalence fixes it. This boundary, obtained by gauging $\mathbb Z_2$, forms Type~IIb pairs with the other two boundaries, as for $[\mathcal T/\mathbb Z_4]$ and $[\mathcal T/\mathbb Z_2]$ in $[\mathcal T/((\mathbb Z_2\times\mathbb Z_2)\rtimes\mathbb Z_4)]$ above.

\paragraph{Type~IIa pairs and categorical families.} In a Type~IIa pair, once $f$ is fixed (up to tensor autoequivalences of $\mathcal C_{jj}$), both universes couple to a common background. These are the categorical families of Brennan--Intriligator \cite{Brennan:2026ira}, with relations of the form \cite[Equation 4.65]{Brennan:2026ira} in which the operation $\mathcal S$ of op.\ cit.\ is the gauging by $A_X=X^\vee\otimes X$ of \eqref{eq:gauging-interface}, for $X\in\mathcal C_{ij}$. In the SymTFT, this non-invertible operation is the fusion of an invertible bulk surface with the symmetry boundary, as for $S_3\supset\mathbb Z_3$ above. If the universes are also isomorphic through an invertible topological interface $\Psi$ outside the image of $F$, with $\mathcal T_j$ on its left and $\mathcal T_i$ on its right, that intertwines their symmetries, then $F(X)\otimes\Psi$ is a self-duality line of $\mathcal T_i$. When $F(X')\otimes\Psi\otimes F(X'')\otimes\Psi$ lies in $F(\mathcal C_{ii})$ for all $X',X''\in\mathcal C_{ij}$ and no $F(X')\otimes\Psi$ does, the lines $F(\mathcal C_{ii})$ and $F(\mathcal C_{ij})\otimes\Psi$ form a $\mathbb Z_2$-graded extension of $\mathcal C_{ii}$ whose odd part is $\mathcal C_{ij}$ transported by $\Psi$. It is of Tambara--Yamagami type \cite{tambara-yamagami} when $\mathcal C_{ii}=\mathrm{Vec}(H)$ for an abelian group $H$ and $X$ gauges all of it. 
Kramers--Wannier duality \cite{Frohlich:2004ef} is the template: $\mathrm{Ising}_m\sqcup \mathrm{Ising}_{-m}$ is a Type~IIa pair whose interface becomes the duality line at $m=0$, where the two universes coincide and $\Psi$ is the identity \cite[Section 3.2.2]{Brennan:2026ira}.

\paragraph{Type~IIb pairs.} In a Type~IIb pair there is no common background for $\mathcal C_{ii}$ and $\mathcal C_{jj}$, the universes cannot be isomorphic compatibly with their symmetries relative to $(\mathcal C,F)$, and no invertible bulk surface relates their symmetry boundaries. Such a pair lies within the general notion of a family in \cite[Equation 3.44]{Brennan:2026ira}, which allows any topological boundary condition of the shared SymTFT. Relative to $(\mathcal C,F)$, however, it admits no relation of the form \cite[Equation 4.65]{Brennan:2026ira}, in which both sides couple to the same background, so the family anomalies of BI are not defined for it. Relative to a smaller symmetry, the same pair can be of Type~IIa, as for $\mathcal T \sqcup \mathcal T/\mathbb Z_2$ relative to $\mathbb Z_2$ rather than $S_3$ (Section~\ref{sect:org}).

\section{Type III from $2$-group-type extensions and RG flows}\label{sec:typeIII}

In this section, we construct Type~III pairs. Section~\ref{sec:trivial}, together with the SymTFT argument of Section~\ref{sec:IIa-IIb}, shows that gauging in a fusion category yields only Type~I and~II pairs, so Type~III pairs need another origin. Some arise without any symmetry relating the universes, such as a $\sigma$-model on a generic disjoint union $X_1\sqcup X_2$, and others, such as the Galois pair of Section~\ref{sec:discussion}, come from a different construction. 
We provide two sources so as to show the examples of this type are not mere artifacts.

In Section~\ref{sec:cohomological}, we present a general construction of $2$-group-type extensions intertwining a $1$-form symmetry $K^{(1)}$ with the remaining symmetry $\mathcal{D}$, known as base change. There, we assume that the fusion category admits a grading by a finite group. In Section~\ref{sect:exact}, we discuss related notions from a perspective of RG flows.

\subsection{Cohomological base change}\label{sec:cohomological}
The data are a fusion category $\mathcal{D}$ with a faithful grading $\mathcal{D}=\bigoplus_{u\in \Upsilon}\mathcal{D}_u$ by a finite group $\Upsilon$, a quotient of the universal grading group $U(\mathcal{D})$ (Section~\ref{sec:review}), a finite abelian group $K$ whose point operators are left invariant by the lines of $\mathcal{D}$, and a class $\beta\in H^3(\Upsilon,K)$ playing the role of a Postnikov class. This class records how the associativity of the lines of $\mathcal{D}$ fails up to point operators of $K$. If the lines of $\mathcal D$ permuted the point operators of $K$, as for a $2$-group with a nontrivial action of $\pi_0$ on $\pi_1$, they would also permute the idempotents of $\mathbb C[K]$ and join the corresponding universes, which would then form Type~I or~II pairs rather than Type~III. We enlarge the junction spaces of $\mathcal{D}$ by the point operators, $\mathrm{Hom}\mapsto\mathrm{Hom}\otimes\mathbb C[K]$, extending composition and fusion of junctions $\mathbb C[K]$-linearly, keep objects and fusion, and twist the associator by $\beta$ evaluated on degrees,
\begin{equation}
\begin{aligned}
\alpha^\beta_{x,y,z} &= \alpha^\mathcal{D}_{x,y,z} \otimes \beta(|x|,|y|,|z|), \\[2pt]
&\quad\text{with }\beta(|x|,|y|,|z|)\in K\subset\mathbb{C}[K]^\times.
\end{aligned}
\end{equation}
The pentagon holds because $\beta$ is a 3-cocycle with values in central group-like elements and $\alpha^\mathcal{D}$ already satisfies it. This is the familiar twist of a graded fusion category by a 3-cocycle on its grading group, with $\mathbb C[K]^\times$ in place of $\mathbb C^\times$, and a coboundary change of $\beta$ is absorbed into the tensor structure. Before the unit is split, this category is not idempotent complete: for $x$ simple in $\mathcal D$, $\mathrm{End}(x)\cong\mathbb C[K]$ contains the idempotents $e_\chi$, whose images are not yet objects. Its Karoubi completion $\mathfrak G_\beta$, which splits the unit by the $|K|$ idempotents $e_\chi$ of $\mathbb C[K]$, decomposes as
\begin{equation}\label{eq:cohomological-typeIII}
\mathfrak G_\beta\simeq\boxplus_{\chi\in\widehat K}\mathcal{D}^{\chi_*\beta},\qquad \chi_*\beta\in H^3(\Upsilon,U(1)),
\end{equation}
where $\mathcal{D}^\omega$ denotes $\mathcal{D}$ with its associator twisted by the pull-back of $\omega$ along the grading. There is a single $1$-form symmetry $K^{(1)}$, whose projectors $e_\chi$ select the summands of \eqref{eq:cohomological-typeIII}.

Let us elaborate on why this process is meaningful from a physical point of view. The map $\chi\mapsto\chi_*\beta$ is a homomorphism $\widehat K\to H^3(\Upsilon,U(1))$, so \textit{the twists carried by the universes are the values of a homomorphism on $\widehat K$ rather than an arbitrary list.} Passing from universe $\chi$ to $\chi'$ is the canonical operation of twisting by $(\chi'\chi^{-1})_*\beta$, which no line of $\mathfrak G_\beta$ realizes. When $\mathcal D^{\chi_*\beta}$ and $\mathcal D^{\chi'_*\beta}$ are not Morita equivalent, this twist changes the SymTFT from $\mathcal Z(\mathcal{D}^{\chi_*\beta})$ to $\mathcal Z(\mathcal{D}^{\chi'_*\beta})$, and it matches the $(-2)$-form operation of \cite{Banerjee:2026ywy}: base change fixes which twist relates the universes, while a physical realization of the operation, as in op.\ cit., is additional data. Notably, the blocks are not necessarily all distinct. When $\chi'\chi^{-1}$ lies in the kernel of $\chi\mapsto\chi_*\beta$, or whenever $\mathcal D^{\chi_*\beta}\simeq\mathcal D^{\chi'_*\beta}$, the two blocks are equivalent, yet the universes remain in different summands, as for the center sectors of $SU(2)$ Yang--Mills theory. Whenever the two blocks are merely Morita equivalent, the SymTFT does not change either, so the operation relating the universes is not a $(-2)$-form one. Instead, an invertible bulk wall joins the two components, as described for $SU(2)$ Yang--Mills theory in item~(iii) of Section~\ref{sect:org}.

\paragraph{Base case: finite $2$-groups.} 
For $\mathcal{D}=\mathrm{Vec}(G)$ graded by $G$, $\mathfrak G_\beta$ is the Karoubi completion of the linearization of the finite $2$-group with $\pi_0=G$, $\pi_1=K$, trivial action of $\pi_0$ on $\pi_1$ and Postnikov class $\beta$. The simplest case is the following. 

Let us consider the case of an anomalous $\Z_2$ $0$-form symmetry together with a $\Z_2$ $1$-form symmetry. Our task is to construct the extension
\begin{equation}
\label{eq:Z2omega}
     \Z^{(1)}_2 \rightarrow \mathfrak{G} \rightarrow \mathrm{Vec}(\Z_2, \omega),
\end{equation}
where $\mathfrak G$ is a multi-fusion category. Here and below, a map from $\mathfrak G$ onto one of its summands is a monoidal projection that sends the other summands to zero. It is not faithful, so it is not a tensor functor in the sense of Section~\ref{sec:framework}, and the sequence is an extension only in this looser sense. 

As explored in \cite{Perez-Lona:2025ncg} for fusion categories, such a projection $\mathfrak{G} \rightarrow \mathrm{Vec}(\Z_2, \omega)$ requires resolving the anomaly $\omega$ of the $\Z_2$ $0$-form symmetry. We first do so at the level of finite $2$-groups, and afterwards linearize to obtain the desired extension.

An anomaly resolution 
\cite{Tachikawa:2017gyf,Wang:2017loc,Robbins:2021xce,Perez-Lona:2025ncg}
means finding a finite $2$-group extension $\mathbb{G}$ fitting in a short exact sequence
\begin{equation}
    1 \to B\Z_2 \xrightarrow{\imath} \mathbb{G} \xrightarrow{\pi} \Z_2 \to 1,
\end{equation}
such that the $\Z_2$-anomaly trivializes in $\mathbb{G}$ as
\begin{equation}\label{eq:z2z2-anomaly-pullback-trivialization}
    \left[\pi^* \omega\right] = 0 \in H^3_{\rm sing}\left(B\mathbb{G},U(1)\right),
\end{equation}
where $H^3_{\rm sing}\left(B\mathbb{G},U(1)\right)$ is the singular cohomology group that controls the 't Hooft anomalies in two dimensions. Since $H^1(B^2\mathbb Z_2,U(1))=0$ and $H^2(B^2\mathbb Z_2,U(1))=\widehat{\mathbb Z}_2$, the Leray--Serre spectral sequence of $B^2\mathbb Z_2\to B\mathbb G\to B\mathbb Z_2$ shows that \eqref{eq:z2z2-anomaly-pullback-trivialization} holds precisely when $\omega$ lies in the image of the transgression $H^0(\mathbb Z_2,\widehat{\mathbb Z}_2)\to H^3(\mathbb Z_2,U(1))$, that is, when
\begin{equation}
    [\omega] = [\sigma_*\beta],
\end{equation}
for a character $\sigma \in \widehat{\Z}_2 = {\rm Hom}(\Z_2,U(1))$, where $[\beta]\in H^3(\Z_2,\Z_2)$ is the Postnikov class.

This is satisfied by the finite $2$-group $\mathbb G$ whose Postnikov class is the unique nontrivial class $[\beta] \in H^3(\Z_2,\Z_2) \cong \Z_2$. A representative $3$-cocycle has a unique nontrivial value
\begin{equation}
\beta:\Z_2 \times \Z_2 \times \Z_2 \to \Z_2, \quad \beta(z,z,z) = k,
\end{equation}
for $z$ and $k$ the generators of $\pi_0=\Z_2$ and $\pi_1=\Z_2$. Composing with the unique nontrivial character $\sigma=\chi_-\in \widehat{\Z}_2$ yields a representative $3$-cocycle for the unique nontrivial anomaly $3$-class in $H^3(\Z_2,U(1))$
\begin{equation}
   \left( \chi_- \circ \beta(z^i,z^j,z^k) \right)= (-1)^{ijk} \in U(1).
\end{equation}
Intuitively, this is an anomaly resolution because the anomaly phases of $\Z_2$ get absorbed by the $2$-group structure of $\mathbb{G}$.

Linearizing $\mathbb G$ and Karoubi completing with respect to the two idempotents of $\mathbb C[\mathbb Z_2]$ is the construction of Section~\ref{sec:cohomological} with $\mathcal D=\mathrm{Vec}(\mathbb Z_2)$, and $\chi_{-*}\beta=\omega$, so
\begin{equation}\label{eq:vecz2-multi-fusion}
\mathfrak G\simeq\mathrm{Vec}(\mathbb Z_2)\boxplus\mathrm{Vec}(\mathbb Z_2,\omega).
\end{equation}
Either summand can serve as the ``quotient''\footnote{By this, we mean the remaining symmetry after RG flow (see also Section~\ref{sec:discussion}).}: the projection onto $\mathrm{Vec}(\mathbb Z_2,\omega)$ exhibits $\mathfrak G$ as the extension of the anomalous symmetry described above, while the projection onto $\mathrm{Vec}(\mathbb Z_2)$ is the projection onto the summand of the trivial character, $\mathcal D^{1_*\beta}=\mathcal D$.
These two projections, $P_+:\mathfrak G\to\mathrm{Vec}(\mathbb Z_2)$ and $P_-:\mathfrak G\to\mathrm{Vec}(\mathbb Z_2,\omega)$, project onto the two universes, labeled by their anomaly class, in agreement with the SymTFT analysis of \cite{Banerjee:2026ywy}.

We can also gain some insight from the three-dimensional SymTFT bulk perspective. There, the anyons are described by the Drinfeld center of the symmetry category at the boundary. In the present case, since the multi-fusion category \eqref{eq:vecz2-multi-fusion} is a block diagonal sum, the Drinfeld center construction distributes:
\begin{equation}
    \mathcal{Z}(\mathfrak{G}) \simeq \mathcal{Z}\left({\rm Vec}\left(\Z_2\right)\right) \boxplus \mathcal{Z}\left({\rm Vec}\left(\Z_2,\omega\right)\right).
\end{equation}
The bulk therefore decomposes into two components, the toric code and the double semion, with four anyons each. Since $\mathfrak G$ is decomposable, no topological interface in the image of $F$ relates the universes. In the SymTFT, every topological wall between the toric code and the double semion is totally reflective. Such walls correspond to Lagrangian algebras of the folded theory $\mathrm{TC}\boxtimes\overline{\mathrm{DS}}$, which is pointed, so they are supported on Lagrangian subgroups. The bosons of $\mathrm{TC}\boxtimes\overline{\mathrm{DS}}$ are products of $\{1,e,m\}$ and $\{1,s\bar s\}$, with $s\bar s$ the boson of the double semion. A Lagrangian subgroup has four elements, all bosons. If its toric-code components included both $e$ and $m$, it would contain an element whose toric-code component is the fermion $f$, which is not a boson. Its toric-code components therefore lie in $\{1,e\}$ or in $\{1,m\}$, and it is one of the factorized subgroups $\{1,e\}\times\{1,s\bar s\}$ and $\{1,m\}\times\{1,s\bar s\}$. Thus, unlike the center sectors of $SU(2)$ Yang--Mills theory, the two bulk components are not joined by any invertible wall, consistent with $\mathrm{Vec}(\mathbb Z_2)$ and $\mathrm{Vec}(\mathbb Z_2,\omega)$ not being Morita equivalent. This agrees with \cite[Section 4.2]{Banerjee:2026ywy}, where the wall between the toric code and the double semion arises from quarter gauging the $\mathbb Z_4$ SymTFT. See Appendix \ref{app:qg} for further details.

\paragraph{A non-pointed example.} 
Take $\mathcal{D}=\mathrm{Rep}(D_4)\simeq\mathrm{TY}(\mathbb Z_2\times\mathbb Z_2,\lambda,+)$ \cite{Perez-Lona:2023djo}, with $\lambda$ its symmetric nondegenerate bicharacter, and its universal $\mathbb Z_2$ grading (Section~\ref{sec:review}): $\mathcal{D}_{\rm ad}$ is generated by $m\otimes m=\bigoplus_x x$, so the invertible lines are even and $m$ is odd. Let $K=\mathbb Z_2$ and let $\beta$ be the generator of $H^3(\mathbb Z_2,\mathbb Z_2)$. The twisted associator differs from that of $\mathcal{D}$ only in the component with all three entries odd, $\alpha^\beta_{m,m,m}=k\,\alpha_{m,m,m}$, with $k$ the generator of $K$. In the universe of the nontrivial character, the Frobenius--Schur indicator flips sign, giving $\mathrm{TY}(\mathbb Z_2\times\mathbb Z_2,\lambda,-)\simeq\mathrm{Rep}(Q_8)$, so \eqref{eq:cohomological-typeIII} yields
\begin{equation}
\label{eq:non-pointed_cohom}
\mathfrak{G}\simeq\mathrm{Rep}(D_4)\boxplus\mathrm{Rep}(Q_8).
\end{equation}

The two summands are not Morita equivalent \cite[Example 5.4]{Naidu2007LagrangianSA}, so here even the SymTFTs differ, and neither is anomalous as both admit fiber functors. A parent SymTFT whose two universes carry exactly these boundary symmetries was constructed in \cite[Section 3.6]{Banerjee:2026ywy}. A parent with $\mathrm{Rep}(SD_{16})$ symmetry that flows to either summand is constructed in Section~\ref{sect:exact}.

In any theory whose symmetry contains such an extension, universe $\chi$ carries a symmetry containing $\mathcal{D}^{\chi_*\beta}$. Group-like examples arise in orbifolds by $2$-groups \cite{Pantev:2022kpl} and in the SymTFT realizations of \cite[Sections 3.2 and 3.6]{Banerjee:2026ywy}. For details of this example, we refer the reader to Appendix~\ref{D4/Q8}.

\subsection{Exact sequences of tensor categories and RG flows} \label{sect:exact} 

\begin{figure*}[t]
\centering
\begin{minipage}[t]{.48\textwidth}\centering
\begin{tikzpicture}[x=.93cm,y=.82cm]
  \fill[qgparent] (0,0) rectangle (6,3);
  \fill[qgupper] (3,1.78) rectangle (6,3);
  \fill[qglower] (3,0) rectangle (6,1.22);
  \draw[qgline] (0,0) rectangle (6,3);
  \draw[qgline] (3,3)--(3,1.78)--(6,1.78);
  \draw[qgline] (3,0)--(3,1.22)--(6,1.22);
  \node[qgtext] at (1.55,1.5) {$\mathfrak{Z}_G$};
  \node[qgtext] at (4.5,2.35) {$\mathfrak{Z}_H$};
  \node[qgtext] at (4.5,.65) {$\mathfrak{Z}_K$};
  \node[qgsmall] at (4.5,1.5) {$\mathfrak{Z}_G$};
  \node[qgsmall,anchor=north] at (0,-.14) {$\mathfrak{B}^{\operatorname{sym}}$};
  \node[qgsmall,anchor=north] at (6,-.14) {$\mathfrak{B}^{\operatorname{phys}}$};
  \node[qgsmall,anchor=south] at (3,3.08) {$\mathcal I_H$};
  \node[qgsmall,anchor=north] at (3,-.10) {$\mathcal I_K$};
  \node[qgrg] at (6,1.78) {};
  \node[qgrg] at (6,1.22) {};
\end{tikzpicture}
\end{minipage}\hfill
\begin{minipage}[t]{.48\textwidth}\centering
\begin{tikzpicture}[x=.93cm,y=.82cm]
  \fill[qgparent] (0,0) rectangle (3,3);
  \fill[qgupper] (3,1.5) rectangle (6,3);
  \fill[qglower] (3,0) rectangle (6,1.5);
  \draw[qgline] (0,0) rectangle (6,3);
  \draw[qgline] (3,0)--(3,3);
  \draw[qgfused] (3,1.5)--(6,1.5);
  \node[qgmeet] at (3,1.5) {};
  \node[qgrg] at (6,1.5) {};
  \node[qgtext] at (1.5,1.5) {$\mathfrak{Z}_G$};
  \node[qgtext] at (4.5,2.27) {$\mathfrak{Z}_H$};
  \node[qgtext] at (4.5,.75) {$\mathfrak{Z}_K$};
  \node[qgsmall,anchor=south] at (4.5,1.53) {$\mathcal I$};
  \node[qgsmall,anchor=north] at (0,-.14) {$\mathfrak{B}^{\operatorname{sym}}$};
  \node[qgsmall,anchor=north] at (6,-.14) {$\mathfrak{B}^{\operatorname{phys}}$};
\end{tikzpicture}
\end{minipage}
\caption{Simultaneous quarter gauging and the fusion of condensation interfaces. Shrinking the intermediate parent region brings $\mathcal I_H$ and $\mathcal I_K$ together. Their composite $\mathcal I=\mathcal I_H^\vee\boxtimes_{\mathfrak{Z}_G}\mathcal I_K$ is the dashed blue interface separating the two descendant topological orders.}
\label{fig:qg-fusion}
\end{figure*}
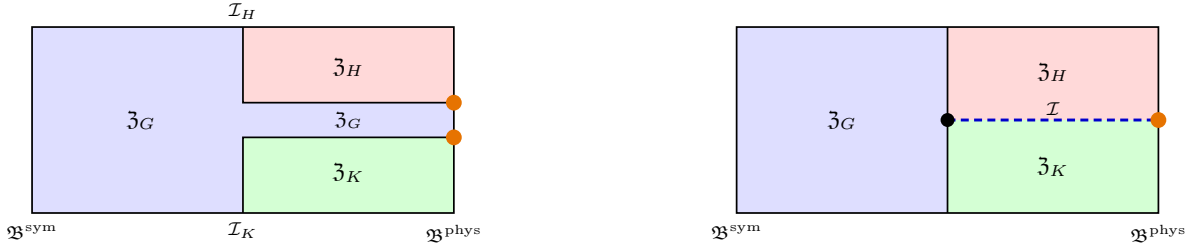

We now construct Type~III pairs from exact sequences of tensor categories, which arise in the quarter-gauging picture of \cite{Banerjee:2026ywy} (summarized in Appendix~\ref{app:qg}). The basic idea is that the universes of a decomposing theory arise as the IR phases of two or more relevant deformations of the same UV theory, related by quarter gauging (Figure~\ref{fig:qg-fusion}). This different physical process results in a different exact sequence in which the multi-fusion category symmetry sits. The figure shows the two IR phases in adjacent regions of spacetime, separated by a wall of the SymTFT, which is not a topological interface of the two-dimensional theory. The decomposing theory is instead the direct sum of the two sandwiches, and the wall is a bulk surface between components as in item~(iii) of Section~\ref{sect:org}.

The data are a fusion category $\mathcal{B}$ (the symmetry of the UV theory), a collection of central commutative algebras $\{(A_i, \sigma_i)\}_{i\in I}$, i.e.\ $A_i\in \mathcal{B}$ is a haploid ($\mathrm{Hom}(\mathbbm 1,A_i)=\mathbb C$) algebra equipped with a half-braiding $\sigma_i$ such that $(A_i, \sigma_i)$ is \'etale (commutative and separable) in $\cal{Z(B)}$ \cite{davydov2011wittgroupnondegeneratebraided} (condensable algebras in $\mathcal{Z}(\mathcal{B})$). Denote the generated fusion subcategories by $\mathcal{N}_i = \langle A_i \rangle$ (symmetries acting trivially in the IR theory), $i\in I$.
Here the index set $I$ plays the role of $\widehat{K}$, the set of universe labels.
In this way, there are $|I|$ exact sequences of fusion categories
\begin{equation}
    \mathcal{N}_i \xrightarrow{\iota_i} \mathcal{B} \xrightarrow{ q_{i}} \mathcal{D}_i,
\end{equation}
where $q_i = -\otimes A_i$ is the free $A_i$-module functor and $\mathcal{D}_i$ is the fusion category of $A_i$-modules in $\mathcal{B}$, with monoidal structure $\otimes_{A_i}$ \cite{Bruguieres2011}. Following \cite{Bruguieres2011, Antinucci:2025fjp}, we require each $q_i$ to be \textit{normal}, equivalently, each $A_i$ to be \textit{self-trivializing}, i.e., $A_{i}\otimes A_{i} \cong A_{i}^{n_i}$ as $A_i$-modules, with $n_i=\mathrm{FPdim}\,A_i$. Normality identifies the kernel to be $\ker q_i=\mathcal N_i=\langle A_i\rangle$. Physically, each $i$ labels an RG flow and $q_i$ relates the symmetries in the UV and IR. 

From this, one defines a multi-fusion category
\begin{equation}
    \mathfrak{G}:=\boxplus_{i} \mathcal{D}_i,
\end{equation}
that fits into the sequence
\begin{equation}
\label{eq:exact_seq_multi-fusion}
        \boxplus_{i} \mathcal{N}_i \xrightarrow{\boxplus_i\iota_i} \boxplus_{i} \mathcal{B} \xrightarrow{\boxplus_i q_i} \mathfrak{G}.
\end{equation}
This sequence is exact only summand by summand.

\paragraph{Realization as decomposition.}
The types of Section~\ref{sect:org} are defined relative to a pair $(\mathfrak{G}, F)$ acting on a theory. We supply the decomposing theory $\mathfrak{T}$ and an action functor $F$ here using the quarter-gauging sandwich.

Recall that a Type~III pair has a decomposed SymTFT bulk 
\begin{equation}
    \mathcal{Z}(\mathfrak{G}) \simeq \boxplus_{i\in I} \mathcal{Z}(\mathcal{D}_i)
\end{equation}
with symmetry boundary and physical boundary
\begin{equation}
    \boxplus_i \mathfrak{B}^{\operatorname{sym}}_{\mathcal{D}_i}, \qquad \boxplus_{i}\mathfrak{B}^{\operatorname{phys}}_i,
\end{equation}
where $\mathfrak{B}^{\operatorname{sym}}_{\mathcal{D}_i}$, the symmetry boundary carrying the lines $\mathcal{D}_i$, is obtained by fusing the symmetry boundary of the parent with the $(A_i, \sigma_i)$-condensation interface, and $\mathfrak{B}^{\operatorname{phys}}_i$ is the physical boundary of the IR region in which $(A_i,\sigma_i)$ is condensed. The theory $\mathfrak{T}$ is the resulting compactified sandwich, and its action functor $F:\mathfrak{G}\to\operatorname{Top}(\mathfrak{T})$ is the tautological action of the boundary lines.

In the quarter-gauging description, 
the universes are the condensations in a common parent $\cal Z(B)$, labeled by the choice of $(A_i, \sigma_i)$, see Figure~\ref{fig:qg-fusion},
\begin{equation}
    \mathcal{Z}(\mathcal{D}_{i}) \simeq \mathcal{Z}(\mathcal{B})^{\operatorname{loc}}_{(A_i, \sigma_i)}.
\end{equation}
No topological interface in the image of $F$ relates them.

We pause here to compare \eqref{eq:exact_seq_multi-fusion} with the sequences of Section~\ref{sec:cohomological}. There, the multi-fusion category $\mathfrak{G}$ sits in the middle of $K^{(1)} \rightarrow \mathfrak{G} \rightarrow \mathcal{D}$, being an extension (in the looser sense defined there) of $\mathcal{D}$, while in \eqref{eq:exact_seq_multi-fusion}, $\mathfrak{G}$ sits on the right, as the quotient.
In Section~\ref{sec:cohomological}, the middle term is the symmetry of the decomposing theory itself, and the sequence restricts that symmetry to one universe, while here it is the symmetry of the UV theory, and the sequence exhibits an IR symmetry as a quotient. 

Sequence \eqref{eq:exact_seq_multi-fusion} is a direct sum of separate sequences, with one copy of $\mathcal B$ per flow. A single UV symmetry would instead act through the diagonal $X\mapsto\bigoplus_i q_i(X)$, whose kernel is $\bigcap_i\ker q_i$. This is $\mathcal{N}_i \cong\mathcal{N}$ when all the $\mathcal N_i$ coincide as subcategories of $\mathcal B$, as in the pointed example below, but it can also be trivial, as in the $\Rep(SD_{16})$ example, whose three subcategories meet only in $\Vect$.

\paragraph{Comparing to the 2-group extension.}
In the 2-group-type extension, $K$ plays two roles at once. It is both the kernel of the projection and the algebra of point operators. Their sizes coincide because the universes are labeled by $\widehat{K}$ and $|\widehat{K}| = |K|$. 
Here, these two roles are separated into the index set $I$ and the generated fusion subcategories $\mathcal{N}_i$. 

The point operators are given by the index set $I$, forming the algebra
\begin{equation}
    \End_{\mathfrak{G}}(\mathbbm{1}) = \bigoplus_{i\in I} \End(\mathbbm{1}_i) = \Fun(I).
\end{equation}
In the $\mathrm{Rep}(SD_{16})$ example below, for instance, $|I|=3$, while each $\mathcal N_i\simeq\mathrm{Rep}(\mathbb Z_2)$ has two simple objects.

\paragraph{A pointed example.}
Take $\mathcal{B} = \Vect(\Z_4)$ and $A=\C_1\oplus \C_{a^2}$, with $a$ a generator of $\Z_4$, the regular algebra of $\mathcal{N} = \Vect(\Z_2) \hookrightarrow \Vect(\Z_4)$. There are two inequivalent half-braidings $\sigma_{\pm}$ on $A$. 
Writing the anyons of the $\mathbb Z_4$ toric code $\mathcal{Z}(\Vect(\Z_4))$ as $e^r m^s$, these two algebras can be decomposed into anyons as \cite{Antinucci:2025fjp}
\begin{equation}
    (A, \sigma_{+}) = 1 \oplus m^2, \quad (A, \sigma_{-}) = 1 \oplus e^2 m^2.
\end{equation}
The two free-module functors are
\begin{equation}
    \begin{split}
        q_{+} & : \mathcal{B} \rightarrow \Vect(\Z_2),\\
        q_{-} & : \mathcal{B} \rightarrow \Vect(\Z_2, \omega),
    \end{split}
\end{equation}
where $\omega$ is the nontrivial element of $H^3(\Z_2, U(1))\cong \Z_2$.
The multi-fusion category agrees with that in the 2-group case, being
\begin{equation}
    \mathfrak{G} \simeq \Vect(\Z_2) \boxplus \Vect(\Z_2, \omega).
\end{equation}
This is the pointed example \eqref{eq:vecz2-multi-fusion}, so its two universes form a Type~III pair.

\paragraph{A non-pointed example.}
Take $\mathcal{B} = \Rep(SD_{16})$, where $SD_{16}$ is the semidihedral group of order 16,
\begin{equation}
    SD_{16} = \langle r, s \mid r^8 = s^2 = 1, srs^{-1} = r^3 \rangle.
\end{equation}
It has three normal subgroups of order 8: $\langle r\rangle \cong \Z_8$, $N_D = \langle r^2, s \rangle \cong D_4$, the order 8 dihedral group, and $N_{Q} = \langle r^2, rs \rangle \cong Q_8$ the quaternion group. For each $N \in \{\langle r \rangle, N_{D}, N_{Q}\}$, $SD_{16}/N\cong \Z_2$. There are three restriction functors $\operatorname{Res}^{SD_{16}}_{N}: \Rep(SD_{16}) \rightarrow \Rep(N)$, fitting in the exact sequences
\begin{align}
    \Rep(\Z_2) \xrightarrow{p_N^*} \Rep(SD_{16}) \xrightarrow{\operatorname{Res}^{SD_{16}}_{N}} \Rep(N),
\end{align}
where $p_N^*$ is the pullback along the projection $p_N:SD_{16} \rightarrow SD_{16}/N$. In the notation above, $A_N=\mathrm{Fun}(SD_{16}/N)$ with its canonical half-braiding, and $\mathcal N_N=\Rep(SD_{16}/N)\simeq\Rep(\Z_2)$. The resulting multi-fusion category is
\begin{equation}
    \mathfrak{G} \simeq \Rep(\Z_8) \boxplus \Rep(D_4) \boxplus \Rep(Q_8).
\end{equation}

We can choose a different index set $I$ by removing the algebra corresponding to $\langle r \rangle \cong \Z_8$ from $I$.  One instead obtains the multi-fusion category
\begin{equation}
    \mathfrak{G} \simeq \Rep(D_4) \boxplus \Rep(Q_8),
\end{equation}
which is the non-pointed example~\eqref{eq:non-pointed_cohom}, so its two universes form a Type~III pair.

In the examples above, the subcategories $\mathcal N_i=\langle A_i\rangle$ are all equivalent to $\mathrm{Rep}(\mathbb Z_2)\simeq\mathrm{Vec}(\mathbb Z_2)$ as fusion categories, although in the $\mathrm{Rep}(SD_{16})$ example they are different subcategories of $\mathcal B$. The quarter-gauging picture also applies to collections of algebras with $\mathcal N_i\not\simeq\mathcal N_j$ for some $i,j\in I$.

\section{The correspondence with Brennan--Intriligator families}\label{sec:families}

In this section, we relate the structures of the previous sections to the families of Brennan--Intriligator \cite{Brennan:2026ira}. 

\subsection{Review of Brennan--Intriligator}
\label{sec:family-of-qft}

The authors of \cite{Brennan:2026ira} proposed a rough classification of families of quantum field theories.  Part of the purpose of this article is to describe how families of universes arising from decomposition (in two-dimensional theories) are described in their language. In this subsection, we briefly review the pertinent ideas of their classification.

The basic example of a family arises from explicit symmetry breaking. Consider a QFT with symmetry $G^{(0)}$, broken by a relevant deformation 
\begin{equation}
    S \supset \lambda \int d^dx ~\mathcal{O}(x),
\end{equation}
where $\mathcal{O}$ transforms nontrivially under $G^{(0)}$, so that the deformation initiates a symmetry-breaking renormalization group flow. The symmetry can be formally restored by promoting $\lambda$ to a spacetime-dependent background field (a spurion) that transforms under $G^{(0)}$ so as to compensate the transformation of $\mathcal{O}$. The deformed theories therefore form a $G^{(0)}$-\textit{equivariant}, or \textit{invertible}, family of QFTs,
\begin{equation}\label{eq:invertible-family-definition}
    \mathfrak{T} = \{ \mathcal{T}_\lambda \}, \qquad \lambda \in X,
\end{equation}
indexed by a parameter $\lambda$ in a space $X$\footnote{Heuristically, we may also think of $X$ as the space of QFTs in this scenario.} with a natural $G^{(0)}$ action,
\begin{equation}\label{eq:group-action-on-family}
    g: \lambda \longmapsto g \cdot \lambda , \qquad \mathcal{T}_\lambda \cong \mathcal{T}_{g \cdot \lambda}, \qquad g\in G^{(0)}.
\end{equation}
A fixed $\mathcal{T}_\lambda$ thus keeps only the stabilizer of $\lambda$ as a symmetry. The remaining elements of $G^{(0)}$ relate theories at different points of $X$. 

We read \cite{Brennan:2026ira} as describing
three kinds of families, distinguished by the identifications one makes to construct the moduli space from the analogue of a Teichm\"uller space.
\begin{enumerate}
    \item[(A)] Ordinary families of QFTs, where the identifications form a group acting as in \eqref{eq:group-action-on-family}. The running example is a family of $SU(N)$ Yang--Mills theories parameterized by their theta angle.
    \item[(B)] Higher families, discussed in \cite[Section 3.1]{Brennan:2026ira}, in which higher categorical identifications are invoked when constructing the moduli space.  (Among the finite decompositions considered in this article, we have found no analogue of these.)
    \item[(C)] Categorical (non-invertible) families, discussed in \cite[Section 3.2]{Brennan:2026ira}, in which non-invertible symmetries are invoked when constructing the moduli space. To be precise, the theories are identified by a topological manipulation $\mathcal{S}$ such as discrete or twisted gauging \cite{Choi:2022zal},
    \begin{equation}
        \mathcal{S}\, \mathcal{T}_\lambda \cong \mathcal{T}_{\lambda'},
    \end{equation}
    which, unlike a group action, need not have an inverse. In the general definition of BI \cite[Section 3.2]{Brennan:2026ira}, the partition function is a section over the parameters together with all topological boundary conditions of the SymTFT. The family relations whose anomalies they study return the same symmetry, so that both sides couple to the same background.
\end{enumerate}
Types (A) and (B) are both ``invertible'' families.  

Furthermore, each kind of family can carry a family anomaly \cite[Section 4]{Brennan:2026ira}, in which the partition functions are sections of a nontrivial line bundle over the space of couplings and background fields. The isomorphism ${\cal T}_\lambda \cong {\cal T}_{\lambda'}$ then holds in the absence of background fields, while in their presence the partition functions may agree only up to a background-field-dependent phase,
\begin{equation}
    Z_\lambda = \Phi \, Z_{\lambda'} .
\end{equation}
As for an 't Hooft anomaly, $\Phi$ is the boundary variation of a symmetry-protected topological (SPT) phase one dimension higher. Following \cite[Section 4]{Brennan:2026ira}, we call it a \textit{family anomaly}. More commonly, these are known as anomalies in the space of coupling constants \cite{Cordova:2019jnf, Cordova:2019uob}. Likewise, if a categorical relation $\mathcal{S}\, \mathcal{T}_\lambda \cong \mathcal{T}_{\lambda'}$ that returns the same symmetry holds only up to a phase, so that $\mathcal{S}\, Z_\lambda = \Phi \, Z_{\lambda'}$, then $\Phi$ is a \textit{categorical family anomaly}.

We propose that the notion of families of QFTs, together with structures such as family anomalies, admits a complementary formulation for discrete subfamilies in terms of decomposition and lower-form symmetries \cite{Hellerman:2006zs,Sharpe:2014tca, Sharpe:2019ddn, Sharpe:2021srf, Sharpe:2022ene, Lin:2025oml, Yu:2024jtk, Banerjee:2026ywy, Komargodski:2020mxz, Cherman:2020cvw, Cherman:2021nox, Tanizaki:2019rbk, Robbins:2020msp, Lin:2022xod, Nguyen:2021naa, Robbins:2021xce, Pantev:2022kpl, Pantev:2022pbf, Perez-Lona:2023llv, Santilli:2024dyz, Najjar:2024vmm, Robbins:2025urk, Perez-Lona:2025add, Nguyen:2021yld, Caldararu:2010ljp, Pantev:2023dim}. For example, the process of promoting coupling constants to background fields used in the construction of an invertible family has an interpretation in terms of turning on the background gauge field for the $(-1)$-form symmetry \cite{Aloni:2024jpb,Yu:2024jtk,Lin:2025oml}. In the BI families, we saw that one would move in the parameter space from one theory to another by the action of a broken $G^{(0)}$ symmetry or by some topological manipulation. In contrast, in the context of decomposition, we would transition across the universes by the action of $(-1)$-form symmetries \cite{Yu:2024jtk,Lin:2025oml,Najjar:2024vmm,Aloni:2024jpb,Perez-Lona:2025add}, or $(-2)$-form symmetries \cite{Banerjee:2026ywy}. For the categorical families, a notion of $(-1)$-form symmetry that acts by discrete gauging was introduced in \cite{Robbins:2025urk}. Moreover, decomposing theories where the universes are related via orbifolding/discrete gauging has already been discussed in the literature \cite{Robbins:2022wlr}. 

These observations suggest a complementary construction of a family of QFTs. Rather than beginning with a parameter space of theories and relations among its points, we begin with a single parent theory with a $(d-1)$-form symmetry, whose universes play the role of the members of the family. A finite $(d-1)$-form symmetry realizes only discrete subfamilies, such as a $\mathbb{Z}_N$ orbit of $\theta$, and in this article we restrict to the finite case (a continuous family requires an infinite $(d-1)$-form symmetry, such as $\mathbb{Z}^{(d-1)}$, whose characters form $U(1)$).

The higher-charge Schwinger model provides a concrete Lagrangian example of this construction, in which a mixed anomaly is reflected in the permutation of universes by the axial symmetry; see Appendix~\ref{app:schwinger}.

Notably, the two approaches differ in the flat gauging of $(-1)$-form symmetries \cite{Lin:2025oml,Perez-Lona:2025add,Perez-Lona:2026mlj,PerezLona:inprep}. Viewing the transition from one member to another as an action of $(-1)$-form symmetry\footnote{
In the Brennan--Intriligator families, the authors specifically avoid identifying the transition across the members as $(-1)$-form symmetries \cite[Footnote 2]{Brennan:2026ira}.} 
opens up the possibility of flat gauging of $(-1)$-form symmetry, and hence of a local quantum field theory carrying the quantum dual $(d-1)$-form symmetry. Therefore, in the decomposition approach to families, we are assuming the existence of a well-defined local quantum field theory whose partition function is essentially the sum of the partition functions of the constituent universes.\footnote{Such a theory violates \textit{cluster decomposition}, and this observation was one of the initial motivations for decomposition \cite{Hellerman:2006zs}.}

\subsection{Comparison to classification by types}

$G^{(0)}$-families are realized by Type~I summands (item (ii) of Section~\ref{sect:org}), categorical families by Type~IIa pairs and families in the general sense of \cite[Equation 3.44]{Brennan:2026ira} by Type~IIb pairs (Section~\ref{sec:IIa-IIb}), and invertible family anomalies by the monoidal structure of the action permuting universes, which we derive in Section~\ref{sec:family-anomalies}. Type~III pairs that share a SymTFT but that no topological interface of $\mathfrak T$ relates, such as the center sectors $SO(3)_\pm$ of $SU(2)$ Yang--Mills theory, can realize families over a discrete parameter, summed by the flat gauging of a $(-1)$-form symmetry (item~(iii) of Section~\ref{sect:org} and the paragraph below). Type~III pairs with different SymTFTs have no counterpart in Brennan--Intriligator, whose general families involve the topological boundary conditions of a single SymTFT \cite[Equation 3.44]{Brennan:2026ira}.

\paragraph{Type~III pairs and discrete parameters.} A shared SymTFT alone does not make a Type~III pair a family over a parameter: as item~(iii) of Section~\ref{sect:org} notes, $\mathcal T\sqcup \mathcal T/G$, $G$ nonabelian, with $\mathrm{Vec}(G)\boxplus\mathrm{Rep}(G)$ is of Type~III only because the symmetry omits the gauging interfaces. To separate the two situations, we write $\mathcal T_i$ and $\mathcal T_j$ as sandwiches with symmetry boundaries $B_i,B_j$ and physical boundaries $P_i,P_j$, over components of the SymTFT joined by an invertible wall $W$. Then $\mathcal T_j$ is also the sandwich over the component of $\mathcal T_i$ with symmetry boundary $B_jW$ and physical boundary $W^{-1}P_j$. If $W^{-1}P_j\cong P_i$, the two universes share their bulk and their physical boundary, and the boundary-changing lines between $B_i$ and $B_jW$, which are nonzero as in Section~\ref{sec:IIa-IIb}, are topological interfaces outside the image of $F$. A Type~III pair that no topological interface of $\mathfrak T$ relates, such as $SO(3)_\pm$ (Section~\ref{sec:dictionary}), therefore has $W^{-1}P_j\not\cong P_i$ for every choice of $W$. Physically, its members differ in their physical boundary, as members of a family can differ in their couplings. The converse does not hold, because a topological interface of $\mathfrak T$ need not intertwine the symmetries of the two universes.

\subsection{Family anomalies}\label{sec:family-anomalies}
Consider a Type~I summand $\mathrm{Mat}_n(\mathcal D)$ and an action $\Phi:\mathrm{Vec}(G\times\mathcal G,\omega)\to\mathrm{Mat}_n(\mathcal D)$ in which $G$ permutes the universes transitively, with stabilizer $G_{i_0}$ of a base universe $i_0$, and $\mathcal G$, the preserved symmetry of Brennan--Intriligator, moves no universe. When $\mathcal D=\mathrm{Vec}$ (for instance for universes with no further symmetry, such as the vacua of a gapped theory), such an action is a module category over $\mathrm{Vec}(G\times\mathcal G,\omega)$ whose simple objects are the universes. By \cite[Theorem 3.1]{Ostrik:2002ohv}, it is $\mathcal M(G_{i_0}\times\mathcal G,\eta)$ with $d\eta=\omega|_{G_{i_0}\times\mathcal G}$. The anomaly restricted to $G_{i_0}\times\mathcal G$ must therefore vanish. An anomalous $G$ can fix a universe only if its anomaly restricts trivially to the stabilizer, while for a free action the pure $G$ part is unconstrained and the anomaly is matched by permuting universes, as for spontaneous breaking.

In particular, $[\omega]|_{\mathcal G}$ is trivial, and we choose a representative with $\omega|_{\mathcal G}=1$. The mixed part of $\omega$ then enters through the loop transgression \cite{Willerton:2005}: for $g\in G$, the transgression of $\omega$ along $g$, restricted to $\mathcal G$, which commutes with $g$, is a 2-cocycle $\iota_g\omega$ on $\mathcal G$. The map $g\mapsto[\iota_g\omega]$ is, up to a convention-dependent sign, the image of $[\omega]$ in the term $E_2^{1,2}=\mathrm{Hom}(G,H^2(\mathcal G,U(1)))$ of the Lyndon--Hochschild--Serre spectral sequence of $G\times\mathcal G\to G$, and hence a homomorphism $G\to H^2(\mathcal G,U(1))$. Reading off the internal endomorphisms of the simple objects of the restriction of $\mathcal M(G_{i_0}\times\mathcal G,\eta)$ to $\mathcal G$ gives the $\mathcal G$-SPT of each universe,
\begin{equation}\label{eq:spt-shift}
\mathcal T_{g\cdot i}\;\cong\;\mathcal T_i\otimes\mathrm{SPT}_{\mathcal G}\big([\iota_g\,\omega]\big),\qquad g\in G .
\end{equation}
Since $[\omega]$ restricts trivially to $G_{i_0}\times\mathcal G$ and $g\mapsto[\iota_g\omega]$ is a homomorphism to an abelian group, $[\iota_h\omega]$ is trivial for $h$ in every stabilizer, so the right-hand side depends only on $g\cdot i$. For general $\mathcal D$, the same holds if we assume $\omega|_{\mathcal G}=1$, which no longer follows from the module-category argument: the restrictions of $\Phi$ to $\mathcal G$ in universes $i$ and $g\cdot i$ are tensor functors $\mathrm{Vec}(\mathcal G)\to\mathcal D$ whose monoidal structures differ by $\iota_g\omega$. This is the family anomaly of \cite[Equation 1.1]{Brennan:2026ira} for $d=2$, with the $U(1)$-valued $d\theta$ of their inflow \cite[Equation 1.2]{Brennan:2026ira} replaced by a $G$-valued 1-class. In a continuous family, the anomaly forces gaplessness, symmetry breaking or a transition \cite[Section 4.4]{Brennan:2026ira}. In a decomposing theory, it is matched without any transition, by universes carrying different $\mathcal G$-SPTs. The underlying mathematics, gapped phases as module categories \cite{Thorngren:2019iar}, is standard. What is new is its reading in terms of the universes of a single theory. The family anomaly is carried by $\omega$ through the monoidal structure of $\Phi$, and it is invisible in $\mathcal C$. The remaining mixed part of $\omega$, with one leg in $\mathcal G$ and two in $G$, means that junctions of permutation lines carry $\mathcal G$-charge. This is a fractionalization rather than the background shifts of the higher families of \cite[Section 3.1]{Brennan:2026ira}, for which we have found no finite analogue.

The extension class of Section~\ref{sec:trivial} plays a different role. Invertible quantum lines are one-dimensional representations $\chi$ of $\Gamma$, which relate $\alpha$ and $\alpha\,\chi|_K$, preserving $G_\alpha$, and shift the discrete torsion by $(\chi|_K)\circ c=d(\chi\circ s)$, a coboundary. Universes whose data $(G_\alpha,[\nu_\alpha])$ are not conjugate, in particular universes with different torsion, are related by no invertible object of $\mathcal C$ (Section~\ref{sec:trivial}), only by non-invertible interfaces, as for $\Gamma=D_4$ with $K$ the center (Section~\ref{sec:typesI-II}). The extension class is therefore categorical family data, not an invertible family anomaly.

Table~\ref{tab:family-decomposition-dictionary} summarizes the correspondence.

\begin{table*}[t]
\centering
\small
\setlength{\tabcolsep}{6pt}
\renewcommand{\arraystretch}{1.3}
\begin{tabular}{|p{0.32\textwidth}|p{0.60\textwidth}|}
\hline
\textbf{Family of QFTs} & \textbf{Realization through decomposition} \\ \hline
$G^{(0)}$-equivariant family & Type~I universes, with invertible objects in every $\mathcal C_{ij}$; the family structure is a tensor
functor $\mathrm{Vec}(G)\to\mathcal C$ permuting units. In the SymTFT: one and the same symmetry boundary. \\ \hline
Categorical family with a relation as in \cite[Equation 4.65]{Brennan:2026ira} & Type~IIa: universes related by a simple $X\in\mathcal C_{ij}$ with
$\mathcal C_{ii}\simeq\mathcal C_{jj}$, and $\mathcal S$ is the gauging by $X^\vee\otimes X$. In the SymTFT: an invertible bulk
surface relating the two symmetry boundaries. \\ \hline
Family only in the general sense of \cite[Equation 3.44]{Brennan:2026ira} & Type~IIb: universes related by a gauging that changes the symmetry; boundaries in different
orbits of $\mathrm{Aut}^{br}(\mathcal Z(\mathcal D))$, \eqref{eq:IIa-IIb-symtft}. \\ \hline
Family anomaly & Type~I universes permuted transitively: the monoidal structure of the permuting action shifts the $\mathcal G$-SPT,
\eqref{eq:spt-shift}; it is invisible in $\mathcal C$ itself. \\ \hline
Family over a discrete parameter whose members no topological interface relates & Can be realized by Type~III pairs with a shared SymTFT that no topological interface of $\mathfrak T$ relates, summed by the
flat gauging of a $(-1)$-form symmetry; the center sectors $SO(3)_\pm$ of $SU(2)$ Yang--Mills theory. \\ \hline
(No analogue in \cite{Brennan:2026ira}; $(-2)$-form operation of \cite{Banerjee:2026ywy}) & Type~III pairs with different SymTFTs, such
as universes whose blocks $\mathcal D^{\chi_*\beta}$ are not Morita equivalent, \eqref{eq:cohomological-typeIII}:
$\mathrm{Vec}(\mathbb Z_2)\boxplus\mathrm{Vec}(\mathbb Z_2,\omega)$, $\mathrm{Rep}(D_4)\boxplus\mathrm{Rep}(Q_8)$. \\ \hline
\end{tabular}
\caption{Families of QFTs and their realization through decomposition.  The left column gives families as described by Brennan--Intriligator \cite{Brennan:2026ira}, and the right column lists relations between pairs of elements of those families.}
\label{tab:family-decomposition-dictionary}
\end{table*}

\section{Discussion and outlook}\label{sec:discussion}

In this article, we described finite symmetries of decomposing two-dimensional theories as multi-fusion categories with an action functor. We used this description to classify pairs of universes, to derive the symmetry inherited by orbifolds with trivially acting subgroups, to construct Type~III pairs, and to relate these structures to the families of BI \cite{Brennan:2026ira}. Several directions remain open.

\paragraph{Beyond finite topological defects.} The assumption that $\mathrm{Top}(\mathfrak T)$ is a multi-tensor category excludes theories with infinitely many universes, such as two-dimensional Yang--Mills theory with its spin universes. Since all statements of this article concern only the image of a finite symmetry $(\mathcal C,F)$, the assumption can be weakened to a monoidal $\mathrm{Top}(\mathfrak T)$ with commutative $\mathrm{End}(\mathbbm 1)$. The price is that $\mathbbm 1=\bigoplus_i\mathbbm 1_i$ and the decomposition $L=\bigoplus_{i,j}\mathbbm 1_i\otimes L\otimes\mathbbm 1_j$ of a line become possibly infinite sums, and that arguments such as the spectral one of Section~\ref{sec:dictionary} require duals for the interfaces they use. Even this does not cover decompositions over a continuous parameter, such as those arising in the next paragraph, whose algebra of topological point operators has no primitive idempotents. There, direct sums are replaced by direct integrals and universes by a spectral decomposition, which we leave for future work.

\paragraph{Continuous families.}

In two-dimensional Maxwell theory, there is a $U(1)^{(1)}$ electric center symmetry. Moreover, there is a magnetic $U(1)^{(-1)}$ symmetry. There is a mixed anomaly between these two.
Because of this anomaly, gauging either symmetry extends the other, and the two gaugings below are alternatives.

Gauging the magnetic $(-1)$-form symmetry produces a $\mathbb{Z}^{(1)}$ symmetry, which extends $U(1)^{(1)}$ into an $\mathbb{R}^{(1)}$ symmetry \cite{Perez-Lona:2025add},
\begin{equation}
    0\rightarrow \mathbb{Z}^{(1)} \rightarrow \mathbb{R}^{(1)} \rightarrow U(1)^{(1)} \rightarrow 0 .
\end{equation}
In other words, topologically gauging the $(-1)$-form symmetry modifies the gauge group of the theory, giving an $\mathbb{R}$ gauge theory.

Similarly, gauging the electric $U(1)^{(1)}$ symmetry produces a $\mathbb{Z}^{(-1)}$ symmetry, which extends $U(1)^{(-1)}$ into an $\mathbb{R}^{(-1)}$ symmetry,
\begin{equation}
    0\rightarrow \mathbb{Z}^{(-1)} \rightarrow \mathbb{R}^{(-1)} \rightarrow U(1)^{(-1)} \rightarrow 0 .
\end{equation}
The $\mathbb R$ gauge theory obtained by gauging the magnetic symmetry is the setting of continuous decomposition. It has a continuum of ``universes'', labeled by the real electric flux, and the members of a continuous family are realized through them. In this article, we have restricted ourselves to finite symmetries, and the continuous case lies outside the framework developed here. It is the subject of a second part of this work, in which we study continuous families of theories and their realization through continuous decomposition.

\paragraph{Exact sequences of multi-fusion categories.} In the base-change examples of Section~\ref{sec:cohomological}, in contrast with Types~I and~II, the ``quotient'' map sends the other summands to the zero object rather than to copies of the monoidal unit $\mathbbm 1$. These projections therefore do not define exact sequences in the sense of \cite{Bruguieres2011} or of \cite{etingof2017exact}. It would be interesting to formulate a notion of exactness adapted to decomposable multi-fusion categories that covers these extensions.

\paragraph{Nonabelian trivially acting subgroups.} The derivation of Section~\ref{sec:trivial} extends with primitive central idempotents of $\mathbb C[K]$, equivalently irreducible representations of $K$, in place of characters, the discrete torsions becoming Clifford obstruction classes. Working out the resulting interfaces explicitly is left for future work.

\paragraph{Unitary and non-unitary universes.} The Fibonacci category $\mathrm{Fib}$ and the Yang--Lee category $\mathrm{YL}$ are Galois conjugate: in a suitable gauge their $F$-symbols lie in $\mathbb Q(\sqrt5)$, and base change along $\mathbb Q(\sqrt5)\otimes_{\mathbb Q}\mathbb C\cong\mathbb C\times\mathbb C$ gives $\mathrm{Fib}\boxplus\mathrm{YL}$. Since $U(\mathrm{Fib})$ is trivial (Section~\ref{sec:review}), this is not an instance of \eqref{eq:cohomological-typeIII}. The pair was proposed as a target for $(-2)$-form symmetries in \cite[Section 3.6]{Banerjee:2026ywy}. The two Drinfeld centers are inequivalent, with global dimensions $\big((5\pm\sqrt5)/2\big)^2$, and neither is a Dijkgraaf--Witten theory. A physical realization is left for future work.

\begin{acknowledgments}

Proofreading and presentation refinement were assisted by Anthropic Claude. We used GPT 6.1 Sol to generate figures. However, we are responsible for all results. S.R.~and D.R.~would like to gratefully acknowledge support from the Simons Center for Geometry and Physics, Stony Brook University, at which some of the research for this paper was performed. S.R.~thanks the organizers of Lotus \& Swamplandia 2026 and Global Categorical Symmetries 2026 for the hospitality during part of this work. S.R.~thanks IHP Paris, the Department of Physics at Virginia Tech and the Theory group at UPenn for the hospitality during part of this work. Y.G.~is supported by the UKRI Frontier Research Grant, underwriting the ERC Advanced Grant ``Generalized Symmetries in Quantum Field Theory and Quantum Gravity''.   E.S.~was partially supported by NSF grant PHY-2609668.

\end{acknowledgments}

\appendix

\section{A lightning review of Karoubi completion}
\label{Karoubi}
In this appendix, we briefly review Karoubi completion. For details, see \cite[Section 6]{buehler2009exactcategories}, \cite{etingof2015tensor,Orlov_2011}.

For an arbitrary category $\mathcal{C}$, an idempotent $e: X\to X$ is a morphism such that $e \circ e = e$.
The idempotent $e$ is said to split if there exist an object $Y$ and maps
\begin{equation}
    Y\xrightarrow{\iota}X\xrightarrow{p}Y
\end{equation}
such that 
\begin{equation}
    \iota \circ p=e,\qquad p\circ \iota=\mathrm{id}_Y.
\end{equation}
In other words, the idempotent splits if it can be factored into a retraction ($p$) followed by a section ($\iota$). We call $Y$ a retract of $X$. It plays the role of the image of $e$.

In an arbitrary category, idempotents need not split.
The Karoubi completion builds a larger category where formal images are forced to exist, and idempotents split.

For example, let $\mathcal{C}$ be the category of even-dimensional complex vector spaces and linear maps.
Then $\mathbb{C}^2$ is an object, the morphism
\begin{equation}
    e \: = \:  \left( \begin{array}{cc} 1 & 0 \\ 0 & 0 \end{array} \right) 
\end{equation}
is idempotent, 
but its one-dimensional image is not an object of $\mathcal{C}$.
Karoubi completion repairs this mismatch.

A category in which every idempotent already splits is called idempotent complete, or Karoubian.

The Karoubi envelope ${\rm Kar}(\mathcal{C})$ is a category with
\begin{itemize}
    \item objects $(X,e)$ where $X\in \mathcal{C},\, e\in {\rm End}_{\mathcal{C}}(X),\, e^2=e$,
    \item morphisms $u: (X,e)\to (Y,f)$ are maps $u:X\to Y$ in ${\cal C}$ satisfying $u=f\circ u\circ e$, equivalently $f\circ u=u=u\circ e$. In particular, the identity morphism of $(X,e)$ is $e$.
\end{itemize}

The original category embeds fully faithfully, via
\begin{equation}
J:\mathcal{C}\to {\rm Kar}(\mathcal{C}),\, X\mapsto (X,\mathrm{id}_X),
\end{equation}
so that
\begin{equation}
\mathrm{Hom}_{{\rm Kar}(\mathcal{C})}(JX,JY)=\mathrm{Hom}_{\mathcal{C}}(X,Y).
\end{equation}

Crucially, every idempotent now splits.  For any idempotent $e$,
there are maps
 \begin{equation}
     JX\xrightarrow{e}(X,e)\xrightarrow{e}JX,
 \end{equation}
 whose composite $JX\to JX$ is $e$ and whose composite $(X,e)\to(X,e)$ is $e=\mathrm{id}_{(X,e)}$. Further, it is straightforward to show that 
 \begin{equation}
{\rm Kar}({\rm Kar}(\mathcal{C}))\simeq{\rm Kar}(\mathcal{C}).
\end{equation}

 Now, suppose $\mathcal{C}$ is monoidal. The tensor product extends to the completion as 
 \begin{equation}
   (X,e)\otimes (Y,f)=(X\otimes Y, e\otimes f),  
 \end{equation}
 which is well defined because
 \begin{equation}
     (e\otimes f)^2=e\otimes f.
 \end{equation}
 In the category of topological line defects, an object $X$ corresponds to a line defect. A morphism $a:X\to Y$ corresponds to a topological junction. The idempotent $e: X\to X$ is a projector inserted on $X$. An object $(X,e)$ is the summand line selected by the projector.

 For decomposition, the relevant category is one that contains the topological point operators but in which the unit has not yet been split, such as the $\mathbb C[K]$-linear category of Section~\ref{sec:cohomological}. There, $\mathrm{End}(\mathbbm 1)$ contains orthogonal idempotents
 \begin{equation}
     \Pi_i \Pi_j=\delta_{ij}\Pi_i,\qquad \sum_i \Pi_i=\mathrm{id}_{\mathbbm 1}.
 \end{equation}
 After splitting,
 \begin{equation}
     \mathbbm 1\cong \bigoplus_i \mathbbm 1_i,\qquad \mathbbm 1_i=(\mathbbm 1,\Pi_i).
 \end{equation}
 The $\Pi_i$ are the topological point operators that select the universes, and $\mathbbm 1_i$ is the identity line of universe $i$.

Also, fusion and multi-fusion categories are already idempotent complete. Their abelian, semisimple structures already contain the images of projectors. Hence
 \begin{equation}
     {\rm Kar}(\mathcal{C})\simeq \mathcal{C}
 \end{equation}
 for such a category.

 In a nutshell, Karoubi completion turns an idempotent endomorphism into an actual object representing its image.

\section{Details of derivations}\label{app: details_der.}

In this appendix, we collect some of the details of the derivations used in the main text. 

\subsection{Direct computation for $S_3\supset\mathbb Z_3$}\label{app:S3}

This subsection gives the direct computation of $\mathrm{Bimod}_A(\mathrm{Vec}(\mathbb Z_2))$ for $A=q(\mathbb C[S_3])$ used in Section~\ref{sec:typesI-II}. We first establish the equivalence with the $\mathbb Z_2$-equivariantization of $\mathrm{Bimod}_{A_0}(\mathrm{Vec})$ and classify its simple objects. We then derive the fusion rules and determine the quantum symmetry functor \eqref{eqn:RepS3functor}.

We recast graded $A$-bimodules in terms of bimodules of the even part $A_0$ of $A$. Since the short exact sequence
$\Zth \to S_3 \to \Zt$ splits, the degree-one component $A_1$ contains an involution. Fixing a transposition $\tau \in A_1$ with $\tau^2 = 1$, we can decompose $A$ as $A = A_0 \oplus A_0\tau$, where $A_0 = \C[\Zth]$. Let $\kappa := \mathrm{Ad}_{\tau}|_{A_0}$ denote the induced algebra automorphism, which acts by inversion on $\Zth$. This exhibits $A$ as the crossed product
\begin{equation}\label{eq:crossed}
    A \;\cong\; A_0 \rtimes_\kappa \Zt ,
\end{equation}
equipped with the canonical crossed-product grading in $\mathrm{Vec}(\Z_2)$. This construction is independent of the chosen transposition: any other choice is of the form $c\tau$ with $c\in\Zth$, and $\mathrm{Ad}_c$ is trivial on the commutative algebra $A_0$. The isomorphism~\eqref{eq:crossed} is the fulcrum of the analysis below.

Since $\tau$ is invertible, $A$ is strongly graded, $A_1=A_0\tau=\tau A_0$, and likewise $M_1=M_0\tau=\tau M_0$ for any graded $A$-bimodule $M$. Conjugation by $\tau$ gives an invertible map
\begin{equation}\label{eq:sigma}
    \sigma_M \colon M_0 \longrightarrow M_0, \quad \sigma_M(m) := \tau \cdot m \cdot \tau^{-1},
\end{equation}
which is well-defined on $M_0$ because $\deg(\tau) = 1 \in \Z_2$. Now, notice that
\begin{equation}\label{eq:sigma-equivariance}
    \sigma_M(a \cdot m \cdot b) = \kappa(a) \cdot \sigma_M(m) \cdot \kappa(b), \quad \sigma_M^2 = \mathrm{id}_{M_0},
\end{equation}
which means that $\sigma_M$ defines an isomorphism of $A_0$-bimodules ${}_{\kappa}(M_0)_{\kappa} \xrightarrow{\sim} M_0$ that squares to the identity. But this is precisely a \textit{$\Zt$-equivariant structure} on $M_0$. The ambient $\mathrm{Vec}(\Z_2)$ grading ties the left and right actions together, leaving only this diagonal $\Zt$-action. 

Conversely, given an equivariant $A_0$-bimodule $(M_0, \sigma)$, we can induce the graded $A$-bimodule $A \otimes_{A_0} M_0$. Its grading is inherited from $A$, its left $A$-action is canonical, and the right action of $\tau$ is determined by $(1 \otimes m) \cdot \tau := \tau \otimes \sigma(m)$. The condition $\sigma^2 = \mathrm{id}$ guarantees associativity of the right action, and the multiplication map $a \otimes m \mapsto a \cdot m$ is an isomorphism: it is surjective by strong grading and injective by matching degree-zero components. The assignment $M \mapsto (M_0, \sigma_M)$ thus yields an equivalence of abelian categories
\begin{equation}
    \mathrm{Bimod}_{A}(\mathrm{Vec}(\Z_2)) \;\cong_{\mathrm{ab.cat.}}\; \big(\mathrm{Bimod}_{A_0}(\mathrm{Vec})\big)^{\Zt}.
\end{equation}

We now show that this is furthermore an equivalence of \textit{tensor} categories. For this, we show that it respects the relative tensor product
\begin{equation}
    M \otimes_A N := \frac{M \otimes N}{\left( m\cdot a \otimes n - m \otimes a\cdot n \right)}
\end{equation}
upon restriction to degree zero. Because the relations are homogeneous, $(M \otimes_A N)_0$ is the quotient of $(M_0 \otimes N_0) \oplus (M_1 \otimes N_1)$ by the degree-zero balancing relations. The canonical map $p \colon M_0 \otimes_{A_0} N_0 \to (M \otimes_A N)_0$ is clearly surjective, as the second summand is redundant: for any $m \in M_1$ and $n \in N_1$, we have $m \otimes n = (m\tau^{-1}) \otimes (\tau n)$ with $m\tau^{-1} \in M_0$ and $\tau n \in N_0$. 

To show that $p$ is injective, we define the linear map
\begin{gather}
    r \colon (M_0 \otimes N_0) \oplus (M_1 \otimes N_1) \longrightarrow M_0 \otimes_{A_0} N_0, \\ 
    r|_{M_0 \otimes N_0} = \pi,\quad
    r(m \otimes n) := (m\tau^{-1}) \otimes (\tau n),
\end{gather}
for $\pi: M_0 \otimes N_0\to M_0\otimes_{A_0}N_0$ the quotient map, and verify that it vanishes on all relations $\left(ma \otimes n - m\otimes an\right)$ of degree zero across the four possible degree triples $(|m|, |a|, |n|)$:
\begin{itemize}
    \item $(0,0,0)$: Follows immediately from $A_0$-balancing in $M_0 \otimes_{A_0} N_0$.
    \item $(0,1,1)$: Setting $a = a_0\tau$, we obtain $r(ma \otimes n) = ma_0 \otimes \tau n = m \otimes a_0\tau n = r(m \otimes an)$.
    \item $(1,1,0)$: Setting $a = \tau a_0$ yields $\tau a=\tau^{-1}a=a_0$, since $\tau^2=1$, hence $r(m \otimes an) = m\tau^{-1} \otimes a_0n = m\tau^{-1}a_0 \otimes n = ma \otimes n = r(ma \otimes n)$.
    \item $(1,0,1)$: Using the identity $ma\tau^{-1} = m\tau^{-1}\kappa(a)$, we find $r(ma \otimes n) = m\tau^{-1} \otimes \kappa(a)\tau n = m\tau^{-1} \otimes \tau an = r(m \otimes an)$.
\end{itemize}
Thus $r$ descends to a well-defined inverse to $p$, and so it is an isomorphism
\begin{equation}\label{eq:degree-zero-tensor}
    (M \otimes_A N)_0 \;\cong\; M_0 \otimes_{A_0} N_0 .
\end{equation}
Under this identification, $\sigma_{M \otimes_A N} = \sigma_M \otimes \sigma_N$ because
\begin{equation}
    \tau \cdot (m \otimes n) \cdot \tau^{-1} = (\tau \cdot m) \otimes (n \cdot \tau^{-1}) = (\tau \cdot m \cdot \tau^{-1}) \otimes (\tau \cdot n \cdot \tau^{-1}),
\end{equation}
where the second equality follows by inserting $\tau^{-1}\tau = 1$ and sliding $\tau$ across the balanced tensor product. Because the unit object $A$ maps to $(A_0, \kappa)$, this yields a monoidal equivalence
\begin{equation}  \label{eq:appb:monoidalequiv}
    \mathrm{Bimod}_{A}(\mathrm{Vec}(\Z_2)) \;\simeq\; \big(\mathrm{Bimod}_{A_0}(\mathrm{Vec})\big)^{\Zt},
\end{equation}
identifying the category as a \textit{$\Zt$-equivariantization of a multi-fusion category}.

The simple objects of an equivariantization are induced from the orbits of the action and decorated on the fixed points by the characters of the stabilizer. If $\kappa_*(E)\not\cong E$, the direct sum $E\oplus\kappa_*(E)$ admits a unique equivariant structure up to isomorphism and is simple in $\mathcal C$. If $\kappa_*(E)\cong E$, an isomorphism $u\colon\kappa_*(E)\to E$ can be normalized to $u\circ\kappa_*(u)=\mathrm{id}$, which leaves two choices, the irreducible representations of $\Zt$. Under $E_{ij}\mapsto E_{\bar\imath\bar\jmath}$, there is one fixed point and four free orbits, giving the six simple objects of Section~\ref{sec:typesI-II}.

We now derive the fusion rules summarized in Section~\ref{sec:typesI-II}, using the simple objects listed there.

For the fusion rules, we have that the forgetful functor $\mathcal{C} \to \mathrm{Bimod}_{A_0}(\mathrm{Vec})$ is strong monoidal and faithful, so the underlying object of any tensor product is computed via matrix multiplication in $\mathrm{Mat}_3(\mathrm{Vec})$. This functor is injective on isomorphism classes when restricted to the four free orbits, so that matrix multiplication determines the fusion rules completely whenever the underlying product avoids the fixed point $E_{\one\one}$. When $E_{\one\one}$ does appear, one simply has to keep track of the $\Zt$-action on the corresponding multiplicity space.

All in all, the fusion is:
\begin{equation}
    g \otimes_A g = E_{\zeta\zeta} \oplus E_{\zeta^2\zeta^2} = \mathbbm 1_2, \quad
    X \otimes_A X = 0, \quad
    Y \otimes_A X = \mathbbm 1_2 \oplus g,
\end{equation}
along with $\mathrm{sgn}_1 \otimes_A X = X$, $X \otimes_A g = X$, and $g \otimes_A Y = Y$. The only non-trivial multiplicity occurs in
\begin{equation}
    X \otimes_A Y = (E_{\one\zeta} \otimes_{A_0} E_{\zeta\one}) \oplus (E_{\one\zeta^2} \otimes_{A_0} E_{\zeta^2\one}) \cong E_{\one\one}^{\oplus 2}.
\end{equation}
Here $\tau$ exchanges the two direct summands, making the multiplicity space the regular representation $\C[\Zt] \cong \one \oplus \mathrm{sgn}$. From this we deduce the lifts to the multi-fusion category as $X \otimes_A Y = \mathbbm 1_1 \oplus \mathrm{sgn}_1$.

Finally, we determine the functor \eqref{eqn:RepS3functor}.

To determine functor $\mathcal{F}$, observe that the right action of the idempotents on $V|_{\mathbb{Z}_3} = \bigoplus_{\alpha}V_{\alpha}$ reads $(x\otimes v) e_{\beta} = x e_{\beta\alpha} \otimes v$ for $v\in V_{\alpha}$. Decomposition $\mathbb{C}[S_3] = \mathbb{C}[\mathbb{Z}_3] \oplus \mathbb{C}[\mathbb{Z}_3]\tau$ gives $e_{\alpha} \mathbb{C}[S_3]e_{\beta} = \delta_{\alpha, \beta} \mathbb{C} e_{\alpha} \oplus \delta_{\alpha, \beta^{-1}} \mathbb{C}e_{\alpha} \tau$.
The diagonal block of the gauged universe is
\begin{equation}
    \Pi_R \mathcal{F}(V) \Pi_{R} \cong \mathbb{C}[\mathbb{Z}_2]\otimes V^{\mathbb{Z}_3},
\end{equation}
where $V^{\mathbb{Z}_3}$ is the $\mathbb{Z}_3$-invariant subspace of $V$. 
It is $\mathbb{C}[\mathbb{Z}_2] = \mathbbm 1_1$ for $V=1$; it is $\mathrm{sgn}_1$ for $V=\operatorname{sgn}$; and it vanishes for $V=\rho$.
This implies that there is no $\mathbbm 1_1$ or $\mathrm{sgn}_1$ component of $\mathcal{F}(\rho)$.

Additionally, a degree-zero bimodule map $A \rightarrow \mathcal{F}(V)$ is multiplication by an element $\sum_{x}x\otimes v_x$ supported on $\ker(\operatorname{sgn}) = \mathbb{Z}_3$ and such that $v_{hkh^{-1}}=h\cdot v_k$. Hence
\begin{equation}
    \Hom(\mathbbm 1, \mathcal{F}(V)) = V^{S_{3}} \oplus V^{\mathbb{Z}_3},
\end{equation}
which is of dimensions 2, 1, 0 for $V=1$, $\operatorname{sgn}$, $\rho$, respectively. The first implies the existence of two universes, $\text{End}(\mathbbm 1) = \mathbb{C}^2$, and the last says that $\mathcal{F}(\rho)$ does not have unit in its entries.

Finally, from the fusion rule $\rho\otimes \rho = 1 \oplus \operatorname{sgn} \oplus \rho$, we determine $\mathcal{F}(\rho)$ uniquely: $\rho$ is self-dual, hence $\mathcal{F}(\rho) = m g \oplus nX \oplus n Y$ for some $m,n\in \mathbb{N}$. It fuses with itself as
\begin{equation}
    \mathcal{F}(\rho)\otimes_{A} \mathcal{F}(\rho) = n^2 ( \mathbbm 1_1 \oplus \mathrm{sgn}_1 \oplus g) \oplus (m^2+n^2) \mathbbm 1_2 \oplus mn (X\oplus Y),
\end{equation}
matching against $1 \oplus \operatorname{sgn} \oplus \rho$ forces $m=n=1$.

\subsection{The $\mathrm{Rep}(D_4)$/$\mathrm{Rep}(Q_8)$ example}\label{D4/Q8}

We want to understand the example where we get ${\rm Rep}(D_4), {\rm Rep}(Q_8)$ as two different universes, arising from an extension, see \cite{Banerjee:2026ywy}. Take $\mathcal{D} = \text{Rep}(D_4)$. We start with the following extension
\begin{equation}
    B\mathbb{Z}_2\rightarrow \mathfrak{G}\rightarrow {\rm Rep}(D_4) .
\end{equation}
We are working in a convention where,
\begin{equation}
\begin{aligned}
    \mathrm{TY}(\mathbb{Z}_2^2,\lambda,\tau_+) &\simeq {\rm Rep}(D_4), \\ 
    \mathrm{TY}(\mathbb{Z}_2^2,\lambda,\tau_-) &\simeq {\rm Rep}(Q_8),
\end{aligned}    
\end{equation}
where $\lambda$ is the symmetric nondegenerate bicharacter and $\tau_\pm=\pm\frac12$, whose sign is the Frobenius--Schur indicator of the non-invertible object \cite{Perez-Lona:2023djo}. As a fusion category, ${\rm Rep}(D_4)$ has three (non-trivial) invertible objects 
denoted by $x,y,z$
and there is one non-invertible line, denoted by $m$. The fusion rules are
\begin{equation}
\begin{aligned}
x\otimes y &\cong z,\\
x\otimes m &\cong m\otimes x\cong m,\\ m\otimes m &\cong \textstyle\bigoplus_{a\in\mathbb Z_2^2}a.
\end{aligned}
\end{equation}
If we twist the associator $\alpha$ by $\omega \in H^3(\mathbb{Z}_2,U(1))$, where 
\begin{equation}
    \omega(g,h,k)=(-1)^{ghk},\quad g,h,k\in \mathbb{Z}_2
\end{equation}
then the only modified component is the one involving three copies of the non-invertible line $m$, that is
$$\alpha^\omega_{m,m,m}=-\alpha_{m,m,m}.$$ Equivalently, $\tau$ changes sign, and with it the Frobenius--Schur indicator of the Tambara--Yamagami category. Thus, we have 
\begin{equation}
   {\rm Rep}(D_4)^\omega\cong {\rm Rep}(Q_8) .
\end{equation}

We first extend the morphism spaces by $R$ and denote the extended object by $\widetilde{\mathfrak{G}}$.
\begin{equation}
 \mathrm{Hom}_{\widetilde{\mathfrak{G}}}(X_R,Y_R)=\mathrm{Hom}_{{\rm Rep}(D_4)}(X,Y)\otimes_{\mathbb{C}}R
\end{equation}
where $X_R, Y_R$ are the copies of $X,Y \in {\rm Rep}(D_4)$ viewed in $\widetilde{\mathfrak{G}}$ after extending the Hom-spaces and $R$ is
\begin{equation}
    R=\mathbb{C}[\mathbb{Z}_2]=\mathbb{C}[u]/(u^2-1) .
\end{equation}
The objects and tensor products are the same as in $\mathcal{D}$, the morphisms and their tensor products use multiplication in $R$ \footnote{
(1) A general morphism can be written as $f_0+u f_1,\, f_0,f_1\in \mathrm{Hom}_{\mathcal{D}}(X,Y)$. 
(2) In a $\mathbb{C}$ linear fusion category, the endomorphism algebra of the unit is $\mathbb{C}$, which cannot contain $u$ as an independent generator with $u^2=1$. Extending Hom spaces to $R=\mathbb{C}\{1,u\}$ gives ${\rm End}_{\widetilde{\mathfrak{G}}}(1_R)=R$ where $u$ can twist the associator. 
(3) Note that ${\rm End}_{\widetilde{\mathfrak{G}}}(1_R)=R$\cite[\S 3.1]{etingof2012descentformstensorcategories}, \cite[pp.~3-6]{davydov1997twistingmonoidalstructures}.
}.

We define an $R^\times$-valued $3$-cocycle 
\begin{equation}
    \Omega(g,h,k) := u^{ghk} , \qquad g,h,k\in\mathbb{Z}_2 .
\end{equation}
We now declare the new associator to be
    \begin{equation}\alpha^{\widetilde{\mathfrak{G}}}_{X_g,Y_h,Z_k}=u^{ghk}\alpha^{\mathcal{D}}_{X_g,Y_h,Z_k}.\end{equation}
This shows that the associator is the same except for the case when all three objects are odd
$\alpha^{\widetilde{\mathfrak{G}}}_{m,m,m}=u \alpha^{\mathcal{D}}_{m,m,m}$.

The pentagon identity follows directly using $u^2=1$, which reads
\begin{equation}\Omega(h,k,l)\Omega(g,h+k,l)\Omega(g,h,k)=\Omega(g+h,k,l)\Omega(g,h,k+l) .\end{equation}

Thus $\widetilde{\mathfrak{G}}$ is a genuinely monoidal category, rather than just a set of fusion rules. Also, keeping the same dual object, we can make a choice of the $\mathrm{coev},\mathrm{ev}$ maps to make it rigid.
Thus, $\widetilde{\mathfrak{G}}$ is a free R-linear extension of $\mathcal{D}={\rm Rep}(D_4)$, with objects same as $\mathcal{D}={\rm Rep}(D_4)$, and the Hom-spaces or vector spaces of morphisms are extended from 
$\mathbb{C}$ to $R$; but twist the associator by the 3-cocycle $\Omega$.

Before completion, the sequence is
\begin{equation}
\mathrm{Lin}(B\mathbb{Z}_2)\xrightarrow{\iota}\widetilde{\mathfrak{G}}\xrightarrow{p_+}{\rm Rep}(D_4).
\end{equation}
Here, we have $\iota$ sending the unique object $\star$ of $\mathrm{Lin}(B\mathbb{Z}_2)$ to $1_R$, and the nontrivial automorphism to $u\in \mathrm{Aut}_{\widetilde{\mathfrak{G}}}(1_R)$, and 
\begin{equation}
p_+:\widetilde{\mathfrak{G}}\rightarrow {\rm Rep}(D_4)
\end{equation}
is base change along $\chi_+:R\rightarrow \mathbb{C},\, u\mapsto 1$.

The functor is strong monoidal since $\chi_+(u^{ghk})=1,\, p_+(\alpha^{\widetilde{\mathfrak{G}}})= \alpha^{\mathcal{D}}$.

There is also a second strong monoidal  functor where
\begin{equation}
p_-:\widetilde{\mathfrak{G}}\rightarrow {\rm Rep}(D_4)^\omega
\end{equation} 
is base change along $\chi_-:R\rightarrow \mathbb{C},\, u\mapsto -1$ where $\omega(g,h,k)=(-1)^{ghk}$ and ${\rm Rep}(D_4)^\omega\cong {\rm Rep}(Q_8)$.\footnote{This can be seen more explicitly. We define the base change category $\widetilde{\mathfrak{G}}_{\chi_+}:=\widetilde{\mathfrak{G}}\otimes_{R,\chi_+}\mathbb{C}$ by setting the objects same as $\mathcal{D}$ while the Hom-spaces are $\mathrm{Hom}_{\widetilde{\mathfrak{G}}_{\chi_+}}(X_R,Y_R)=\mathrm{Hom}_{\widetilde{\mathfrak{G}}}(X_R,Y_R)\otimes_{R,\chi_+}\mathbb{C}=\mathrm{Hom}_{\mathcal{D}}(X,Y)$. The associator is also specialized by $u\mapsto 1$, so $\widetilde{\mathfrak{G}}_{\chi_+}\cong {\rm Rep}(D_4)$. Similarly, we get $\widetilde{\mathfrak{G}}_{\chi_-}\cong {\rm Rep}(Q_8)$.}

Now, the multi-fusion category emerges by splitting the unit. The group algebra $R$ has primitive central idempotents \begin{equation}e_+=\frac{1+u}{2},\, e_-=\frac{1-u}{2}.\end{equation} 

Now take the $\mathbb{C}$ linear Karoubi completion \begin{equation}\mathfrak{G}={\rm Kar}(\widetilde{\mathfrak{G}}).\end{equation} Here the unit splits $1_{\mathfrak{G}}=1_+\oplus 1_-,\, 1_{\pm}=(1_R,e_{\pm}\mathrm{id}_{1_R})$. Similarly we have $X=X_+\oplus X_-,\, X_{\pm}=(X_R, e_\pm \mathrm{id}_{X_R})$. 
Since we have $e_\epsilon e_\delta=\delta_{\epsilon,\delta}e_\epsilon$ for $\epsilon,\delta\in \{+,-\}$, we have $\mathrm{Hom}_{\mathfrak{G}}(X_\epsilon, Y_\delta)=\mathrm{Hom}_{{\rm Rep}(D_4)}(X,Y)$ whenever $\epsilon=\delta$, otherwise $0$. Also, we have $X_\epsilon\otimes_{\mathfrak{G}}Y_\delta= (X\otimes Y)_\epsilon$ whenever $\epsilon=\delta$, and $X_\epsilon\otimes_{\mathfrak{G}} Y_\delta=0$ whenever $\epsilon\neq \delta$.

Because $e_{+}e_{-}=0$, the hom-space between the two idempotent sectors is zero. The off-diagonal blocks vanish:
\begin{equation}
    \mathfrak{G}_{+-} = 1_{+}\mathfrak{G}1_{-} = 0, \quad \mathfrak{G}_{-+} = 1_{-}\mathfrak{G}1_{+} = 0.
\end{equation}
This is consistent with the fact that $\text{Rep}(D_4)$ and $\text{Rep}(Q_8)$ are not Morita equivalent \cite[Example 5.4]{Naidu2007LagrangianSA}.
The diagonal blocks have $u=\pm 1$. For the $u=1$ sector
\begin{equation}
    \mathfrak{G}_{++} = 1_{+} \mathfrak{G} 1_{+} \cong \text{Rep}(D_4),
\end{equation}
and for the $u=-1$ sector,
\begin{equation}
    \mathfrak{G}_{--} = 1_{-}\mathfrak{G} 1_{-} \cong \text{Rep}(D_4)^{\omega} \cong \text{Rep}(Q_8).
\end{equation}
Thus after performing the Karoubi completion, we obtain the decomposable multi-fusion category $\mathfrak{G}$:
\begin{equation}
    \mathfrak{G} \cong \text{Rep}(D_4) \boxplus \text{Rep}(Q_8) .
\end{equation}

Now, gauging the $B\mathbb{Z}_2$ 1-form symmetry corresponds to condensing the algebra object $A_\pm=1_\pm$. Thus, condensing along $A_+$ gives us the category \begin{equation}\mathrm{Bimod}_{A_+}(\mathfrak{G})\cong {\rm Rep}(D_4)\end{equation} and condensing along $A_-$ gives us the category 
\begin{equation}
\mathrm{Bimod}_{A_-}(\mathfrak{G})\cong {\rm Rep}(Q_8).
\end{equation}

\section{Quarter gauging in SymTFT}\label{app:qg}

This appendix summarizes the quarter-gauging construction of \cite[Section 4]{Banerjee:2026ywy},
which provides the physical origin of the Type~III pairs of Section~\ref{sec:typeIII}. Starting from a single UV theory, two different relevant deformations produce two IR phases whose effective symmetries are the two summands, $\mathrm{Vec}(\mathbb{Z}_2)$ and $\mathrm{Vec}(\mathbb{Z}_2,\omega)$, together with the totally reflective interface between their SymTFTs.

Consider a three-dimensional
SymTFT $\mathfrak{Z}_G=\mathcal Z(\mathrm{Vec}(G))$ on the slab
$M_3=M_2\times[0,1]$, with a symmetry boundary $\mathfrak B^{\rm sym}$ at one end and a physical boundary $\mathfrak B^{\rm phys}$ at the other. A non-anomalous bulk one-form symmetry may be gauged in a region of this slab. Equivalently, one condenses the corresponding condensable algebra in the Drinfeld center $\mathcal Z(\mathrm{Vec}(G))$. Gauging throughout the right half of the slab gives the familiar club-sandwich interface between two topological orders. The quarter-gauging construction makes the condensation region smaller, so that distinct algebras may be condensed in different parts of the \emph{same} parent bulk.

Let $\mathcal A_H$ and $\mathcal A_K$ be two choices of condensable algebras and $\mathfrak{Z}_H$ and $\mathfrak{Z}_K$ denote their respective reduced topological orders. Condensing these simultaneously in the upper and lower right quarters of $M_3$ creates a scenario where we have  $\mathfrak{Z}_H$ and $\mathfrak{Z}_K$ separated by a topological wall $\mathcal I$ of the SymTFT (Figure~\ref{fig:qg-fusion}), which is not a topological interface of the two-dimensional theory. Orienting the condensation interfaces $\mathcal I_H$ and $\mathcal I_K$ from $\mathfrak{Z}_G$ to $\mathfrak{Z}_H$ and $\mathfrak{Z}_K$, this interface is $\mathcal I=\mathcal I_H^\vee\boxtimes_{\mathfrak{Z}_G}\mathcal I_K$. On the left half of $M_3$, we still have $\mathfrak{Z}_G$. We can view the individual condensation to each half as two distinct renormalization group flows from a $G$-symmetric UV theory to an $H$- or $K$-symmetric IR theory. Now, we can collapse the left half and therefore we end up with a scenario where we have the SymTFTs of $\mathfrak{Z}_H$ and $\mathfrak{Z}_K$ separated by an interface obtained via a common UV origin. 

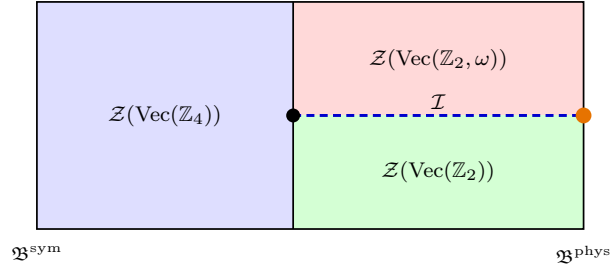
\begin{figure*}[t]
\centering
\begin{tikzpicture}[x=1.13cm,y=1.0cm]
  \fill[qgparent] (0,0) rectangle (3,3);
  \fill[qgupper] (3,1.5) rectangle (6.4,3);
  \fill[qglower] (3,0) rectangle (6.4,1.5);
  \draw[qgline] (0,0) rectangle (6.4,3);
  \draw[qgline] (3,0)--(3,3);
  \draw[qgfused] (3,1.5)--(6.4,1.5);
  \node[qgmeet] at (3,1.5) {};
  \node[qgrg] at (6.4,1.5) {};
  \node[qgtext] at (1.5,1.5)
     {$\mathcal Z(\mathrm{Vec}(\mathbb Z_4))$};
  \node[qgtext] at (4.7,2.24)
     {$\mathcal Z(\mathrm{Vec}(\mathbb Z_2,\omega))$};
  \node[qgtext] at (4.7,.76)
     {$\mathcal Z(\mathrm{Vec}(\mathbb Z_2))$};
  \node[qgtext,anchor=south] at (4.7,1.54) {$\mathcal I$};
  \node[qgsmall,anchor=north] at (0,-.18) {$\mathfrak{B}^{\operatorname{sym}}$};
  \node[qgsmall,anchor=north] at (6.4,-.18) {$\mathfrak{B}^{\operatorname{phys}}$};
\end{tikzpicture}
\caption{Quarter gauging in the $\mathbb Z_4$ SymTFT. Condensation of $\mathbf1\oplus e^2m^2$ in the upper region produces the double semion model, and condensation of $\mathbf1\oplus m^2$ in the lower region produces the toric code. The dashed line denotes the fused interface $\mathcal I$ of \eqref{eq:qg-zfour-wall}, which is totally reflective.}
\label{fig:qg-zfour}
\end{figure*}

Consider the parent symmetry $G=\mathbb Z_4$. Its SymTFT is the $\mathbb Z_4$ quantum double $\mathcal Z(\mathrm{Vec}(\mathbb Z_4))$, whose anyons may be labeled $e^r m^s$ with $r,s\in\mathbb Z_4$. Consider the two condensable algebras
\begin{equation}\label{eq:qg-algebras}
\mathcal A_{-}=\mathbf1\oplus e^2m^2, \qquad \mathcal A_{+}=\mathbf1\oplus m^2.
\end{equation}
Condensing $\mathcal A_{-}$ in the upper-right region produces $\mathcal Z(\mathrm{Vec}(\mathbb Z_2,\omega))$, the double-semion topological order corresponding to an anomalous $\mathbb Z_2$ symmetry. Condensing $\mathcal A_{+}$ in the lower-right region instead produces $\mathcal Z(\mathrm{Vec}(\mathbb Z_2))$, the toric code, whose $\mathbb Z_2$ symmetry is anomaly-free. The resulting configuration is shown in Figure~\ref{fig:qg-zfour}. Its internal interface is the fusion
\begin{equation}\label{eq:qg-zfour-wall}
 \mathcal I=\mathcal I_{e^2m^2}^\vee\boxtimes_{\mathcal Z(\mathrm{Vec}(\mathbb Z_4))}\mathcal I_{m^2}.
\end{equation}
The condensation conditions also explain why this interface is expected to be \emph{totally reflective}. To see the origin of these conditions, recall the cochain description of the parent theory used in \cite{Banerjee:2026ywy}, with bulk action
\begin{equation}\label{eq:qg-parent-action}
 S_{\mathbb Z_4} = \pi \int_{M_3} \left( a_1\cup\delta b_1 + \widehat a_1\cup\delta\widehat b_1- \frac12 a_1\cup\delta\widehat b_1\right).
\end{equation}
The two one-form symmetry generators condensed in the upper and lower quarters are respectively, $\exp\bigl(i\pi\oint(a_1+\widehat b_1)\bigr)$ and $\exp\bigl(i\pi\oint\widehat b_1\bigr)$. The corresponding condensations impose
\begin{equation}\label{eq:qg-conditions}
\begin{aligned}
  \mathcal I_{e^2m^2}:& \quad a_1+\widehat b_1=0, \\
  \mathcal I_{m^2}:& \quad \widehat b_1=0.
\end{aligned}
\end{equation}
Upon bringing the interfaces together, these conditions require
\begin{equation}\label{eq:qg-junction}
 a_1=0,\qquad \widehat b_1=0.
\end{equation}
In particular, the possible two-dimensional interface term proportional to $\int a_1\cup\widehat a_1$ is then expected to vanish. Together with the fact that the toric code and the double semion model are not related by discrete gauging, this motivates the conclusion that the fused wall is \emph{totally reflective}, as the Lagrangian-subgroup argument of Section~\ref{sec:cohomological} confirms, so that no nontrivial anyon is transmitted from one resulting phase to the other.

\section{The higher-charge Schwinger model}\label{app:schwinger}
\noindent\textit{This appendix was written by Xingyang Yu (Virginia Tech, FirstPrinciples).}

The higher-charge Schwinger model gives a Lagrangian realization of universe permutation by an anomalous symmetry. We work with its bosonized realization and the finite bosonic symmetry specified below. At charge three, gauging charge conjugation produces two universes that form a Type~IIa pair relative to this symmetry, realizing the noneffective $S_3$ example of Section~\ref{sec:typesI-II}. We identify both the multi-fusion category and its action, then give its associators. Throughout, the Dirac fermion is massless, the gauge coupling is finite and nonzero, and $\theta=0$. Charge conjugation acts by the ordinary scalar reflection together with universe permutation, with untwisted bosonic gauging and no independently stacked topological theory. We do not classify fermionic defects or other spin refinements.

\subsection*{Universes and the axial symmetry}

Consider two-dimensional $U(1)$ gauge theory with one Dirac fermion of integer charge $q>1$. The electric $\Z_q^{(1)}$ symmetry is generated by a topological point operator $U$ with $U^q=1$. Its primitive projectors are
\begin{equation}
 p_k=\frac1q\sum_{\ell=0}^{q-1}\zeta^{-k\ell}U^\ell,
 \qquad \zeta=e^{2\pi i/q},\qquad k\in\Z_q.
 \label{eq:sch-projectors}
\end{equation}
They obey $p_kp_j=\delta_{kj}p_k$ and $\sum_kp_k=1$, and resolve the theory into $q$ universes \cite{Cherman:2020cvw,Honda:2021ovk,Nguyen:2021yld,Nguyen:2021naa,Cherman:2021nox}. The residual axial symmetry, modulo fermion parity, has a generator $r$ whose mixed anomaly with $\Z_q^{(1)}$ can be written as
\begin{equation}
 rUr^{-1}=\zeta^{-1}U,
 \qquad rp_kr^{-1}=p_{k+1}.
 \label{eq:sch-anomaly}
\end{equation}
The sign of the phase fixes our convention for $r$ and $U$. Thus the axial symmetry permutes the universes, as also follows from the symmetry-operator algebra in \cite{Anber:2018jdf}.

At zero fermion mass, exact bosonization makes both the equivalence of these universes and the symmetry action explicit. A branch of the bosonized Euclidean action is \cite[Section~3.1]{Cherman:2020cvw}
\begin{equation}
 S_n=\int d^2x\left[\frac{(\partial\phi)^2}{8\pi}
 +\frac{g^2}{8\pi^2}(q\phi-\theta-2\pi n)^2\right],
 \label{eq:sch-branches}
\end{equation}
with $(\phi,n)\sim(\phi+2\pi,n+q)$. Here $n$ labels a branch and is not an independently fluctuating integer field. In the universe $k=n\bmod q$, this identification unwraps the scalar. Setting
\begin{equation}
 \sigma=\frac{1}{2\sqrt\pi}
 \left(\phi-\frac{\theta+2\pi n}{q}\right)
\end{equation}
gives the same massive noncompact scalar theory $\mathcal Q$ in every universe:
\begin{equation}
 S_{\mathcal Q}=\frac12\int d^2x\,
 \left[(\partial\sigma)^2+\mu^2\sigma^2\right],
 \qquad \mu=\frac{qg}{\sqrt\pi}.
 \label{eq:sch-scalar}
\end{equation}
The nonzero scalar mass is generated by the Schwinger mechanism. We retain this finite-coupling theory, including its local excitations.

At $\theta=0$, charge conjugation $C$ and the axial generator act by
\begin{equation}
 \begin{aligned}
 r:(\phi,n)&\longmapsto(\phi+2\pi/q,n+1),\\
 C:(\phi,n)&\longmapsto(-\phi,-n).
 \end{aligned}
 \label{eq:sch-branch-action}
\end{equation}
In the variables $(k,\sigma)$ these become
\begin{equation}
 \begin{aligned}
 r:(k,\sigma)&\longmapsto(k+1,\sigma),\\
 C:(k,\sigma)&\longmapsto(-k,-\sigma).
 \end{aligned}
 \label{eq:sch-action}
\end{equation}
Hence, $C$ includes the internal reflection $L:\sigma\mapsto-\sigma$ of $\mathcal Q$ as well as a permutation of labels. The line $L$ is a non-anomalous $\Z_2$ symmetry defect and acts nontrivially on local observables, including the electric field. This internal action is essential for identifying the result of gauging.

Before gauging, the axial generator has only off-diagonal blocks:
\begin{equation}
 \begin{aligned}
 p_i r p_j&=0\quad\text{unless }i=j+1\pmod3,\\
 r&=E_{10}\oplus E_{21}\oplus E_{02}\quad(q=3).
 \end{aligned}
 \label{eq:sch-axial-blocks}
\end{equation}
Here $E_{ij}$ is the identity-gluing interface from the scalar in universe $j$ to that in universe $i$, defined algebraically below. These invertible interfaces make the three universes Type~I relative to the parent symmetry. The induced defect after gauging need not remain off diagonal.

\subsection*{Charge-conjugation gauging at charge three}

Set $q=3$. The action of $C$ on the universe labels has a fixed orbit $\{0\}$ and a free orbit $\{1,2\}$. Gauging the stabilizer on each orbit therefore gives
\begin{equation}
 \mathfrak T_3/\Z_2^C\cong
 (\mathcal Q/\Z_2^L) \sqcup \mathcal Q,
 \label{eq:sch-gauged-universes}
\end{equation}
where $\mathfrak T_3$ denotes the charge-three Schwinger model.
We denote these two universes by $\mathcal T_1=\mathcal Q/\Z_2^L$ and $\mathcal T_2=\mathcal Q$, matching the labels $\mathbbm 1_1$ and $\mathbbm 1_2$ of Table~\ref{tab:S3-fusion}. The fixed orbit retains the reflection orbifold, whereas the free orbit has trivial stabilizer. We normalize the three scalar copies identically and use the standard finite-group orbifold normalization. The comparison with the noneffective $S_3$ orbifold then involves a common Euler counterterm, computed below.

To specify the finite symmetry being computed, resolve the axial and charge-conjugation defects by the projectors. With indices in $\Z_3$, define
\begin{equation}
 E_{ij}=p_i r^{i-j}p_j,\qquad
 L_j=p_j r^{2j}Cp_j.
\end{equation}
The first defect identifies the scalar in universes $j$ and $i$, and the second reflects the scalar in universe $j$. Their products obey
\begin{equation}
 \begin{aligned}
 E_{ij}E_{k\ell}&=\delta_{jk}E_{i\ell},\\
 L_iE_{ij}&=E_{ij}L_j,\qquad L_j^2=p_j.
 \end{aligned}
\end{equation}
The category generated by these defects and their summands is therefore
\begin{equation}
 \mathcal C_{\rm pre}\simeq\Mat_3(\mathcal D),
 \qquad \mathcal D=\Vect(\Z_2),
 \label{eq:sch-parent-category}
\end{equation}
with eighteen simples $E_{ij}L_j^\epsilon$, $\epsilon=0,1$. The untwisted associator follows from identity gluing between the scalar copies and the ordinary reflection action within each copy.

The algebra implementing $C$ gauging is $\mathcal A=\one\oplus C$. In the order $0,1,2$, it has the matrix presentation
\begin{equation}
 \mathcal A=
 \begin{pmatrix}
 \one\oplus L&0&0\\
 0&\one&L\\
 0&L&\one
 \end{pmatrix},
 \qquad R=\one\oplus L\in\mathcal D,
 \label{eq:sch-gauging-algebra}
\end{equation}
with multiplication induced by $C^2=\one$. Its fixed block is $R$. Conjugating the free-orbit block by $\mathrm{diag}(\one,L)$ gives the ordinary two-by-two matrix algebra, which is Morita equivalent to $\one$. The resulting category of defects is thus
\begin{equation}
 \begin{aligned}
 \mathcal C_{\rm Sch}
 &=\Bimod_{\mathcal A}\bigl(\Mat_3(\mathcal D)\bigr) , 
 \\
 &\simeq\Bimod_{R\oplus\one}(\mathcal D).
 \end{aligned}
 \label{eq:sch-category}
\end{equation}
The tensor products here are relative tensor products over the indicated algebras. To specify the physical action, let $\Phi_{\mathcal Q}:\mathcal D\to\mathrm{Top}(\mathcal Q)$ send the odd simple to the actual reflection defect $L:\sigma\mapsto-\sigma$, with multiplication junction given by ordinary composition, $L^2=\one$. Put $A_1=R$ and $A_2=\one$. An $(A_i,A_j)$-bimodule is realized as an interface from $\mathcal Q/A_j$ to $\mathcal Q/A_i$ by applying $\Phi_{\mathcal Q}$ to its strands and gauging the algebra networks on its two sides. Bimodule intertwiners give the physical junctions. The tensorator is the relative-tensor-product projector. On the $R$ side, it uses the separability element
\begin{equation}
 e_R(\one)=\tfrac12(\one\otimes\one+L\otimes L).
\end{equation}
This defines a monoidal action $F_{\rm link}$ of $\Bimod_{R\oplus\one}(\mathcal D)$ on $(\mathcal Q/\Z_2^L)\sqcup\mathcal Q$, including its junction data. On the Schwinger side, retaining universe labels $0$ and $1$ gives the full corner $\mathrm{diag}(R,\one)$ of \eqref{eq:sch-gauging-algebra}. The omitted label $2$ is connected to $1$ by $L$. Removing this redundant sheet gives the equivalence $E_{\rm Sch}$ of \eqref{eq:sch-category}, with $F_{\rm Sch}\cong F_{\rm link}\circ E_{\rm Sch}$ for the action functor $F_{\rm Sch}:\mathcal C_{\rm Sch}\to\mathrm{Top}(\mathfrak T_3/\Z_2^C)$ obtained by gauging.

For comparison, specify an $S_3$ action on $\mathcal Q$ through the sign quotient $s:S_3\to\Z_2$, with each transposition acting by the same $\Phi_{\mathcal Q}(L)$ and the normal $\Z_3$ acting trivially. Let $A_{S_3}=s_*\C[S_3]\in\mathcal D$, as in Appendix~\ref{app:S3}. The Fourier idempotents $e_0,e_1,e_2$ of the normal $\Z_3$ (the $e_{\one},e_\zeta,e_{\zeta^2}$ of Section~\ref{sec:typesI-II}) obey $\tau e_k\tau^{-1}=e_{-k}$ for a transposition $\tau$. Thus $e=e_0+e_1$ is full and
\begin{equation}
 eA_{S_3}e\cong R\oplus\one.
\end{equation}
The corresponding physical corner has the same reflection and multiplication junctions as above, so its Morita equivalence $E_{S_3}$ also satisfies $F_{S_3}\cong F_{\rm link}\circ E_{S_3}$. After the orbit/Fourier identification of the target theories and the Euler normalization below, the equivalence
\begin{equation}
 E_{S_3}^{-1}E_{\rm Sch}:\mathcal C_{\rm Sch}
 \simeq\Bimod_{A_{S_3}}\bigl(\Vect(\Z_2)\bigr)
 \label{eq:sch-s3-equivalence}
\end{equation}
therefore intertwines their physical actions. This realizes the noneffective $S_3$ decomposition \cite{Robbins:2022wlr} for the scalar theory \eqref{eq:sch-scalar}. The use of the genuine reflection is essential: replacing $\Phi_{\mathcal Q}(L)$ by the identity would instead give $\mathcal Q/\Z_2^{\rm trivial}$ on the fixed orbit. Abstract equivalence of the defect categories alone would not establish the claimed realization.

There are six simple defects. In the notation of Table~\ref{tab:S3-fusion}, $\one_1,\mathrm{sgn}_1$ lie in universe $\mathcal T_1$, $\one_2,g$ lie in universe $\mathcal T_2$, and $X\in\mathcal C_{12}$, $Y=X^\vee\in\mathcal C_{21}$ are the two oriented orbifold interfaces. Here $\mathrm{sgn}_1$ is the quantum $\Z_2$ line and $g=L$ is the scalar reflection. In particular,
\begin{equation}
 XY=\one_1\oplus\mathrm{sgn}_1,\qquad
 YX=\one_2\oplus g.
 \label{eq:sch-interface-fusion}
\end{equation}
The diagonal blocks are $\Rep(\Z_2)$ and $\Vect(\Z_2)$, which are tensor-equivalent, while neither off-diagonal simple is invertible. The universes therefore form a Type~IIa pair relative to this symmetry. Their interfaces have balanced Frobenius--Perron dimension $\sqrt2$. The quantum $\Rep(S_3)$ action has the object images of \eqref{eqn:RepS3functor}. In particular, its two-dimensional representation maps to $g\oplus X\oplus Y$.

The induced axial defect can also be computed directly. Write $e_{\rm Sch}=\mathrm{diag}(\one,\one,0)$ for the preceding Schwinger corner. Free induction gives the $\mathcal A$-bimodule $\mathcal N=\mathcal A\otimes r\otimes\mathcal A$, whose compressed underlying matrix and regular one-sided $R$ actions are
\begin{equation}
 e_{\rm Sch}\mathcal N e_{\rm Sch}
 =\begin{pmatrix}0&R\\R&L\end{pmatrix}.
\end{equation}
Hence
\begin{equation}
 \begin{aligned}
 \mathcal N&\cong X\oplus Y\oplus g,\\
 \mathcal N^2&\cong(\one_1\oplus\one_2)
 \oplus(\mathrm{sgn}_1\oplus\one_2)\oplus\mathcal N.
 \end{aligned}
 \label{eq:sch-induced-axial}
\end{equation}
The $\mathcal T_2$-to-$\mathcal T_2$ path first moves $1\to2$ by $r$ and returns $2\to1$ by the gauged $C$ identification, which reflects $\sigma$, so its image is $g=L$. This reproduces the fusion of the two-dimensional $\Rep(S_3)$ image and exhibits its diagonal summand explicitly.

For clarity, the Euler normalization can be determined independently. On a connected closed surface of genus $h$, let $Z_{\mathcal Q}[b]$ denote the scalar partition function in reflection background $b\in H^1(\Sigma,\Z_2)$. The three-copy theory has partition function $3Z_{\mathcal Q}[0]$ for $b=0$, and $Z_{\mathcal Q}[b]$ otherwise, since only the fixed universe label survives a nontrivial background. Thus
\begin{equation}
 \begin{aligned}
 Z_{\mathfrak T_3/\Z_2^C}
 &=\tfrac12\left(3Z_{\mathcal Q}[0]+\sum_{b\ne0}Z_{\mathcal Q}[b]\right) ,
 \\
 &=Z_{\mathcal Q/\Z_2^L}+Z_{\mathcal Q}.
 \end{aligned}
\end{equation}
A prescribed sign background has $3^{2h}$ flat $S_3$ lifts if $b=0$ and $3^{2h-1}$ if $b\ne0$: in the latter case the surface relation imposes one nonzero linear condition on the $2h$ normal-$\Z_3$ holonomies. Division by $|S_3|=6$ gives
\begin{equation}
 Z_{\mathcal Q/S_3}=3^{2h-2}Z_{\mathfrak T_3/\Z_2^C}
 =3^{-\chi(\Sigma)}Z_{\mathfrak T_3/\Z_2^C}.
 \label{eq:sch-euler-normalization}
\end{equation}
Multiplication of $Z_{\mathcal Q/S_3}$ by $3^{\chi(\Sigma)}$ matches these normalizations. This is a common Euler counterterm, not a change of the monoidal associator. The comparison uses the stated scalar-orbifold normalizations. An absolute normalization inherited from the microscopic continuous-gauge path integral must be fixed separately.

\subsection*{Junctions and associators}

For completeness, we give explicit associators in a fixed junction basis. Use the monoidal model \eqref{eq:appb:monoidalequiv},
\begin{equation}
 \mathcal C_{\rm Sch}\simeq
 \bigl(\Mat_3(\Vect)\bigr)^{\Z_2},
 \qquad \kappa_*(E_{ij})=E_{\bar i\bar j},
\end{equation}
where $\bar0=0$, $\bar1=2$, and $\bar2=1$. Write
\begin{equation}
 (a,b,c,d,x,y)
 =(\one_1,\mathrm{sgn}_1,\one_2,g,X,Y).
\end{equation}
Their equivariant matrix representatives are
\begin{equation}
 \begin{aligned}
 a&=(E_{00},+),& b&=(E_{00},-),\\
 c&=E_{11}\oplus E_{22},& d&=E_{12}\oplus E_{21},\\
 x&=E_{01}\oplus E_{02},& y&=E_{10}\oplus E_{20}.
 \end{aligned}
\end{equation}
Each free orbit carries the swap equivariant structure. This is a tensor-category presentation. Its physical action is fixed by the reflection and gauging construction above.

Take matrix paths $|ijk\rangle$ as an orthonormal basis of tensor products. For the two fixed-output channels in $xy$, choose the junction embeddings
\begin{equation}
 v_{xy}^{a}=\frac{|010\rangle+|020\rangle}{\sqrt2},\qquad
 v_{xy}^{b}=\frac{|010\rangle-|020\rangle}{\sqrt2}.
 \label{eq:sch-junction-basis}
\end{equation}
For a free-output channel, put coefficient $+1$ on the lexicographically first endpoint component and determine the other coefficient by equivariance. For a one-path fixed-output channel, take coefficient $+1$. These rules specify all trivalent junctions, since every allowed binary fusion has multiplicity one.

Define $[F^{uvw}_{z}]_{f,e}$ by expanding the left-associated embedding through $e\subset uv$ in the right-associated basis through $f\subset vw$. Thus rows are right channels and columns are left channels. The complete list of nonidentity blocks in this gauge is
\begin{equation}
 F^{bxd}_{x}=F^{dyb}_{y}=F^{xdy}_{b}=F^{ybx}_{d}=-1,
 \label{eq:sch-f-signs}
\end{equation}
and
\begin{equation}
 F^{xyx}_{x}=F^{yxy}_{y}
 =\frac1{\sqrt2}
 \begin{pmatrix}1&1\\1&-1\end{pmatrix}.
 \label{eq:sch-f-fourier}
\end{equation}
For $F^{xyx}_{x}$, order the left channels as $(a,b)$ and the right channels as $(c,d)$. For $F^{yxy}_{y}$, use $(c,d)$ and $(a,b)$, respectively. All other admissible blocks are the scalar $+1$.

The matrices in \eqref{eq:sch-f-fourier} are the $\Z_2$ Fourier transform: one basis resolves a pair of paths into its even and odd combinations, while the other distinguishes equal and unequal intermediate indices. The signs in \eqref{eq:sch-f-signs} express the pairing of the nontrivial character with the nontrivial group element. Since these matrices are changes of decomposition basis in the equivariant matrix category, they satisfy the pentagon identities. They make explicit the associator data in \eqref{eq:sch-s3-equivalence}.

Finally, a generic fixed nonzero fermion mass breaks the axial symmetry used in \eqref{eq:sch-anomaly}. Axial transformations can still relate theories with different mass phases, but this is a statement about a family of theories. The six-object symmetry computed here applies to the massless-fermion model specified above. Likewise, taking the strict infrared limit can make the scalar reflection ineffective, so the finite-coupling action functor must be retained when interpreting \eqref{eq:sch-category}.

\bibliographystyle{apsrev4-1}
\bibliography{refs}


\end{document}